\documentclass[11pt]{article}

\usepackage{subfig}
\usepackage{multirow}
\usepackage{makecell}
\usepackage[singlelinecheck=false]{caption}
\usepackage{cite}

\newsavebox{\foobox}

\newcommand{\mpi}{\pi}
\newcommand{\mdelta}{\delta}

\usepackage[left=2cm, right=2cm, top=2.5cm, bottom=2.5cm]{geometry}
\usepackage{booktabs}

\usepackage[x11names]{xcolor}
\usepackage{fancyhdr, amssymb, cancel, amsmath, graphicx, pgfplots, tikz}
\usepackage{isomath}

\usetikzlibrary{shadows}

\newcommand{\stylecolor}{blue!50!black}

\usepackage[labelfont={bf,sf, color=\stylecolor}, margin={1.5cm,0cm}]{caption}

\usepackage[colorlinks=true, urlcolor=\stylecolor!70!white, linkcolor=\stylecolor, citecolor=\stylecolor!70!white, hyperindex=true, linktocpage=true]{hyperref}

\usepackage[explicit]{titlesec}

\newcommand*\sectionlabel{}
\titleformat{\section}
  {\gdef\sectionlabel{}
   \Large\bfseries\scshape}
  {\gdef\sectionlabel{\thesection }}{0pt}
  {\begin{tikzpicture}[remember picture,overlay]
       \end{tikzpicture}
  }
\titlespacing*{\section}{0pt}{0pt}{0pt}

\newcommand*\subsectionlabel{}
\titleformat{\subsection}
  {\gdef\subsectionlabel{}
   \large\bfseries\scshape}
  {\gdef\subsectionlabel{\thesubsection  }}{0pt}
  {\begin{tikzpicture}[remember picture]
    	\draw (-0.15, 0) node[left] {\color{\stylecolor} \textsf{\subsectionlabel}};
	\draw (0.15, 0) node[right] {\color{\stylecolor} \textsf{#1}};
	\fill[color=\stylecolor] (-0.05, -0.23) rectangle (0.05, 0.23);
       \end{tikzpicture}
  }
\titlespacing*{\subsection}{-4pt}{10pt}{0pt}

\newcommand*\subsubsectionlabel{}
\titleformat{\subsubsection}
  {\gdef\subsubsectionlabel{}
   \bfseries\scshape}
  {\gdef\subsubsectionlabel{\thesubsubsection.\ \  }}{0pt}
  {\begin{tikzpicture}[remember picture]
    	\draw (0, 0) node[left] {\color{\stylecolor} \textsf{\subsubsectionlabel}};
	\draw (0, 0) node[right] {\color{\stylecolor} \textsf{#1}};
       \end{tikzpicture}
  }
\titlespacing*{\subsubsection}{-4pt}{7pt}{0pt}

\pgfplotsset{every axis legend/.append style={at={(1.02,1)},anchor=north west}}

\begin{document}

\allowdisplaybreaks

\pagestyle{fancy}
\renewcommand{\headrulewidth}{0pt}
\fancyhead{}

\fancyfoot{}
\fancyfoot[C] {\textsf{\textbf{\thepage}}}

\begin{equation*}
\begin{tikzpicture}
\draw (\textwidth, 0) node[text width = \textwidth, right] {\color{white} easter egg};
\end{tikzpicture}
\end{equation*}

\begin{equation*}
\begin{tikzpicture}
\draw (0.5\textwidth, -3) node[text width = \textwidth] {\huge  \textsf{\textbf{Electron hydrodynamics 
 with a polygonal Fermi \\\vspace{0.07 in} surface}} };
\end{tikzpicture}
\end{equation*}
\begin{equation*}
\begin{tikzpicture}
\draw (0.5\textwidth, 0.1) node[text width=\textwidth] {\large \color{black}  \textsf{Caleb Q. Cook and Andrew Lucas}};
\draw (0.5\textwidth, -0.5) node[text width=\textwidth] {\small\textsf{Department of Physics, Stanford University, Stanford, CA 94305, USA}};
\end{tikzpicture}
\end{equation*}
\begin{equation*}
\begin{tikzpicture}
\draw (0, -12.6) node[right, text width=0.5\paperwidth] {\texttt{calebqcook@gmail.com}};
\draw (0, -13.1) node[right, text width=0.5\paperwidth] {\texttt{ajlucas@stanford.edu}};
\draw (\textwidth, -13.1) node[left] {\textsf{\today}};
\end{tikzpicture}
\end{equation*}
\begin{equation*}
\begin{tikzpicture}
\draw[very thick, color=\stylecolor] (0.0\textwidth, -5.75) -- (0.99\textwidth, -5.75);
\draw (0.12\textwidth, -6.25) node[left] {\color{\stylecolor}  \textsf{\textbf{Abstract:}}};
\draw (0.53\textwidth, -6) node[below, text width=0.8\textwidth, text justified] {\small Recent experiments have observed hints of hydrodynamic electron flow in a number of materials, not all of which have an isotropic Fermi surface.  We revisit these experiments in $\mathrm{PdCoO}_2$, a quasi-two-dimensional material whose Fermi surface is a rounded hexagon, and observe that the data appears quantitatively consistent with a non-hydrodynamic interpretation.  Nevertheless, motivated by such experiments, we develop a simple model for the low temperature kinetics and hydrodynamics of a two-dimensional Fermi liquid with a polygonal Fermi surface.  A geometric effect leads to a finite number of additional long-lived quasihydrodynamic ``imbalance" modes and corresponding qualitative changes in transport  at the ballistic-to-hydrodynamic crossover.  In the hydrodynamic limit, we find incoherent diffusion and a new dissipative component of the viscosity tensor arising from the explicit breaking of rotational invariance by the Fermi surface.  Finally, we compute the conductance of narrow channels across the ballistic-to-hydrodynamic crossover and demonstrate a modification of the Gurzhi effect that allows for non-monotonic temperature and width dependence in the channel conductance.};
\end{tikzpicture}
\end{equation*}

\tableofcontents

\begin{equation*}
\begin{tikzpicture}
\draw[very thick, color=\stylecolor] (0.0\textwidth, -5.75) -- (0.99\textwidth, -5.75);
\end{tikzpicture}
\end{equation*}

\titleformat{\section}
  {\gdef\sectionlabel{}
   \Large\bfseries\scshape}
  {\gdef\sectionlabel{\thesection }}{0pt}
  {\begin{tikzpicture}[remember picture]
	\draw (0.2, 0) node[right] {\color{\stylecolor} \textsf{#1}};
	\draw (0.0, 0) node[left, fill=\stylecolor,minimum height=0.27in, minimum width=0.27in] {\color{white} \textsf{\sectionlabel}};
       \end{tikzpicture}
  }
\titlespacing*{\section}{0pt}{20pt}{5pt}

\section{Introduction to Electronic Hydrodynamics}
In recent years, experiments have uncovered evidence for the hydrodynamic flow of electrons in ultra-pure metals \cite{molenkamp, bandurin, crossno, mackenzie, levitov1703, felser,bakarov, bandurin18}.  Electron hydrodynamic flow occurs when momentum-relaxing collisions of electrons with impurities, phonons, or other electrons are significantly slower than momentum-conserving electron-electron collisions \cite{gurzhi}: see the recent review \cite{lucasreview17}.  A number of recent theoretical works have attempted to understand the consequences of such hydrodynamic flow \cite{hkms, andreev, succiturb, tomadin, lucas3, scaffidi, hartnoll1704}, which include negative nonlocal resistance \cite{polini, levitovhydro, torre, levitov1806} and super-ballistic flows in narrow constrictions \cite{levitov1607}.  Furthermore, hydrodynamics has been proposed as a sensible mechanism to explain existing transport mysteries in experiments: viscous effects \cite{alekseev} may explain negative magnetoresistance in GaAs \cite{kwwest}, and mysterious $T^2$ resistivity in low density $\mathrm{SrTiO}_3$ \cite{behnia,stemmer} may be explained by hydrodynamic flows through certain inhomogeneous media \cite{lucasRFB}.

Much of the existing work on electron hydrodynamics focuses on simple models with rotationally-invariant Fermi surfaces.   However, some of the metals in which evidence for hydrodynamic electron flow has been observed, such as $\mathrm{PdCoO}_2$ \cite{ong2010, mackenzie16} and $\mathrm{WP}_2$ \cite{felser}, have Fermi surfaces that are highly anisotropic. In principle, if the electronic mean free path were infinitesimally small, the equations governing electronic transport would simply be hydrodynamic equations with the same symmetry as the Fermi surface. In practice, the strongest evidence for electron hydrodynamics comes from experiments with materials in which momentum-conserving scattering rates are not parametrically larger than ballistic scattering rates (e.g. boundary scattering in narrow channels). Accurately modelling transport in such systems therefore requires consideration of both ballistic and hydrodynamic effects \cite{lucasreview17}. 

The purpose of this paper is to give a simple and experimentally motivated example of how the physics at the ballistic-to-hydrodynamic crossover can depend sensitively on the Fermi surface.  We will consider two-dimensional metals with simply connected Fermi surfaces that are ``reasonably well" approximated by regular polygons. We will also assume inversion symmetry, which requires that the Fermi surface be symmetric and hence excludes polygons with an odd number of sides; we will therefore restrict our discussion to \emph{even-sided} polygonal Fermi surfaces.\footnote{The physically relevant polygons are squares and hexagons, which are crystallographically allowed Fermi surfaces.}  The flatness of the Fermi surface along each edge of the polygon has dramatic consequences and leads to a ``quasihydrodynamic" \cite{lucas1810} regime with new long-lived ``imbalance modes".  This quasihydrodynamic regime can modify or destroy the signatures of hydrodynamics at the ballistic-to-hydrodynamic crossover that occur in metals with circular Fermi surfaces. 

For example, the Gurzhi effect \cite{gurzhi} predicts that, for hydrodynamic electron flow in a narrow channel with a circular Fermi surface, the low-temperature channel conductance increases monotonically with temperature and scales with the cube of the channel width; by contrast, we find via numerical solutions of the Boltzmann equation with a polygonal Fermi surface a strong modification of the Gurzhi effect that allows for \emph{non-monotonic} dependence of the channel conductance on the temperature and channel width. In addition, we find unexpected new hydrodynamic effects in our model. Most importantly, we find a new kind of dissipative viscosity, which we call ``rotational viscosity," that opposes rotations of the fluid and arises from the \emph{explicitly} broken rotational invariance of a polygonal Fermi liquid.  We note that this rotational viscosity is distinct from both the non-dissipative Hall viscosity \cite{avron} and the additional viscosities in liquid crystals which \emph{spontaneously} break rotational invariance.  We also find that a polygonal Fermi liquid exhibits incoherent conductivity \cite{hartnoll1}, which arises from the broken Galilean invariance of the polygonal Fermi surface and allows for a charge current to flow even in the absence of momentum.

Our study of hydrodynamics with a polygonal Fermi surface was initially motivated by a recent experiment \cite{mackenzie} on  $\mathrm{PdCoO}_2$, a quasi-two-dimensional material with a rounded hexagon Fermi surface \cite{ong2010}.  This experiment \cite{mackenzie} studied electronic transport in a narrow channel and reported some signatures of hydrodynamic electron flow but not others. We begin with a discussion of this experimental data in Section \ref{sec:exp}, where we argue that all temperature dependence in the data is consistent with conventional ohmic and ballistic effects.  Such considerations call into question a hydrodynamic interpretation of the $\mathrm{PdCoO}_2$ channel flow data, but leave open the theoretical question of a true ballistic-to-hydrodynamic crossover in a material with an anisotropic, e.g. polygonal, Fermi surface.   In Sections \ref{sec:kinetic} and \ref{sec:hydro}, we will develop the  kinetic theory and hydrodynamics of electrons with a polygonal Fermi surface, and discuss elementary properties of the resulting fluid.   Finally, in Section \ref{sec:channel} we solve the Boltzmann equation for these polygonal Fermi liquids in a narrow channel, as studied experimentally in \cite{mackenzie}. 

\section{Electron Hydrodynamics in Delafossites?}
\label{sec:exp}

\begin{figure}[t]
\centering
\subfloat[Reported channel conductance $G$ for $\mathrm{PdCoO}_2$ as a function of temperature $T$ and channel width $w$. We have divided the conductance $G$ by $w^2$ in an attempt to highlight the putative crossover from ballistic transport ($G\sim w^2$) to hydrodynamic transport ($G\sim w^3$).]{
    \includegraphics[width=.48\textwidth]{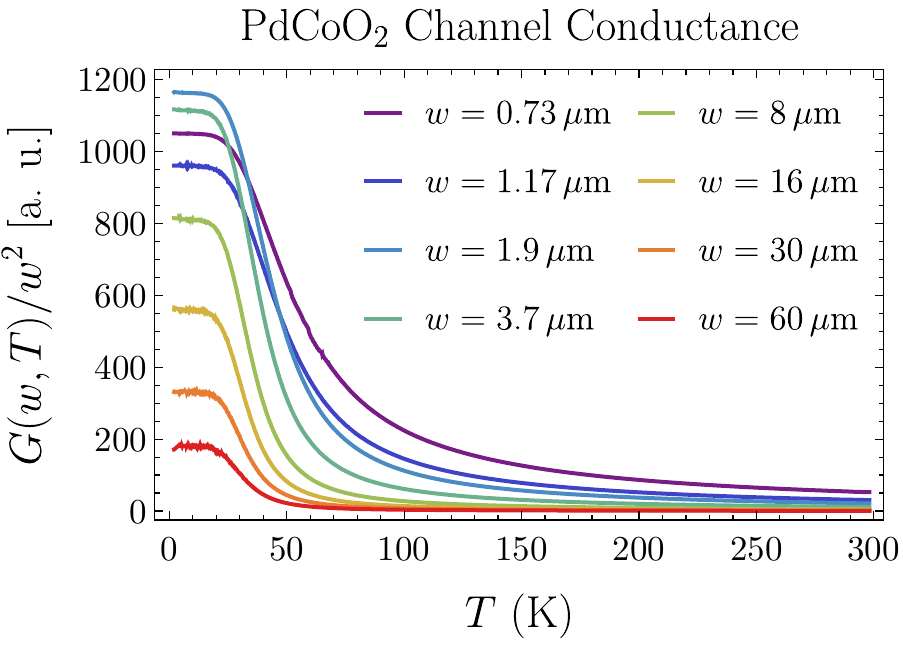}
    \label{fig:channelCon}
}
\subfloat[Comparison of PdCoO$_2$ channel conductance data normalized by its zero-temperature value (solid) to the simple model given by Eq. (\ref{eq:GLfit}), with a single parameter $A$ fit over all curves, i.e. independently of temperature and channel width (dotted). Data for larger channel widths not shown for ease of comparison.]{
    \includegraphics[width=.48\textwidth]{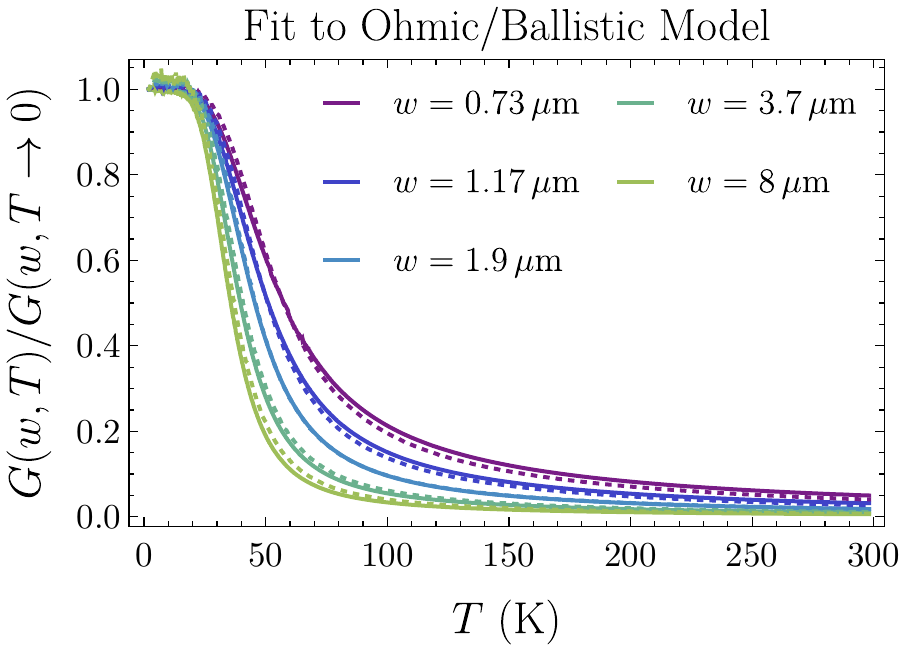}
    \label{fig:ohmBallistic}
}
\caption{Electrical conductance of narrow channels of $\mathrm{PdCoO}_2$.  Experimental data is plotted with permission using the full data set of \cite{mackenzie}.}
\label{fig:data}
\end{figure}

We begin by revisiting the evidence for hydrodynamic electron flow in the delafossite $\mathrm{PdCoO}_2$ \cite{mackenzie}. In Figure \ref{fig:channelCon}, we plot the full dependence of the channel conductance $G$ on the channel width $w$ and temperature $T$ in narrow channels of $\mathrm{PdCoO}_2$, as reported in \cite{mackenzie}.

We first check if the temperature dependence of the channel conductance data can be explained by ``conventional" momentum-relaxing processes. In this effort, we apply the Matthiesen rule and add the resistances due to ohmic scattering in the bulk and ballistic scattering at the channel walls, yielding the simple model
\begin{equation}
    \frac{1}{GL} = \frac{\rho_{\mathrm{bulk}}(T)}{w}+ \frac{A}{w^2}  
    \label{eq:GLfit}
\end{equation}
for the inverse channel conductance (in the above, $L$ is the length of the channel, assumed to be much larger than any other length scale in the problem). The first term (\ref{eq:GLfit})  represents the resistance arising from ohmic momentum-relaxing processes occurring in the middle of the channel; the scaling $w^{-1}$ is simply understood as a geometric ``parallel resistor" effect -- the wider the channel, the smaller the resistance per length.   The second term arises from ballistic effects: if momentum is relaxed largely at the boundaries, then in the absence of bulk collisions we would obtain an inverse conductance scaling as $w^{-2}$ due to a wall-to-wall scattering rate $\gamma\sim v_\text{F} w^{-1}$ enhanced by the same $w^{-1}$ ``parallel resistor" effect mentioned above. The coefficient $A$ in this term encodes details about how precisely quasiparticles scatter off the walls and can in general be quite complicated, but the for the sake of simplicity we will assume it to be constant.

Using the reported $T\rightarrow 0$ momentum-relaxing mean free path of 20 $\mu$m for PdCoO$_2$ \cite{mackenzie}, we note that transport in the widest, $w=60$ $\mu\text{m}$ channel will be dominated at all temperatures by bulk scattering. Thus we may to good approximation estimate the bulk resistivity $\rho_{\mathrm{bulk}}(T)$ for $\mathrm{PdCoO}_2$ from the resistivity reported in the $w=60$ $\mu\text{m}$ channel. We may then compare our model (\ref{eq:GLfit}) against the data in Figure \ref{fig:channelCon} using the constant $A$ as our single fit parameter. The result of this fit is shown in Figure \ref{fig:ohmBallistic}, which demonstrates that all temperature dependence in the conductance data is quantitatively well described by our toy model (\ref{eq:GLfit}), even with the extremely crude assumption of constant $A$.  This suggests that all temperature dependence in this data arises from thermally activated momentum-relaxing processes.

The channel width dependence of the reported channel conductance data in PdCoO$_2$ \cite{mackenzie} is much more unusual. From Figure \ref{fig:channelCon}, we see that as $T\to0$ the conductance $G/w^2$ does not decrease monotonically with increasing channel width $w$, as predicted by (\ref{eq:GLfit}).  So while the $T$ dependence of the data appears rather conventional, the low-temperature $w$ dependence of $G$ is neither ohmic ($G\sim w$) nor ballistic ($G\sim w^2$) and thus quite unconventional as $T\rightarrow 0$. This unconventional reported $w$-scaling of the channel conductance, while not fully hydrodynyamic ($G\sim w^3$), was cited as the primary evidence for identifying a possible hydrodynamic crossover in PdCoO$_2$ \cite{mackenzie}. However, the fact that this unconventional $w$-scaling of $G(w)$ occurs \emph{only} at very low temperature calls into question a hydrodynamic interpretation. Indeed, the Fermi temperature of $\mathrm{PdCoO}_2$ is approximately $T_{\mathrm{F}} \approx 3\times 10^4$ K \cite{mackenzie16}, and so the unconventional $w$-scaling of the conductance is seen to persist down to roughly $T\approx 1 \text{ K} \approx 10^{-4}T_{\mathrm{F}}$. At this fraction of the Fermi temperature, electron-electron scattering should be completely negligible, even in a nearly perfect polygonal Fermi liquid; since the hydrodynamic regime is only reached when electron-electron
scattering events dominate momentum relaxing scatter events, it is therefore unlikely that this unconventional $w$-scaling of the conductance is hydrodynamic in origin.


In a conventional Fermi liquid, the only scattering mechanism present at such low temperatures is impurity scattering.  It has been noted \cite{usui} that electrons scatter faster on the same edge of the Fermi surface than they do between different edges, due to spin-momentum locking.  However, this effect is not so strong as to suppress inter-edge scattering by a parametric amount.  We are unsure if impurity scattering alone could fully explain the unusual $w$ dependence in $G(w)/w^2$.  A final possibility is that the ballistic boundary conditions, and/or the approximate angle that the Fermi surface makes relative to the channel, picked up some weak $w$ dependence in the experiment of \cite{mackenzie}.

Although electronic transport in PdCoO$_2$ -- or delafossites more generally -- may not be hydrodynamic, it remains an open question what effects anisotropy in the Fermi surface may have on a true ballistic-to-hydrodynamic crossover. Indeed, recent experimental work has demonstrated that circular models of the Fermi surface are not sufficient to explain experimental observations \cite{dgg18}. So while the cartoon model of a perfectly polygonal Fermi surface we describe in the remainder of the paper may suffer from a few drawbacks -- including possible instability to charge density wave formation\footnote{It appears, however, that such charge density wave instabilities may be rather fine-tuned \cite{mazin}. In the context of $\mathrm{PdCoO}_2$, it seems that such instabilities do not occur at the relevant electron density \cite{mackenzie16}.} and formally infinite quasiparticle scattering rates\footnote{In practice, we regularize these scattering rates by introducing a small ``rounding" of the otherwise flat edges of the polygonal Fermi surface; see Appendix \ref{app:coll}.} -- this polygonal model nonetheless probes the hydrodynamic limit of strong anisotropy in the Fermi surface. In reality, in materials such as $\mathrm{PdCoO}_2$ the rounded hexagonal Fermi surface lies somewhere ``in between" these two extreme limits of a perfect circle and a prefect polygon.

\section{Kinetic Theory with a Polygon Fermi Surface}\label{sec:kinetic}

In this section, we will develop a simple kinetic theory for the electrons in a material with an even-sided polygon Fermi surface, within linear response out of equilibrium.  Our focus will be on developing the formalism suitable for transport computations in materials with these Fermi surfaces where momentum-conserving electron-electron scattering cannot be ignored.  


\subsection{The  Boltzmann Equation}

We seek a description of transport in systems weakly
perturbed away from thermal equilibrium. In Fermi liquids with weak interactions and long lived quasiparticles,
such a description is given by a Boltzmann equation that dictates
the time evolution of the single particle distribution function
$f\left(\mathbf{x},\mathbf{p}\right)$. For quasiparticles with dispersion relation $\epsilon_\mathbf{p}$ in the presence of an external
force $\mathbf{F}$, the Boltzmann equation for $f$ reads \cite{kamenev}
\begin{equation}
\label{eq:fullBoltz}
 \partial_{t}f+\mathbf{v}_{\mathbf{p}}\cdot\partial_{\mathbf{x}}f+\mathbf{F}\cdot\partial_{\mathbf{p}}f=\mathcal{C}\left[f\right],
\end{equation}
where $\mathbf{v}_{\mathbf{p}}=\partial_{\mathbf{p}}\epsilon_{\mathbf{p}}$
is the quasiparticle velocity, and the collision integral term \emph{$\mathcal{C}\left[f(\mathbf{x}^\prime, \mathbf{p}^\prime)\right](\mathbf{x},\mathbf{p})$} accounts for the effects of multi-particle collisions. Eq. (\ref{eq:fullBoltz}) can be derived from the Schwinger-Keldysh
formalism as a controlled expansion when the following two conditions are satisfied:
(\emph{i}) the length- and momentum-scales over which $f\left(\mathbf{x},\mathbf{p}\right)$
varies satisfy $\left|\mathrm{\mdelta}\mathbf{x}\right|\cdot\left|\mathrm{\mdelta}\mathbf{p}\right|\gg\hbar$,
and (\emph{ii}) quasiparticles are well-defined, which qualitatively means
that all scattering rates (the eigenvalues of the linearized collision
operator; see below) are small compared to $k_{\text{B}}T/\hbar$.
In this limit, the collision operator is typically well-approximated
by a small number of Feynman diagrams, though we will not explicitly
calculate any such diagrams in this paper.  We will also neglect renormalization of $\epsilon_{\mathbf{p}}$ over its bare value due to quantum fluctuations, though this can be accounted for in a more sophisticated treatment \cite{kamenev}.

If all collisions between fermionic quasiparticles are spatially local
two-body scattering events (e.g. screened Coulomb interactions), then the collision
integral only has non-trivial momentum dependence and can be written
as the difference of transition probabilities for scattering processes that populate
and vacate the state of momentum $\mathbf{p}$:
\begin{equation}
\mathcal{C}[f](\mathbf{p}) = \int\frac{\mathrm{d}^d \mathbf{q}\mathrm{d}^d \mathbf{q}^\prime \mathrm{d}^d \mathbf{p}^\prime}{(2\mpi\hbar)^{3d}}\left(W_{\mathbf{p}^{\prime}\mathbf{q}^{\prime}\to\mathbf{p}\mathbf{q}}-W_{\mathbf{p}\mathbf{q}\to\mathbf{p}^{\prime}\mathbf{q}^{\prime}}\right)
\label{eq:2body}
\end{equation}
For two-body scattering, the transition probability $W_{\mathbf{p}^{\prime}\mathbf{q}^{\prime}\to\mathbf{p}\mathbf{q}}$
is given to leading order in the quasiparticle interaction strength
by
\begin{equation}
W_{\mathbf{p}^{\prime}\mathbf{q}^{\prime}\to\mathbf{p}\mathbf{q}}
=
\left|\mathcal{M}_{\mathbf{p}\mathbf{q}\mathbf{p}^\prime\mathbf{q}^\prime}\right|^{2}f\left(\mathbf{p}^{\prime}\right)f\left(\mathbf{q}^{\prime}\right)\left[1-f\left(\mathbf{p}\right)\right]\left[1-f\left(\mathbf{q}\right)\right]\mdelta\left(\epsilon_{\mathbf{p}^{\prime}}+\epsilon_{\mathbf{q}^{\prime}}-\epsilon_{\mathbf{p}}-\epsilon_{\mathbf{q}}\right)\mdelta\left(\mathbf{p}^{\prime}+\mathbf{q}^{\prime}-\mathbf{p}-\mathbf{q}\right)
\end{equation}
where $\mathcal{M}$ is the relevant scattering
matrix element determined by the microscopic quantum theory and
$f$ and $1-f$ are the probabilities that initial states are occupied and final states are unoccupied, respectively. 

At thermal equilibirum with $\mathbf{F}=\mathbf{0}$, we expect a
local Fermi-Dirac distribution
\begin{equation}
\label{eq:localFD}
    f^{0}\left(\mathbf{x},\mathbf{p}\right)=n_{\text{F}}\left(\lambda_{\mathbf{x}}^{a}X_{\mathbf{p}}^{a}\right)
\end{equation}
to be a solution of the Boltzmann equation, where $n_{\text{F}}(z)=1/(1+\mathrm{e}^{z})$, $X_{\mathbf{p}}^{a}$
are the single particle contributions to the conserved quantities of the many-body Hamiltonian labeled by $a$, 
and $\lambda_{\mathbf{x}}^{a}$ are the corresponding (spatially-varying) conjugate thermodynamic variables. Assuming spatial translation invariance, these conserved quantities include charge, momentum and energy, given respectively by:
\begin{equation}
    X_{\mathbf{p}}^{a}=\left(1,\mathbf{p},\epsilon_{\mathbf{p}}\right)^a. \label{eq:Xiboltz}
\end{equation}
Indeed, combining (\ref{eq:localFD}) and (\ref{eq:Xiboltz}), it is easy to see that the collision integral (\ref{eq:2body}) vanishes.  The hydrodynamic equations then arise from integrating the Boltzmann equation over $\mathbf{p}$, weighted by each of the (\ref{eq:Xiboltz}): \begin{equation}
\partial_{t}\rho^{a}+\nabla\cdot\mathbf{J}^{a} =0
\end{equation}
where the conserved densities are \begin{equation}
    \rho^{a} \equiv\int\frac{\mathrm{d}^{d}\mathbf{p}}{\left(2\mpi\hbar\right)^{d}}X_{\mathbf{p}}^{a}f^{0}\left(\mathbf{x},\mathbf{p}\right)
\end{equation}
and the associated currents are \begin{equation}
    \mathbf{J}^{a} \equiv \int\frac{\mathrm{d}^{d}\mathbf{p}}{\left(2\mpi\hbar\right)^{d}}\mathbf{v}_{\mathbf{p}}X_{\mathbf{p}}^{a}f^{0}\left(\mathbf{x},\mathbf{p}\right).
\end{equation}

Our focus in this paper will be the linearized Boltzmann equation near thermal equilibrium at a fixed temperature $T$ and chemical potential $\mu$.  This thermal distribution function is \begin{equation}
    f^0(\mathbf{p}) = \frac{1}{1+\mathrm{e}^{\beta (\epsilon-\mu)}}
\end{equation}
where $\beta = 1/k_{\mathrm{B}}T$.    Following \cite{hartnoll1705}, we introduce the following notation for linearizing the kinetic equations.  Firstly, we suppose that the distribution function $f$ takes the form \begin{equation}
    f = f^0(\mathbf{p}) -\frac{\partial f^{0}}{\partial\epsilon_{\mathbf{p}}}\Phi(\mathbf{x},\mathbf{p}) + \cdots \label{eq:Phi}
\end{equation}
where $\Phi$ denotes the perturbation of the distribution function within linear response (terms at $\mathrm{O}(\Phi^2)$ will be neglected).   We will discuss $\Phi$ rather than $f-f^0$ as the former is less singular.

We take $k_{\mathrm{B}}T\ll \mu$, so that the system behaves as a conventional Fermi liquid.  In this regime, it is generally acceptable to write \cite{lucasreview17} \begin{equation}
    -\frac{\partial f^{0}}{\partial\epsilon_{\mathbf{p}}} = \mdelta(\epsilon_{\mathbf{p}} - \mu) + \mathrm{O}\left(\frac{k_{\mathrm{B}}T}{\mu}\right)
\end{equation}
as the distribution $\Phi(\mathbf{x},\mathbf{p})$ is generally non-singular as $T\rightarrow 0$.  Defining the ket \begin{equation}
\left|\Phi\right\rangle \equiv\int\mathrm{d}^{d}\mathbf{x}\,\mathrm{d}^{d}\mathbf{p}\;\Phi\left(\mathbf{x},\mathbf{p}\right)\left|\mathbf{x}\mathbf{p}\right\rangle,
\end{equation}
the matrices \begin{subequations}
\begin{align}
\mathsf{W}\left|\mathbf{x}\mathbf{p}\right\rangle  & \equiv\int\mathrm{d}^{d}\mathbf{x}^{\prime}\,\mathrm{d}^{d}\mathbf{p}^{\prime}\;\frac{\mdelta\mathcal{C}}{\mdelta f}\left[f^{0}\left(\mathbf{x}^{\prime},\mathbf{p}^{\prime}\right)\right]\left(\mathbf{x^{\prime}},\mathbf{p}\right)\mdelta\left(\mathbf{x}-\mathbf{x}^{\prime}\right)\left|\mathbf{x}^{\prime}\mathbf{p}^{\prime}\right\rangle ,\\
\mathsf{L}\left|\mathbf{x}\mathbf{p}\right\rangle  & \equiv-\int\mathrm{d}^{d}\mathbf{x}^{\prime}\,\mathrm{d}^{d}\mathbf{p}^{\prime}\;\left(\mathbf{v}_{\mathbf{p}}\cdot\partial_{\mathbf{x}}\right)\mdelta\left(\mathbf{x}-\mathbf{x}^{\prime}\right)\mdelta\left(\mathbf{p}-\mathbf{p}^{\prime}\right)\left|\mathbf{x}^{\prime}\mathbf{p}^{\prime}\right\rangle .
\end{align}
\end{subequations}
representing the linearized collision operator and streaming operators respectively, and the inner product 
\begin{equation}
\left\langle \mathbf{x}^{\prime}\mathbf{p}^{\prime}\left|\mathbf{x}\mathbf{p}\right.\right\rangle \equiv\frac{1}{\left(2\mpi\hbar\right)^{d}V}\left(-\frac{\partial f^{0}}{\partial\epsilon_{\mathbf{p}}}\right)\mdelta\left(\mathbf{x}-\mathbf{x}^{\prime}\right)\mdelta\left(\mathbf{p}-\mathbf{p}^{\prime}\right),  \label{eq:innerproduct}
\end{equation}
with $V$ the spatial volume of the system, we recast the Boltzmann equation as an infinite dimensional linear system: 
\begin{equation}
    \label{eq:matrixBoltz}
    \left(\partial_{t}+\mathsf{W}+\mathsf{L}\right)\left|\Phi\right\rangle = 0.
\end{equation}
We now discuss a few properties of $\mathsf{L}$
and $\mathsf{W}$. Firstly, we notice via integration by
parts that the streaming operator $\mathsf{L}$ satisfies
\begin{equation}
\left\langle \mathbf{x}^{\prime}\mathbf{p}^{\prime}\left|\mathsf{L}\right|\mathbf{x}\mathbf{p}\right\rangle =-\left\langle \mathbf{x}\mathbf{p}\left|\mathsf{L}\right|\mathbf{x}^{\prime}\mathbf{p}^{\prime}\right\rangle 
\end{equation}
and is therefore an anti-symmetric matrix. Furthermore, we will assume
time-reversal invariance and inversion symmetry in our kinetic theory;
the former implies that the linearized collision operator $\mathsf{W}$
satisfies $\text{\ensuremath{\left\langle \mathbf{p}^{\prime}\left|\mathsf{W}\right|\mathbf{p}\right\rangle }=\ensuremath{\left\langle -\mathbf{p}\left|\mathsf{W}\right|-\mathbf{p}^{\prime}\right\rangle }}$,
while the latter implies that $\left\langle \mathbf{p}^{\prime}\left|\mathsf{W}\right|\mathbf{p}\right\rangle =\left\langle -\mathbf{p}^{\prime}\left|\mathsf{W}\right|-\mathbf{p}\right\rangle $
(here we have suppressed the spatial indices). Combining these equalities,
we conclude that 
\begin{equation}
\label{eq:symmetricW}
    \left\langle \mathbf{x}^{\prime}\mathbf{p}^{\prime}\left|\mathsf{W}\right|\mathbf{x}\mathbf{p}\right\rangle =\left\langle \mathbf{x}\mathbf{p}\left|\mathsf{W}\right|\mathbf{x}^{\prime}\mathbf{p}^{\prime}\right\rangle.
\end{equation}
Hence $\mathsf{W}$ is symmetric.  Finally, $\mathsf{W}$ has null vectors
$\left|\mathsf{X}\right\rangle $ associated with conservation laws \cite{hartnoll1705}.
We define vectors
\begin{equation}
\left|\mathsf{X}^a(\mathbf{x})\right\rangle\equiv\int\mathrm{d}^{d}\mathbf{p}\; X^a\left(\mathbf{p}\right) |\mathbf{xp}\rangle ,
\end{equation}
which have the property that \begin{equation}
   \langle \mathsf{X}^a |\Phi\rangle = \int \frac{ \mathrm{d}^d\mathbf{p}}{(2\mpi \hbar)^d V} X^a(\mathbf{p})\Phi(\mathbf{x},\mathbf{p}) \left(-\frac{\partial f^0}{\partial \epsilon_{\mathbf{p}}}\right) = \frac{\rho^a(\mathbf{x})}{V},
\end{equation}
namely that they encode (up to the normalization of the inner product) the parts of the local distribution function which correspond to conserved quantities.  Since (\ref{eq:localFD}) has to solve the Boltzmann equation for for any $\lambda^a$, we conclude that $\lambda^a |X^a\rangle$ must be an exact solution of (\ref{eq:matrixBoltz}).  Thus,  
\begin{equation}
\mathsf{W}\left|\mathsf{X}^a(\mathbf{x})\right\rangle =0.
\end{equation}

Finally, we note that the inner product (\ref{eq:innerproduct}) allows us to approximately ignore all $\mathbf{p}$ dependence of $\Phi$, except for the value of $\Phi$ along the Fermi surface itself.  In fact, with the exception of the $T$ dependence of the scattering rates (matrix elements of $\mathsf{W}$), it is acceptable to completely neglect all dynamics beyond the ``wobbling" of the Fermi surface itself, which is captured by the value of $\Phi$ exactly on the Fermi surface.

\subsection{A Separation of Time Scales}
\label{sec:2times}

The next two sections apply the general formalism above to the problem of interest, where the Fermi surface of the Fermi liquid is a polygon.  In this section, we will discuss the most subtle point, arising in the behavior of $\mathsf{W}$.  In particular, we will find a hierarchy of two-body quasiparticle scattering rates, arising from the polygonal geometry of the Fermi surface. 

\begin{figure}[t]
\includegraphics[width=.35\textwidth]{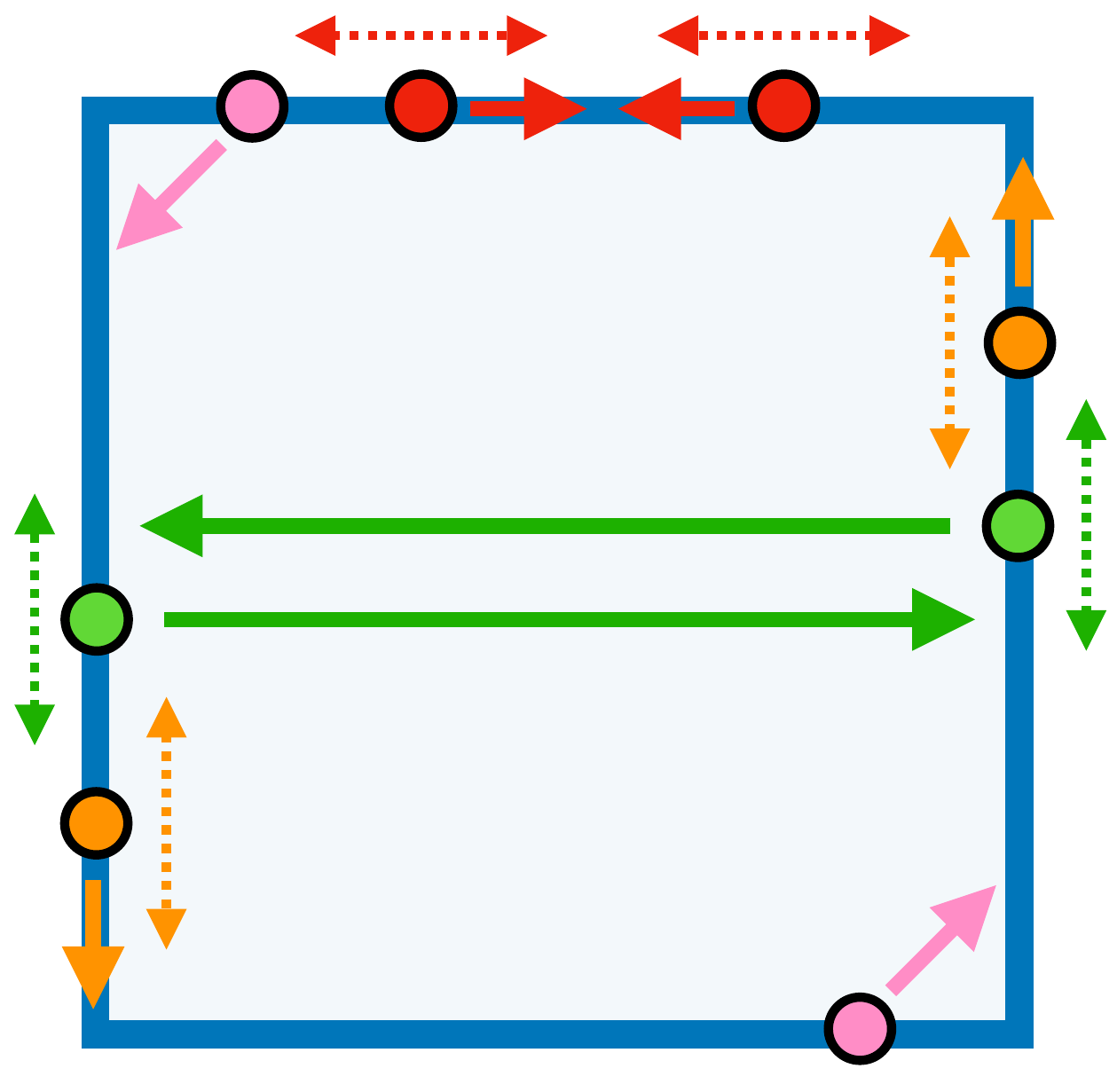}
\centering
\caption{A depiction of the two-body scattering events on an even-sided polygonal Fermi surface (blue) that are allowed by energy- and momentum-conservation. Only scattering of the pink kind can change the total number of quasiparticles on a Fermi surface edge; all other scattering events can change the particle density \emph{at a point}, but leave the particle number on each edge invariant. Additionally, the phase space for non-pink type scattering is extremely large due to the freedom to, for fixed collisional momentum transfer, independently ``slide" the initial particles along a given Fermi surface edge, as illustrated by dashed arrows. By contrast, the phase space for pink type scattering is far more restricted for fixed momentum transfer.  Note that, due to the indistinguishability of fermions, some of these scattering events (e.g. green and orange) are equivalent and enter $\mathsf{W}$ in identical ways.}
\label{fig:2times}
\end{figure}

Consider energy- and momentum-conserving two-body scattering events on a polygonal Fermi surface, as depicted in the case of a square Fermi surface in Figure \ref{fig:2times}. We identify two classes of such scattering events: those that conserve the net number of quasiparticles on each Fermi surface edge, and those that do not\footnote{The net quasiparticle number $\int\mathrm{d}p_{m}\;\Phi(p_{m})$ on a Fermi surface edge $m$ is simply the quasiparticle excitation distribution $\Phi$ intregrated over the momenta $p_m$ constituting that edge; this may equivalently be thought of as the net charge excitation on a Fermi surface edge.}. Scattering events that alter particle number \emph{at a Fermi surface point} but conserve particle number on each edge are of the form
\begin{equation}
    \left[\begin{array}{c}
\mathbf{p}\\
\mathbf{q}
\end{array}\right]\to\left[\begin{array}{c}
\mathbf{p}^{\prime}\\
\mathbf{q}^{\prime}
\end{array}\right]=\left[\begin{array}{c}
\mathbf{p}-\mathbf{k}_{\parallel}\\
\mathbf{q}+\mathbf{k}_{\parallel}
\end{array}\right]
\end{equation}
with  the collisional momentum transfer $\mathbf{k}_\parallel$ parallel to edge(s) on which the initial quasiparticles $\mathbf{p}, \mathbf{q}$ lie (the green scattering in Figure \ref{fig:2times} is equivalent to the orange scattering, which has this property). The allowed phase space for scattering events of this type is quite large due to the fact that, for fixed momentum transfer $\mathbf{k}_\parallel$, the initial quasiparticles $\mathbf{p}, \mathbf{q}$ possess a ``sliding" freedom in that they can - \emph{independently} - lie anywhere along a given Fermi surface edge; see Figure (\ref{fig:2times}). Due to this ``sliding" freedom and associated extensive allowed phase space, we conclude that the scattering rate $\gamma_\text{f}$ associated with collisions that relax particle number at a point, but conserve total edge particle number, is extremely large.

In contrast, two-body scattering events that alter edge particle number must be more fine-tuned. For an initial quasiparticle $\mathbf{p}$ and fixed collisional momentum transfer $\mathbf{k}$, there is only a single  quasiparticle $\mathbf{q}$ with which $\mathbf{p}$ can scatter in such a way so as to simultaneously conserve energy, conserve momentum, and alter edge particle number; see Figure (\ref{fig:2times}). This reduced allowed phase space  (in contrast to edge particle conserving collisions) is similar to the contrast between head-on scattering, with large allowed phase space, and small-angle scattering, with much smaller allowed phase space, for quasiparticles on a circular Fermi surface \cite{ledwith1,ledwith2}. Scattering events that can alter the particle number on a Fermi surface edge are associated with a much smaller scattering rate $\gamma_\text{s}\ll\gamma_\text{f}$. 

In the case of a perfect polygonal Fermi surface with exactly flat edges, the fast scattering rate $\gamma_\text{f}$ associated with two-body events that leave edge particle number invariant is singular due to the aforementioned ``sliding" freedom, which generates infinities in the the two-body collision integral (\ref{eq:2body}) due to the alignment of two constraints in the delta functions imposing energy and momentum conservation. One may regularize this calculation by ``rounding out" the edges of the polygonal surface into circular arcs with a degree of curvature $\alpha \ll 1$ (or equivalently, radius of curvature $R\sim\alpha^{-1}$ much larger than a Fermi surface edge; see Figure \ref{fig:2timesA}).  In Appendix \ref{app:coll} we estimate the regulated scattering rates and find 
\begin{equation}
    \frac{\gamma_{\text{f}}}{\gamma_{\text{s}}}\sim \alpha^{-1}.
\end{equation}
  Thus for ``nearly flat" $\alpha \ll 1$ Fermi surface edges, we find a hierarchy of decay rates $\gamma_\text{s}\ll\gamma_\text{f}$, with the precise magnitude of this hierarchy determined by the degree to which the Fermi surface deviates from a perfect polygon. 

A similar distinction between inter- and intra-edge scattering of electrons off of impurities has been made in the specific case of the ``nearly" hexagonal Fermi surface of $\text{PdCoO}_2$ \cite{usui}. We point out, however, that the hieararchy we identify here is much more dramatic, as $\gamma_{\mathrm{f}}/\gamma_{\mathrm{s}}$ can be arbitrarily large, in contrast to the case of electron-impurity scattering, where the enhancement is by an O(1) factor. 

Before moving on, we briefly address scattering events involving more than two quasiparticles. For example, three-body scattering can decay the approximate ``imbalance mode" in graphene \cite{foster}, which also arises due to kinematic constraints on two-body scattering \cite{lucasreview17}.  Three-body scattering also decays ``imbalance" modes of our model, including the number density on a fixed edge.\footnote{For example, consider two particles on the left edge of Figure \ref{fig:2times} sliding down, allowing a third particle on the bottom edge to move to the top edge, thus altering the net quasiparticle number on the bottom and top edges. A sliding freedom for this process on the left edge means that the scattering rate associated with this process, like $\gamma_\text{f}$ discussed in the text, also has a formally infinite value in the limit $\alpha \rightarrow 0$ of flat Fermi surface edges.}  In spite of their large allowed phase space, however, such scattering events are subleading in the quasiparticle scattering matrix $\mathcal{M}$, both in the coupling constant strength and in the power of $T/T_{\mathrm{F}}$ which arises.   If we take the limit $T\rightarrow 0$ before $\alpha \rightarrow 0$, we may neglect three-body (and beyond) scattering events in the collision integral.

\subsection{Two-Time Model for Linearized Collision Operator}

We now begin our construction of a phenomenological model for the linearized collision operator $\mathsf{W}$ appearing in (\ref{eq:matrixBoltz}). To achieve this goal, we first introduce a convenient basis for the quasiparticle excitations $\Phi$ with which to construct the matrix $\mathsf{W}$.

\begin{figure}[t]
\includegraphics[width=.45\textwidth]{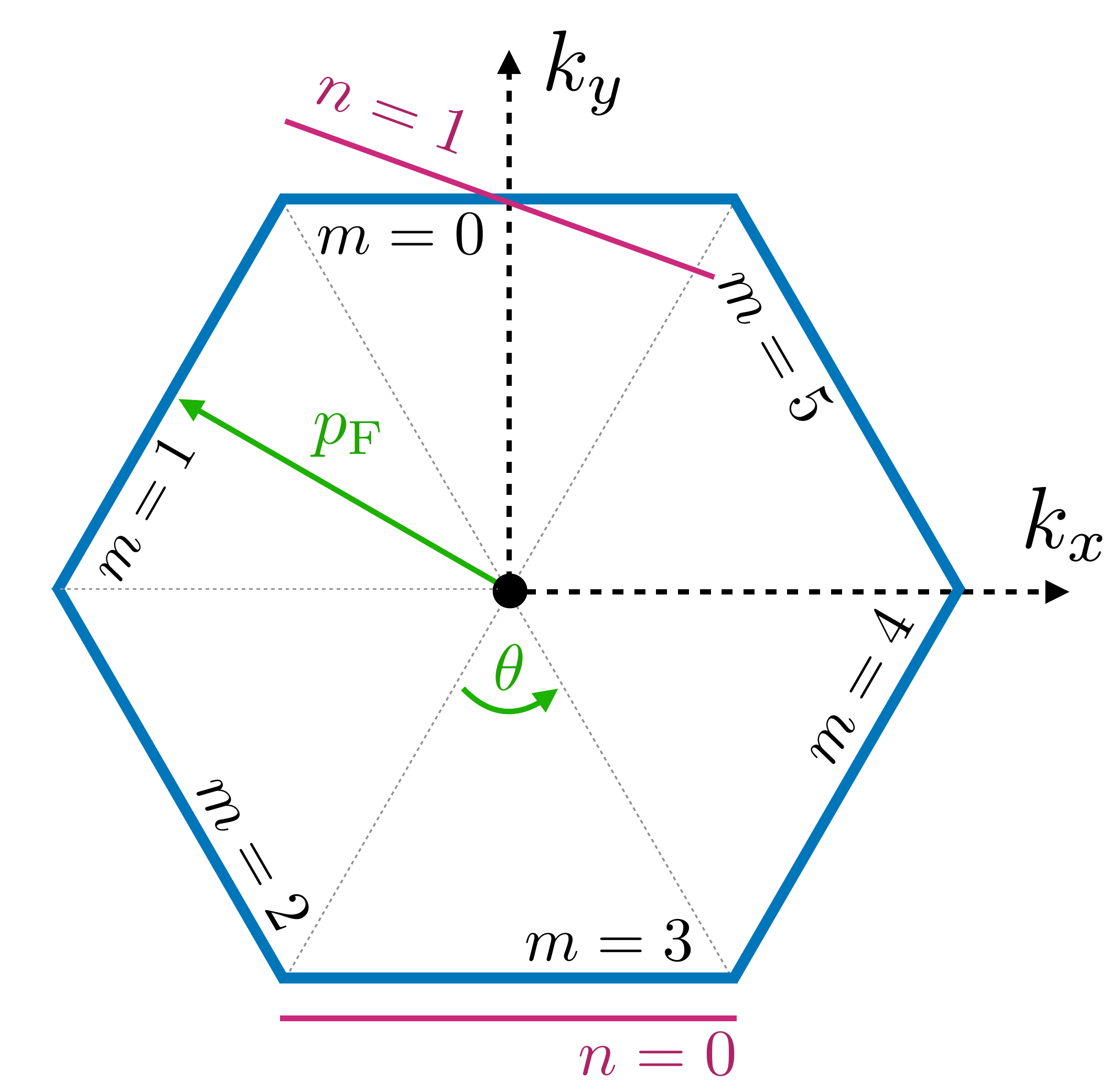}
\centering
\caption{Model Fermi surface in the hexagonal $M=6$ case. For general (even) $M$, we align the $m=0$ edge so as to be bisected by the $k_{y}$-axis, and increase the edge label $m$ going counter-clockwise around the polygon. Here $\theta=2\mpi/M$ denotes the symmetry angle of the polygon.}
\label{fig:hexModes}
\end{figure}

On a finite domain, any sufficiently smooth function may be written as a weighted sum of Legendre polynomials. We employ such an expansion for $\Phi$ on each edge of the $M$-gon Fermi surface via
\begin{equation}
\label{eq:legExpansion}
    \Phi\left(\mathbf{p}\right)=\sum_{m=0}^{M-1}\sum_{n=0}^{\infty}\Phi_{mn}\left|n,m\right\rangle ,
\end{equation}
where
\begin{equation}
\label{eq:legVector}
    \left|n,m\right\rangle 
    \equiv
    \sqrt{\frac{2n+1}{2}}\mathrm{L}_{n}\left(\frac{p_{m}}{p_{\text{L}}}\right)
\end{equation}
is vector (in the vector space of smooth functions) representing the $n$-th Legendre polynomial $\mathrm{L}_n$ 
of the $m$-th Fermi surface edge momentum $p_m$. In the above we have also introduced the length $2p_\text{L}\equiv 2p_\text{F}\tan(\mpi/M)$ of each Fermi surface edge, so that $p_m/p_\text{L}\in[-1,1]$ parameterizes the momentum-coordinate along the $m$-th edge, increasing counter-clockwise; see Figure \ref{fig:hexModes}. The Legendre mode vectors $|n,m\rangle$ (\ref{eq:legVector}) have also been suitably normalized so as to satisfy the orthonormality condition
\begin{equation}
   \left\langle n^{\prime},m^{\prime}\right.\left|n,m\right\rangle =\mdelta_{mm^{\prime}}\int_{-1}^{1}\mathrm{d}x\;\sqrt{\frac{2n^{\prime}+1}{2}}\mathrm{L}_{n^{\prime}}\left(x\right)\cdot\sqrt{\frac{2n+1}{2}}\mathrm{L}_{n}\left(x\right)=\mdelta_{mm^{\prime}}\mdelta_{nn^{\prime}}.
\end{equation}

The Legendre basis possesses two properties that will prove very useful for us. Firstly, we note that only $n=0$ Legendre modes carry a nonzero number of quasiparticles on a Fermi surface edge; all higher-order $n\geq1$ excitations have zero net quasiparticles on an edge. This follows from the fact that higher $n\geq1$ Legendre modes are orthogonal to the constant $n=0$ mode and therefore vanish when integrated over an edge. Per our discussion in Section \ref{sec:2times}, we will require that modes with nonzero edge density decay at a rate $\gamma_\text{s}$ slower than all other, non-conserved modes, which instead decay at a rate $\gamma_\text{f}\gg\gamma_\text{s}$. Thus, in terms of our linearized collision operator $\mathsf{W}$, these considerations impose the constraint
\begin{equation}
\label{eq:decayConstraint}
    \left\langle n,m\left|\mathsf{W}\right|n,m\right\rangle \sim
    \begin{cases}
    \gamma_{\text{s}} & \left(n=0\right)\\
    \gamma_{\text{f}} & \left(n\geq1\right)
    \end{cases}.
\end{equation}

The second useful property of our Legendre basis is the fact that only $n=0$ and $n=1$ Legendre modes carry a nonzero amount of momentum; all higher-order $n\geq2$ excitations have zero net momentum. This follows from the fact that Fermi surface edge momenta is parameterized by a line, which is a linear combination $n=0$ and $n=1$ Legendre modes; thus any integral summing the momentum of a higher $n\geq2$ mode over an edge will vanish due to the orthogonality of the Legendre modes. The quantities that we wish to conserve in our kinetic theory, namely the total particle density
\begin{equation}
\label{eq:densityMode}
    \left|N\right\rangle =n_{0}\sum_{m=0}^{M-1}\left|0,m\right\rangle
\end{equation}
and the $x$- and $y$-components of the total momentum
\begin{equation}
    \left|P_{i}\right\rangle =\left|P_{i}^{0}\right\rangle +g_{M}\left|P_{i}^{1}\right\rangle,
\end{equation}
therefore lie entirely in the $2M$-dimensional subspace spanned by $n=0,1$ modes on each of the $M$  Fermi surface edges. In the above, we have introduced the equilibirium particle density $n_0$, the vectors $|P_i^n\rangle$ which specify how Legendre modes of order $n$ contribute to the $i$-th component of the momentum, given explicitly by
\begin{subequations}
\begin{align}
    \left|P_{x}^{0}\right\rangle &=-p_{\text{F}}\sum_{m=0}^{M-1}\sin\left(m\theta\right)\left|0,m\right\rangle ,\quad\left|P_{x}^{1}\right\rangle =-p_{\text{F}}\sum_{m=0}^{M-1}\cos\left(m\theta\right)\left|1,m\right\rangle \\\left|P_{y}^{0}\right\rangle &=+p_{\text{F}}\sum_{m=0}^{M-1}\cos\left(m\theta\right)\left|0,m\right\rangle ,\quad\left|P_{y}^{1}\right\rangle =-p_{\text{F}}\sum_{m=0}^{M-1}\sin\left(m\theta\right)\left|1,m\right\rangle,
\end{align}
\end{subequations}
and the geometrical factor $g_M=\tan(\mpi/M)/\sqrt{3}$ that relates how $n=0,1$ modes differentially contribute to the momentum. Imposing the charge- and momentum-conservation constraints
\begin{equation}
\label{eq:conserveConstraint}
    \mathsf{W}\left|N\right\rangle =\mathsf{W}\left|P_{x}\right\rangle =\mathsf{W}\left|P_{y}\right\rangle =0
\end{equation}
therefore only requires that $\mathsf{W}$ act non-trivially in the $n=0,1$ subspace; for higher modes, the linearized collision operator can simply act as $\mathsf{W}_{n\geq2}\sim\gamma_\text{f}\,\mathsf{1}$ and satisfy the required constraints (\ref{eq:decayConstraint}) and (\ref{eq:conserveConstraint}).  

The simplest $2M\times 2M$ linearized collision operator $\mathsf{W}$ that satisfies these constraints is
\begin{equation}
\label{eq:Wmatrix}
    \mathsf{W}=\mathsf{P}\mathsf{W}_{0}\mathsf{P}+\gamma_{\text{s}}\mathsf{P}^{\prime}
\end{equation}
where we have introduced the operators
\begin{subequations}\begin{align}
    \mathsf{W}_{0}&=\gamma_{\text{s}}\sum_{m=0}^{M-1}\left|0,m\right\rangle \left\langle 0,m\right|+\gamma_{\text{f}}\sum_{m=0}^{M-1}\left|1,m\right\rangle \left\langle 1,m\right|,\\\mathsf{P}&=\mathsf{1}-\frac{|N\rangle\langle N|}{\langle N|N\rangle}-\sum_{i=x,y}\sum_{n=0,1}\frac{|P_{i}^{n}\rangle\langle P_{i}^{n}|}{\langle P_{i}^{n}|P_{i}^{n}\rangle},\\\mathsf{P}^{\prime}&=\sum_{i=x,y}\frac{|\widetilde{P}_{i}\rangle\langle\widetilde{P}_{i}|}{\langle\widetilde{P}_{i}|\widetilde{P}_{i}\rangle}.
\end{align}\end{subequations}
The operators $\mathsf{P}, \mathsf{P}^\prime$ are in fact projection operators, with $\mathsf{P}$ projecting out the total particle density and each of the $n=0,1$ modes of the momentum \emph{individually}, and $\mathsf{P}^\prime$ projecting onto the modes
\begin{equation}
    |\widetilde{P}_{i}\rangle=g_{M}|P_{i}^{0}\rangle-|P_{i}^{1}\rangle 
\end{equation}
orthogonal to the momentum: $\langle P_{i}|\widetilde{P}_{j}\rangle=0$. 

First, we note that the matrix $\mathsf{W}$ (\ref{eq:Wmatrix}) is symmetric, as required by time-reversal invariance and reflection symmetry (\ref{eq:symmetricW}). Next, since the projection operators $\mathsf{P}$, $\mathsf{P}^\prime$ both vanish on the desired conserved modes $|N\rangle$, $|P_x\rangle$, $|P_y\rangle$, so too does the matrix $\mathsf{W}$ (\ref{eq:Wmatrix}). Finally, the term $\mathsf{PW_0P}$ ensures that the decay constraint (\ref{eq:decayConstraint}) is satisfied, but has the unphysical property that it conserves the orthogonal momentum since $\mathsf{P}|\tilde{P}_{i}\rangle=0$; this is remedied by adding a term $\gamma_\text{s}\mathsf{P}^\prime$ that causes the orthogonal momentum $|\tilde{P}_{i}\rangle$ to instead decay at the rate $\gamma_\text{s}$, which is chosen since $|\tilde{P}_{i}\rangle$ contains $n=0$ parts. Explicit constructions of $\mathsf{W}$ (\ref{eq:Wmatrix}) are given in Appendix \ref{app:Wmats} for both a square and hexagon Fermi surface. Additionally, in Table \ref{table:dihedralB} of Appendix \ref{app:dihedral} we list the complete eigenvector/eigenvalue decomposition of $\mathsf{W}$ for general (even) $M$, which we now briefly summarize.

The collision matrix $\mathsf{W}$ (\ref{eq:Wmatrix}) has the eigenvalue zero with multiplicity $3$; these correspond by construction to the $3$ conserved modes $\{|N\rangle$, $|P_x\rangle$, $|P_y\rangle\}$. In conventional fluid dynamics (relevant for us later, when we take the hydrodynamic limit of our kinetic theory), momentum density is written as a velocity field, which amounts to the a simple rescaling
\begin{equation}
    |V_{i}\rangle=c_{M}\frac{v_{\text{F}}}{p_{\text{F}}}|P_{i}\rangle,
\end{equation}
where $c_M$ is an $\mathrm{O}(1)$ constant that depends on the particular Fermi surface $M$-gon in question; for the square and hexagon, we have $c_4=4/3$ and $c_6=10/9$, respectively.\footnote{The factor $c_M$ makes the continuity equation (charge conservation) take the canonical form.}

Next, $\mathsf{W}$ has the eigenvalue $\gamma_\text{s}$ with multiplicity $(M-1)$, corresponding to $(M-1)$ ``slow" modes. These slow modes include $(M-4)$ ``spin-$k$" modes $\{|Q_{-}^{k}\rangle, |Q_{+}^{k}\rangle\}$ labeled by integer $k\in\{2,\ldots,(M/2-1)\}$, so-named due to the way they transform when they polygonal Fermi surface is rotated: a spin-$k$ mode first returns to itself (i.e. transforms trivially) when the Fermi surface is rotated through the minimal angle $2\mpi/k = M\theta/k$ (recall that $\theta=2\mpi/M$ is the symmetry angle of the $M$-gon). These spin modes are discussed in detail in Appendix \ref{app:dihedral}, but the punchline is that for each $k$, the two spin-$k$ mode ``components" $|Q_{\pm}^{k}\rangle$ will be repackaged into an rank-$k$ tensor $|Q_{i_1\cdots i_k}\rangle$ that transforms naturally under continuous Fermi surface rotations.  These modes do not arise for the square Fermi surface, but do arise for higher even $M$, including the hexagon. For the hexagon in particular, the two spin-$2$ modes are given explicitly by
\begin{subequations}
\begin{align}
    |Q_{-}^{2}\rangle	\equiv|Q_{-}\rangle=-c_{M}\frac{v_{\text{F}}}{p_{\text{F}}}\sum_{m=0}^{5}\sin\left(2m\theta\right)|0,m\rangle,
    \\
    |Q_{+}^{2}\rangle	\equiv|Q_{+}\rangle=+c_{M}\frac{v_{\text{F}}}{p_{\text{F}}}\sum_{m=0}^{5}\cos\left(2m\theta\right)|0,m\rangle,
\end{align}
\end{subequations}
which first return to themselves when the hexagonal Fermi surface is rotated through the minimal angle $\mpi=3\theta$. These spin-2 modes are repackaged into the traceless symmetric rank-$2$ tensor
\begin{equation}
    |Q_{ij}\rangle\equiv\frac{1}{\sqrt{2}}\left[\begin{array}{cc}
    -|Q_{+}\rangle & |Q_{-}\rangle\\
    |Q_{-}\rangle & |Q_{+}\rangle
    \end{array}\right]_{ij}
\end{equation}
that transforms naturally under continuous rotations of the Fermi surface; see Appendix \ref{app:dihedral} and in particular Eq. (\ref{eq:spin2}) for details.

The remaining $3$ slow modes are those orthogonal to the $3$ conserved modes, namely the orthogonal momentum $|\widetilde{P}_{i}\rangle$ which we also rescale into velocity fields
\begin{equation}
    |\widetilde{V}_{i}\rangle=c_{M}\frac{v_{\text{F}}}{p_{\text{F}}}|\widetilde{P}_{i}\rangle,
\end{equation}
and the orthogonal number density
\begin{equation}
    |\widetilde{N}\rangle = n_{0}\sum_{m=0}^{M-1} (-1)^m |0,m\rangle.
\end{equation}
We highlight in particular the mode $|\widetilde{N}\rangle$, which is invariant under reflections and alternates sign under discrete
rotations of the Fermi surface by the symmetry angle $\theta=2\mpi/M$. This means that the mode $|\widetilde{N}\rangle$ in fact first returns to itself after a Fermi surface rotation of $2\theta=2\mpi/(M/2)$, and thus with respect to continuous Fermi surface rotations it is more natural to regard $|\widetilde{N}\rangle$ as a ``spin-$M/2$" mode with an associated rank-$M/2$ tensor $|\widetilde{N}_{i_1\cdots i_{M/2}}\rangle$. The details of how this spin-$M/2$ tensor structure is determined are also spelled out in Appendix \ref{app:dihedral}.

Finally, $\mathsf{W}$ has the eigenvalue $\gamma_\text{f}$ with multiplicity $(M-2)$, corresponding to $(M-2)$ ``fast" modes. In our study of the quasihydrodynamic-to-hydrodynamic crossover in this model, these fast modes will not be dynamical due to the short timescales $\tau_\text{f}\sim \gamma_\text{f}^{-1}$ on which they decay. As such, these fast modes will only serve to give rise to viscous and diffusive effects for the slow and conserved modes in the long time scale, large length scale effective theories of the model, i.e. the quasihydrodynamic and hydrodynamic regimes.

\subsection{The Streaming Operator}
Finally, we specify the streaming operator $\mathsf{L}$ in our reduced $2M$-dimensional Legendre mode basis. We begin by observing that the polygonal Fermi surface detailed above has necessarily constrained the form of $\epsilon(\mathbf{p})$.  For simplicity, we will choose $\epsilon(\mathbf{p})$ to be defined piecewise in such a way that the Fermi velocity $v_{\mathrm{F}}$ is uniform along each edge of the Fermi surface:  
\begin{equation}
   \mathbf{v}(\mathbf{p}) \equiv v_{\mathrm{F}}\, \hat{\mathbf{n}}(\mathbf{p}).
   \label{eq:simpleV}
\end{equation}
While no actual dispersion relation $\epsilon(\mathbf{p})$ is \emph{this} simple, we do note that for PdCoO$_2$ in particular the quasiparticle velocity $\mathbf{v}(\mathbf{p})$ is in fact of roughly constant magnitude along each edge of the approximately hexagonal Fermi surface, up to the rounded corners \cite{mackenzie16}. The primary effect of these rounded Fermi surface corners is to allow for a continuum of quasiparticle velocity directions, which are excluded by (\ref{eq:simpleV}) but can lead to dramatic effects in the ballistic regime \cite{dgg18}. However, for the purpose of studying a mathematically tractable model of the ballistic-to-hydrodynamic crossover, Eq. (\ref{eq:simpleV}) is a reasonable simplification. 

Indeed, the constant magnitude quasiparticle velocity (\ref{eq:simpleV}) greatly simplifies our calculations, leading to following action of $\mathsf{L}$ in the Legendre basis:\footnote{We have suppressed spatial indices in defining the Legendre basis $|n,m\rangle$, but keep in mind they do carry spatial dependence coming from the phase space vectors $|\mathbf{x}\mathbf{p}\rangle$.}
\begin{equation}
    \mathsf{L}|n,m\rangle=v_{\text{F}}\Big[-\sin\left(m\theta\right)\partial_{x}+\cos\left(m\theta\right)\partial_{y}\Big]|n,m\rangle.
    \label{eq:Lmatrix}
\end{equation}
The key simplification of assuming constant quasiparticle velocity magnitude $|\mathbf{v}(\mathbf{p})|=v_\text{F}$ on the Fermi surface is that $\mathsf{L}$ (\ref{eq:Lmatrix}) and $\mathsf{W}$ (\ref{eq:Wmatrix}) are now both block diagonal, with the $n=0,1$ sector of the Boltzmann equation (\ref{eq:matrixBoltz}) decoupling from the $n\geq 2$ sector. Since we are ultimately concerned with calculating conductances and thus currents, which as explained above lie in the $n=0,1$ sector, we see that we have therefore successfully reduced the seemingly infinite-dimensional Boltzmann equation (\ref{eq:matrixBoltz}) to a $2M$-dimensional one.

\section{Hydrodynamics with a Polygon Fermi Surface}
\label{sec:hydro}

Having developed our two-time ``relaxation time" approximation for the Boltzmann equation above, we can now derive quasihydrodynamic and hydrodynamic equations of motion for our theory, depending on whether we are interested in physics on time scales $\omega \ll \gamma_{\mathrm{f}}$ or $\omega \ll \gamma_{\mathrm{s}}$, respectively.

\subsection{Integrating Out Modes}
\label{subsec:integration}

\begin{table}
\begin{center}
\begin{tabular}{c|c}
transport regime & length scale hierarchy
\tabularnewline
\hline 
\hline 
ballistic/Knudsen & $\left|\boldsymbol{x}\right|\ll\ell_{\text{f}}\ll\ell_{\text{s}}$
\tabularnewline
\hline 
quasihydrodynamic & $\ell_{\text{f}}\ll \left|\boldsymbol{x}\right|\ll\ell_{\text{s}}$
\tabularnewline
\hline 
hydrodynamic/Poiseuille & $\ell_{\text{f}}\ll\ell_{\text{s}}\ll \left|\boldsymbol{x}\right|$
\tabularnewline
\end{tabular}
\caption{Outline of the transport various regimes of our model at varying length scales  $\left|\boldsymbol{x}\right|$ of interest (for the one-dimensional channel flow problem, this will be the channel width $w$). Here we have also introduced the length scales $\ell_\text{s,f}=v_\text{F} \gamma_\text{s,f}^{-1}\sim \mathsf{W}^{-1}$ associated with the fast- and slow-decaying electronic excitations, arising from the geometric imbalance mode explained in Section \ref{sec:2times}.
}
\label{table:regimes}
\end{center}
\end{table}

In Table \ref{table:regimes}, we define the various transport regimes of our model via the length scale of interest. Moving away from the ballistic regime of our model, in which quasiparticles are infinitely long-lived and $\mathsf{W}\approx0$, and into the quasihydrodynamic and hydrodynamic regimes, in which the collision matrix $\mathsf{W}$ can not be neglected, will require ``integrating out'' the decaying modes that
enter into the theory. In particular, the quasihydrodynamic regime is reached by integrating out the $(M-2)$ fast modes, leaving dynamical the $(M-1)$ slow modes and the $3$ conserved
modes; the hydrodynamic regime is reached by integrating out the $\left(2M-3\right)$ fast and slow
modes, leaving dynamical only the $3$ conserved modes. We now outline how this ``integrating out" procedure is performed. 

Consider a solution $|\Phi\rangle$ of the Boltzmann equation in the
absence of a source: $(\partial_{t}+\mathsf{W}+\mathsf{L})|\Phi\rangle=0$.
If we let $a$ label the modes we wish to leave dynamical and $b$
label the (relatively faster decaying) modes we wish to integrate
out, we can write the Boltzmann equation in a block-diagonal basis
of $\mathsf{W}$ as
\begin{equation}
\left[\partial_{t}+\left(\begin{array}{cc}
\mathsf{W}_{a} & \mathsf{0}\\
\mathsf{0} & \mathsf{W}_{b}
\end{array}\right)+\left(\begin{array}{cc}
\mathsf{L}_{a} & \mathsf{\mathsf{L}}_{ab}\\
-\mathsf{L}_{ab}^{\dagger} & \mathsf{L}_{b}
\end{array}\right)\right]\left(\begin{array}{c}
|\Phi_{a}\rangle\\
|\Phi_{b}\rangle
\end{array}\right)=\left(\begin{array}{c}
0\\
0
\end{array}\right)
\end{equation}
where we have used the fact that the streaming matrix $\mathsf{L}$
is anti-Hermitian in the Fourier basis. Note that obtaining the quasihydrodynamic
equations corresponds to taking $a=(\text{slow and conserved})$ and
$b=\left(\text{fast}\right)$, whereas obtaining the hydrodynamic
equations corresponds to taking $a=(\text{conserved})$ and $b=\left(\text{slow and fast}\right)$.

Now, since we are studying the model on timescales for which the $b$-modes
have effectively decayed away, we have that $\partial_{t}\ll\mathsf{W}_{b}$.
Thus we may to good approximation take
$\partial_{t}\approx0$ in the $b$-sector equation. We then solve the $b$-sector equation for
the modes $|\Phi_{b}\rangle$ and substitute the result into the $a$-sector
equation, which yields 
\begin{equation}
\left[\partial_{t}+\mathsf{W}_{a}+\mathsf{L}_{a}+\mathsf{L}_{ab}\left(\mathsf{W}_{b}+\mathsf{L}_{b}\right)^{-1}\mathsf{L}_{ab}^{\dagger}\right]|\Phi_{a}\rangle=0.
\end{equation}
Now we note that since $\mathsf{W}_{b}\sim\gamma_{b}$ and $\mathsf{L}_{b}\sim v_{\text{F}}k$,
in our assumed regime we have $\mathsf{L}_{b}\ll\mathsf{W}_{b}$.
Thus we may to good approximation take $\ensuremath{\left(\mathsf{W}_{b}+\mathsf{L}_{b}\right)^{-1}\approx\mathsf{W}_{b}^{-1}}$,
so that 
\begin{equation}
\left(\partial_{t}+\mathsf{W}^{\prime}+\mathsf{L}_{a}\right)|\Phi_{a}\rangle=0
\end{equation}
in our regime, with the effective collision integral $\mathsf{W}^{\prime}=\mathsf{W}_{a}+\mathsf{L}_{ab}\mathsf{W}_{b}^{-1}\mathsf{L}_{ab}^{\dagger}$.  This effective collision integral is the origin of diffusive contributions to the (quasi)hydrodynamic equations for the dynamical modes $|\Phi_a\rangle$.

\subsection{The Quasihydrodynamic Limit (with Imbalance Modes)}

\subsubsection{Square Fermi Surface ($M=4$)}
In this subsection, we derive the quasihydrodynamic equations valid in the limit $\partial_t \ll \gamma_{\mathrm{f}}$.   First we do this for the square, which is simpler as there are fewer degrees of freedom to keep track of.   Following the procedure outlined in Section \ref{subsec:integration}, we integrate out 2 fast modes from the $n=1$ sector to obtain 
\begin{subequations}\label{eq:squarequasihydro}\begin{align}
\partial_{t}N+n_0\partial_{i}\left(V_i+\frac{1}{\sqrt{3}}\widetilde{V}_i\right)  &=0
\\
\partial_{t}V_{i}+\frac{3}{8}\frac{v_{\mathrm{F}}^2}{n_0}\partial_{j}\left(\mdelta_{ij}N+\widetilde{N}_{ij}\right)-\frac{v_{\mathrm{F}}^2}{4\gamma_{\mathrm{f}}}\mathbb{P}^\perp_{jikl}\partial_{j}\partial_{k}\left(V_l-\sqrt{3}\widetilde{V}_l\right)  &=0
\\
\partial_{t}\widetilde{N}_{ij}+ n_0 \mathbb{P}^\parallel_{ijkl}\partial_k\left(V_l+\frac{1}{\sqrt{3}}\widetilde{V}_l\right) &=-\gamma_{\text{s}}\widetilde{N}_{ij}
\\
\partial_{t}\widetilde{V}_{i}+\frac{\sqrt{3}}{8}\frac{v_{\mathrm{F}}^2}{n_0}\partial_{j}\left(\mdelta_{ij}N+\widetilde{N}_{ij}\right)+\frac{\sqrt{3} v_{\mathrm{F}}^2}{4\gamma_{\mathrm{f}}}\mathbb{P}^\perp_{jikl}\partial_{j}\partial_{k}\left(V_l-\sqrt{3}\widetilde{V}_l\right) &=-\gamma_{\text{s}}\widetilde{V}_{i}
\end{align}\end{subequations}
where we have introduced the projection tensors
\begin{subequations}\begin{align}
    \mathbb{P}^\perp_{jikl} &= \frac{1}{2}\left(\mdelta_{jk}\mdelta_{il} - \sigma^z_{jk}\sigma^z_{il}\right), \\
    \mathbb{P}^\parallel_{jikl} &=  \sigma^z_{ij}\sigma^z_{kl}.
\end{align}\end{subequations}
which project onto ``parallel" terms and ``perpendicular" parts of a tensor,  respectively: $\mathbb{P}^\perp_{jikl}a_{ji}a_{kl} = a_{xy}^2 + a_{yx}^2$ and $\mathbb{P}^\parallel_{jikl}a_{ji}a_{kl} = (a_{xx}-a_{yy})^2$.
The ``tensor" degree of freedom $\widetilde{N}_{ij}$ is in fact a scalar: \begin{equation}
    \widetilde{N}_{ij} = \widetilde{N} \sigma^z_{ij},
\end{equation}
but on formal grounds, it is more natural to express the equations of motion as (\ref{eq:squarequasihydro}).   

The form of (\ref{eq:squarequasihydro}) is highly constrained by the symmetry of the square Fermi surface.  These equations are written in terms of invariants of the discrete symmetry group of the square Fermi surface.  In Appendix \ref{app:dihedral}, we discuss the representation theory of the dihedral groups and elucidate the structure of (\ref{eq:squarequasihydro}) from a group theoretic perspective.   Let us emphasize that (\ref{eq:squarequasihydro}) is \emph{not} the most general form of the quasihydrodynamic equations.  If we slightly round the corners of the square, new terms which are allowed by symmetry should generically appear in the equations of motion.  We will not fully classify all such allowed terms in this work.

As a simple application of our quasihydrodynamic theory, let us calculate the quasinormal modes of (\ref{eq:squarequasihydro}) in the limit $\gamma_{\mathrm{s}} = 0$. Namely, we look for plane wave solutions of (\ref{eq:squarequasihydro}) where the $x_i$ and $t$ dependence of all variables is $\mathrm{e}^{\mathrm{i}(k_xx+k_yy-\omega t)}$.  This becomes an eigenvalue problem for a $6\times 6$ matrix.  In the limit $\gamma_{\mathrm{s}}\rightarrow 0$, the results become particularly simple:  \begin{subequations}
    \begin{align}
        \omega_{\pm, \text{sound 1}} = \pm \mathrm{i}v_{\mathrm{F}}k_x, \\
        \omega_{\pm, \text{sound 2}} = \pm \mathrm{i}v_{\mathrm{F}}k_y, \\
        \omega_{\text{diff. 1}} = -\mathrm{i}\frac{v_{\mathrm{F}}^2 k_x^2}{\gamma_{\mathrm{f}}}, \\
         \omega_{\text{diff. 2}} = -\mathrm{i}\frac{v_{\mathrm{F}}^2 k_y^2}{\gamma_{\mathrm{f}}},
    \end{align}
\end{subequations}
One can obtain these results explicitly from (\ref{eq:squarequasihydro}), but it is simpler to instead go back to the Legendre basis introduced previously.  The two sound modes, which propagate ballistically in either the $x$ or $y$ direction, come from the $n=0$ modes on the left/right and top/bottom edges, respectively.   The two diffusive modes describe the diffusive decay of the $n=1$ contributions to transverse momentum.    

On a square with rounded corners, if $k_x \rightarrow 0$ at finite $k_y$ (or vice versa), we do not expect the relevant sound mode to become strictly non-dynamical.  Rather, this sound mode will instead decay diffusively, with a decay rate set by the corrections to our toy model (\ref{eq:squarequasihydro}).  

At finite $\gamma_{\mathrm{s}}$, some of the modes described above pick up additional decay channels due to the slow relaxation of imbalance modes.  We do not have an elegant analytic description for this regime, but will describe it numerically in Section \ref{sec:qhydroplot}, after we discuss the hydrodynamic limit of $\partial_t \ll \gamma_{\mathrm{s}}$.

\subsubsection{Hexagonal Fermi Surface ($M=6$)}
We now repeat the analysis above for the hexagonal Fermi surface.  The quasihydrodynamic equations are 
\begin{subequations}
\begin{align}
\partial_{t}N+n_0\partial_{i}\left(V_i+\frac{1}{3}\widetilde{V}_i\right) & =0
\\
\partial_{t}V_{i}+\frac{9v_{\mathrm{F}}^2}{20n_0}\partial_i N + \frac{3v_{\mathrm{F}}}{2\sqrt{10}} \partial_jQ_{ij}-\eta_{jikl}^{\text{f}}\partial_{j}\partial_{k}\left(V_l-3\widetilde{V}_l\right) & =0
\\
\partial_{t}\widetilde{N}_{ijk}-\frac{\sqrt{10}}{6}n_0\lambda_{ijk}^{+}\lambda_{lmn}^{+}\partial_{l}Q_{mn} & =-\gamma_{\text{s}}\widetilde{N}_{ijk}
\\
\partial_{t}\widetilde{V}_{i} + \frac{3v_{\mathrm{F}}^2}{20n_0} \partial_i N + \frac{v_{\mathrm{F}}}{2\sqrt{10}} \partial_j Q_{ij} -3\eta_{jikl}^{\text{f}}\partial_{j}\partial_{k}\left(3\widetilde{V}_l-V_l\right)& =-\gamma_{\text{s}}\widetilde{V}_{i}
\\
\partial_{t}Q_{ij}+ \frac{3v_{\mathrm{F}}}{2\sqrt{10}}\left[\frac{v_{\mathrm{F}}}{n_0}\partial_k\widetilde{N}_{ijk} + \partial_i V_j + \partial_j V_i - \mdelta_{ij}\partial_k V_k - \frac{1}{3} \left(\partial_i \widetilde{V}_j + \partial_j \widetilde{V}_i - \mdelta_{ij}\partial_k \widetilde{V}_k\right) \right] & =-\gamma_{\text{s}}Q_{ij}
\end{align}
\end{subequations}
with
\begin{equation}
\eta_{ijkl}^{\text{f}}=\frac{v_{\mathrm{F}}^2}{40\gamma_{\text{f}}}\left(\mdelta_{ik}\mdelta_{jl} + \mdelta_{il}\mdelta_{jk} - \mdelta_{ij}\mdelta_{kl}+2\epsilon_{ij}\epsilon_{kl}\right),
\end{equation}
and the rank-3 tensor $\lambda^+_{ijk}$ defined in Appendix \ref{app:dihedral}, Eq. (\ref{eq:lambdaR3}), which we note satisfies the useful identity \begin{equation}
    \lambda_{ijm}^{+}\lambda_{mkl}^{+} = \mdelta_{ik}\mdelta_{jl} + \mdelta_{il}\mdelta_{jk} - \mdelta_{ij}\mdelta_{kl}.
\end{equation}

The properties of these equations are rather similar to the square, except that the symmetry group of the hexagon is $\mathrm{D}_{12}$ rather than $\mathrm{D}_8$.  As such, as the ``scalar" imbalance mode $\widetilde{N}_{ijk}$ now comes with three indices rather than two: $\widetilde{N}_{ijk} = \widetilde{N}\lambda^+_{ijk}$ Similarly, there are an additional two imbalance degrees of freedom found in $Q_{ij}$.  The group theoretic understanding of these equations is found in Appendix \ref{app:dihedral}.

The quasinormal modes in the quasihydrodynamic regime are rather similar to the square case above, except that in general there will be 3 sets of propagating modes, each one propagating normal to one pair of edges of the Fermi surfaces.

\subsection{The Hydrodynamic Limit}
We now turn to the hydrodynamic limit of $\partial_t \ll \gamma_{\mathrm{s}}$.  We will first talk about the hexagon this time, as its hydrodynamic limit turns out to be simpler due to the higher symmetry.

\subsubsection{Hexagonal Fermi Surface ($M=6$)}
The hydrodynamic equations describe the dynamics of the exactly conserved quantities:  density and momentum.  As discussed previously, we will work with the more conventional fluid variables of density and velocity.  Within linear response, the hydrodynamic equations are
\begin{subequations}\label{eq:hexhydro}
\begin{align}
    \partial_t N + n_0 \partial_i V_i - D \partial_i \partial_i N &= 0, \\
    \partial_t V_i + \frac{v_{\mathrm{s}}^2}{n_0}\partial_i N - \frac{1}{mn_0}\eta_{jikl}\partial_k V_l &= 0
\end{align}
\end{subequations}
where the incoherent diffusion constant is \begin{equation}
    D = \frac{v_{\mathrm{F}}^2}{20\gamma_{\mathrm{s}}},
\end{equation}the speed of sound is \begin{equation}
    v_{\mathrm{s}} = \sqrt{\frac{9}{20}}  v_{\mathrm{F}}
\end{equation} 
the viscosity tensor is 
\begin{equation}
    \eta_{ijkl} = \eta (\mdelta_{ik}\mdelta_{jl} + \mdelta_{il}\mdelta_{jk} - \mdelta_{ij}\mdelta_{kl}) + \tilde\eta \epsilon_{ij}\epsilon_{kl}
\end{equation}
and the shear viscosity $\eta$ and \emph{rotational viscosity} $\tilde\eta$ are \begin{subequations}
\begin{align}
    \eta &= mn_0 v_{\mathrm{F}}^2\left(\frac{9}{40\gamma_{\mathrm{s}}}+ \frac{1}{40\gamma_{\text{f}}}\right), \\
    \tilde\eta &= \frac{mn_0 v_{\mathrm{F}}^2}{20\gamma_{\text{f}}}.  \label{eq:hexRotVisc}
\end{align}
\end{subequations}
The form of $\eta_{ijkl}$ has been explicitly computed using the microscopic kinetic theory.

There are two key features of these equations which differ from the conventional hydrodynamics found in textbooks \cite{landau}.   Firstly, we observe the presence of an incoherent \cite{hartnoll1} diffusion constant for charge, $D$.   The origin of this effect is the broken Galilean invariance due to the polygonal Fermi surface.  Because the charge current $J_i$ is not equivalent to the momentum $P_i$ (up to an overall prefactor), it is possible to have a charge current which flows in the absence of momentum.  In the quasihydrodynamic language, this mode corresponds to $\widetilde{V}_i$.  Because $\widetilde{V}_i$ is a quasihydrodynamic mode, it decays at rate $\gamma_{\mathrm{s}}$, and integrating this mode out leads to $D\sim 1/\gamma_{\mathrm{s}}$, as explained in Section \ref{subsec:integration}.

Secondly, we observe that there is both a shear viscosity and a \emph{rotational viscosity}.  To the best of our knowledge, the rotational viscosity $\tilde\eta$ has never been named as such, nor has its significance been described previously in the literature.  The rotational viscosity arises due to the \emph{explicit} breaking of rotational invariance by the ionic lattice and the Fermi surface itself.   In fact, previously studied anisotropic models such as \cite{link} do exhibit rotational viscosity (although the effect was not named or elucidated): this effect is not peculiar to our polygonal Fermi surface model.  We emphasize that this rotational viscosity is not the same as the Hall viscosity \cite{avron}, whose tensor structure is \cite{yarom}
\begin{equation}
    \eta^{\mathrm{Hall}}_{ijkl} \sim \epsilon_{ik}\mdelta_{jl} + \epsilon_{il}\mdelta_{jk} + \epsilon_{jk}\mdelta_{il} + \epsilon_{jl}\mdelta_{ik}
\end{equation} 
This tensor structure is dissipationless: $\eta^{\mathrm{Hall}}_{ijij} = 0$, in contrast to the rotational viscosity; furthermore, $\eta^{\mathrm{Hall}}_{ijkl}=\eta^{\mathrm{Hall}}_{jikl}$, in contrast to the antisymmetric contribution to the electronic stress tensor from the rotational viscosity.

  Normally (without parity violation) the viscosity tensor is assumed to have the same symmetries as the elastic moduli tensor in a solid:   \begin{equation}
    \eta_{jikl}= \eta_{ijkl} = \eta_{klij} . \label{eq:viscsymmetry}
\end{equation}
It is the first equality in (\ref{eq:viscsymmetry}) which is violated by rotational viscosity.  The second equality continues to hold.   So one might ask why in a conventional non-disordered elastic solid (which does break rotational invariance, as all crystalline space groups are discrete) the elastic moduli must obey all equalities in (\ref{eq:viscsymmetry}) \cite{landauvol7}.   The reason is that an elastic solid \emph{spontaneously breaks} rotational invariance.  As a consequence, the symmetry of rotational invariance (which enforces angular momentum conservation, and the symmetry of the stress tensor) is not lost, and instead there are ``massless degrees of freedom": global rotations of a solid, which do not cost any energy.   In contrast, the polygonal Fermi surface is held in place by the ionic degrees of freedom which we have not accounted for (in our standard Born-Oppenheimer approximation separating the electronic and ionic degrees of freedom).   On any time scale where the ionic lattice dynamics is negligible, the electronic fluid moves in an environment where rotational invariance is \emph{explicitly} broken by the lattice and angular momentum can be removed by torques applied by static ions.

We note that a rotational viscosity with identical tensor structure also arises in the Ericksen-Leslie theory of liquid crystal hydrodynamics \cite{ericksen,leslie}.  However, liquid crystals \emph{spontaneously} break rotational invariance and therefore angular momentum conservation is not lost; rotational viscosity is allowed only in an interplay between the velocity and order parameter dynamics \cite{pershanprl,pershan}.  We again emphasize that our model explicitly breaks rotational invariance and so unlike a liquid crystal, rotational viscosity and an antisymmetric stress tensor are physical and will generically exist.

\begin{figure}[t]
\includegraphics[width=.45\textwidth]{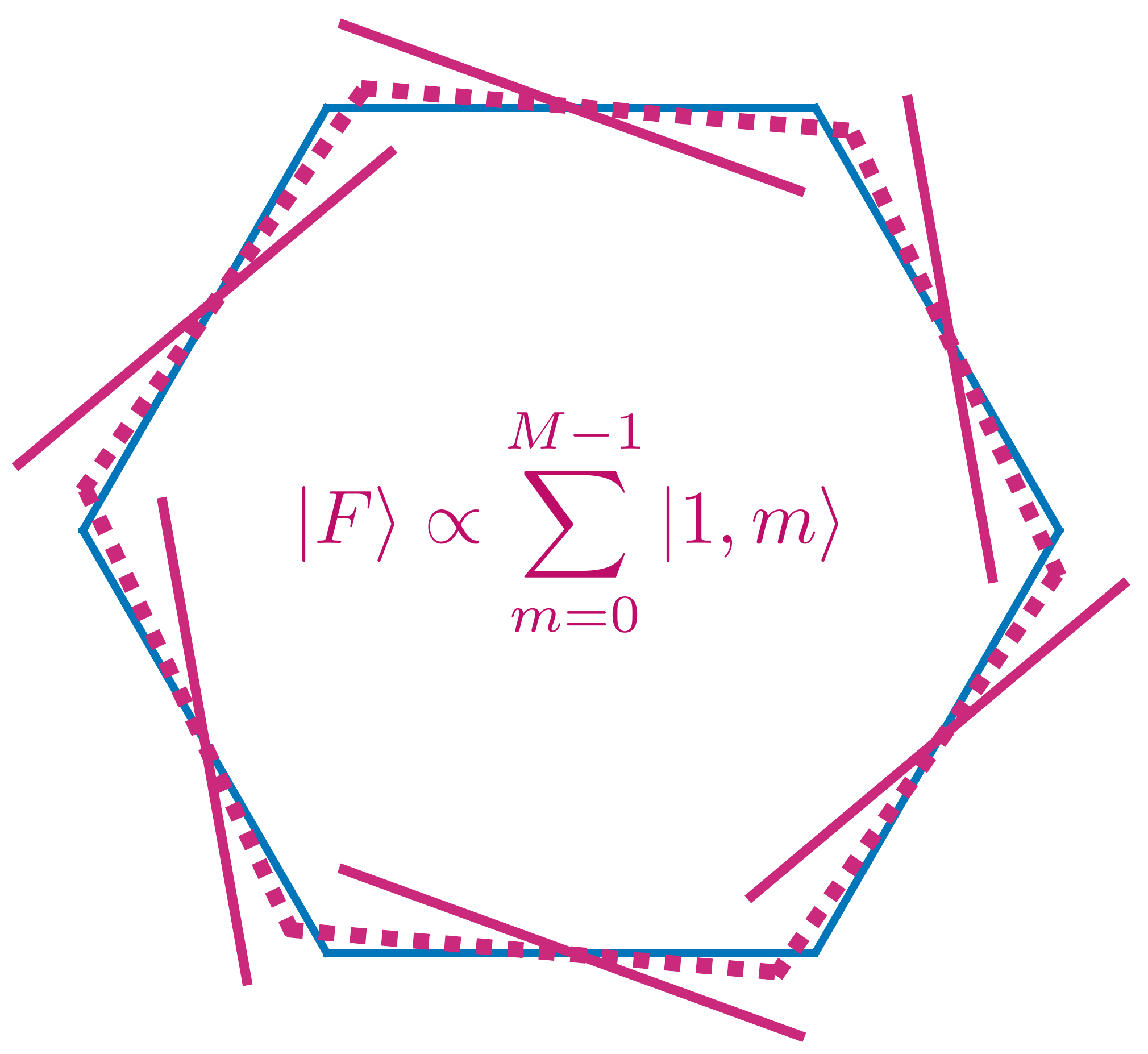}
\centering
\caption{A sketch of a hexagonal Fermi surface (blue) after a rotation by an infinitesimal angle (pink, dashed). Such rotations generate the electronic excitation $|F\rangle$ corresponding to $n=1$ modes on each edge (pink, solid) and thus cost energy. In the hydrodynamic limit, the rapid decay of the mode $|F\rangle$ at the rate $\gamma_\text{f}$ will therefore generate a viscosity $\tilde\eta$ associated with local rotations scaling as $\tilde\eta\sim\gamma_\text{f}^{-1}$. Note here that we have drawn the $n=1$ modes with an exaggerated slope for ease of viewing.}
\label{fig:rotvisc}
\end{figure}

To see that this rotational viscosity is not simply an artifact of our particular model, consider the electronic fluctuation $|F\rangle$ generated when the Fermi surface is rotated by an infinitesimal angle, sketched in Figure \ref{fig:rotvisc}. The mode $|F\rangle$ is a combination of $n=1$ Legendre modes which decays rapidly at the rate $\gamma_\text{f}$. Since viscosity in the (quasi)hydrodynamic limit arises from integrating out fast modes, including $|F\rangle$, we therefore expect that the viscosity associated with local rotations will scale as $\tilde\eta \sim 1/\gamma_{\mathrm{f}}$, which is indeed consistent with our results (\ref{eq:hexRotVisc}, \ref{eq:squareRotVisc}).  In contrast, shear viscosity arises from integrating out quasihydrodynamic slow modes, and so $\eta \sim 1/\gamma_{\mathrm{s}}$.  Hence the rotational viscosity is supressed by factor $(\gamma_\text{s}/\gamma_\text{f})\ll 1$ relative to the shear viscosity, leading to the hierarchy of viscosities $0<\tilde\eta \ll \eta$.\footnote{Note that there is no bulk viscosity:  it is a generic result that the bulk viscosity of a Fermi liquid is suppressed by $(T/T_{\mathrm{F}})^4$ \cite{sykes}.}

Both of the above points regarding incoherent conductivity and viscosity in an electronic fluid are generic.  They will be relevant for essentially all Fermi liquids (and electron fluids more broadly) except for those with nearly circular Fermi surfaces, such as graphene \cite{bandurin, crossno} or GaAs \cite{molenkamp,bakarov}.   It is an important open question to develop practical methods to observe the presence of $D$ and $\tilde\eta$ in an experiment.

\subsubsection{Square Fermi Surface ($M=4$)}
On a square Fermi surface, the hydrodynamic equations of motion take the same form as (\ref{eq:hexhydro}).    What changes are the values of the relevant parameters.  The incoherent diffusion constant is \begin{equation}
    D = \frac{v_{\mathrm{F}}^2}{8\gamma_{\mathrm{s}}},
\end{equation}
the speed of sound is \begin{equation}
    v_{\mathrm{s}} = \sqrt{\frac{3}{8}}v_{\mathrm{F}},
\end{equation}
and the viscosity tensor becomes \begin{equation}
 \eta_{ijkl} = \eta_\parallel \mathbb{P}^\parallel_{ijkl} + \eta_\perp \mathbb{P}^{\perp\mathrm{s}}_{ijkl} +    \tilde\eta \epsilon_{ij}\epsilon_{kl}
\end{equation}
where \begin{equation}
    \mathbb{P}^{\perp\mathrm{s}}_{ijkl} = \frac{1}{2}\left(\mdelta_{ik}\mdelta_{jl} + \mdelta_{il}\mdelta_{jk} - \sigma^z_{ik}\sigma^z_{jl}-\sigma^z_{il}\sigma^z_{jk}\right)
\end{equation}
is a projector onto symmetric shear components of a tensor:  $\mathbb{P}^{\perp\mathrm{s}}_{ijkl} a_{ij}a_{kl} = (a_{xy}+a_{yx})^2$, and the three viscosities are \begin{subequations}
\begin{align}
    \eta_\parallel &= \frac{3v_{\mathrm{F}}^2}{8\gamma_{\text{s}}} mn_0 , \\
    \eta_\perp = \tilde\eta &= \frac{v_{\mathrm{F}}^2}{8\gamma_{\text{f}}} mn_0.
    \label{eq:squareRotVisc}
\end{align}
\end{subequations}
The last equality appears to be a coincidence within our toy model.

In the form we have written it above, the viscosities $\eta_\parallel$ and $\eta_\perp$ are two of the three allowed components of a fourth rank tensor in two dimensions with the symmetries (\ref{eq:viscsymmetry}):  they correspond to longitudinal and transverse shear viscosities.   As $\eta_\parallel$ arises due to the decay of $\widetilde{N}_{ij}$, $\eta_\parallel \sim 1/\gamma_{\mathrm{s}}$, while $\eta_\perp \sim 1/\gamma_{\mathrm{f}}$ as it arises entirely from $n=1$ modes.  $\tilde\eta$ is again the rotational viscosity, and its interpretation is identical to before.

\subsubsection{Quasinormal Modes}\label{sec:qhydroplot}

In either of the above cases, we can discuss the quasinormal mode solutions to the linearized hydrodynamic equations.  We will discuss the square case (without assuming $\eta_\perp = \tilde\eta$) as it is more generic; the hexagonal hydrodynamics follows upon setting $\eta=\eta_\perp=\eta_\parallel$.    The hydrodynamic modes are the usual sound wave, coupling the density $N$ with the longitudinal velocity $k_iV_i$,\footnote{Even in an anisotropic system, the dihedral symmetry group is sufficiently strong to ensure this is the case \cite{lucasRFB}.} together with the diffusion of transverse momentum $k_i \epsilon_{ij}V_j$.  The dispersion relation for the sound modes is \begin{equation}
    \omega = \pm v_{\mathrm{s}}k - \frac{\mathrm{i}}{2}\left(D + \frac{\eta_\perp}{mn_0}  \sin^2(2\varphi)+ \frac{\eta_\parallel}{mn_0}  \cos^2(2\varphi)\right)k^2 + \mathrm{O}\left(k^3\right)
\end{equation}
where $\tan\varphi = k_y/k_x$, and the dispersion relation for the transverse diffusion mode is \begin{equation}
    \omega = -\mathrm{i} \frac{\tilde\eta + \eta_\perp \cos^2(2\varphi) + \eta_\parallel \sin^2(2\varphi)}{mn_0}k^2+ \mathrm{O}\left(k^3\right) .
\end{equation}

One important feature of these equations is the relative anisotropy in the decay rates of the quasinormal modes:  for small angles $\varphi \approx 0$ the sound wave decays much faster than the diffusion mode ($\eta_\parallel \gg \eta_\perp$ in our models), whereas when $\varphi \approx \mpi/4$ the sound wave decays much slower.    Just as important are the new dissipative contributions:  the incoherent charge diffusion constant $D_0$ contributes to the decay of the sound mode, while the rotational viscosity $\tilde\eta$ contributes to the decay of transverse momentum. 

\begin{figure}[t]
\includegraphics[width=.7\textwidth]{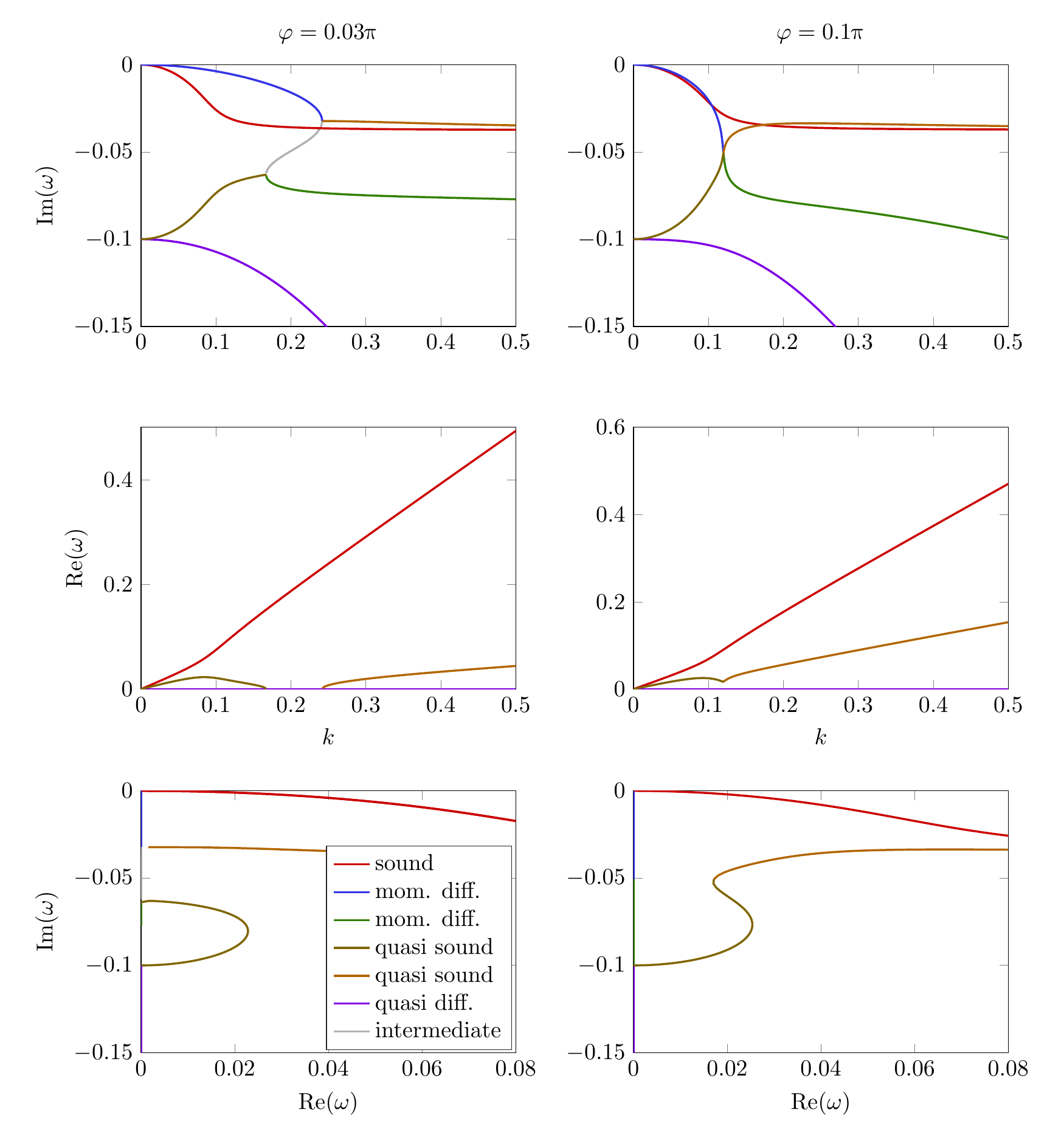}
\centering
\caption{Quasinormal modes across the quasihydrodynamic-to-hydrodynamic crossover on the square.  We take $v_{\mathrm{s}}=\gamma_{\mathrm{f}} = 1$ and $\gamma_{\mathrm{s}} = 0.1$.  Some modes are noted twice in the figure legend:  this arises when these modes are involved in pole collisions, so there may be some ambiguity as to which mode is labeled as which.}
\label{fig:qnms}
\end{figure}

Finally, let us return to the behavior of the quasinormal modes across the transition between the quasihydrodynamic and hydrodynamic regimes.  For simplicity, we will focus on the square ($M=4$), where we argued above that there will be 2 sound and 2 diffusion modes in the quasihydrodynamic regime, and 1 sound and 1 diffusion mode in the hydrodynamic regime.  Figure \ref{fig:qnms} shows the intricate interplay between these modes as a function of the angle $\varphi$ that the wave number $k_i$ makes with the Fermi surface ($\varphi=n\mpi/2$ for $n\in\mathbb{Z}$ implies that $k_i$ is oriented along the Fermi velocity on two of the edges).   Roughly speaking, the hydrodynamic sound mode is a well defined quasinormal mode throughout the entire hydrodynamic and quasihydrodynamic regime.   Although it picks up a finite decay rate $\gamma_{\mathrm{s}}$  there exists a well-behaved ``quasi-diffusion" mode which is non-propagating ($\mathrm{Re}(\omega(k))=0$).  In contrast, the transverse momentum diffusion mode and the other ``quasi-sound" mode have a very curious interplay across the hydrodynamic-to-quasihydrodynamic transition.   Depending on the angle $\varphi$, we observe in Figure \ref{fig:qnms} that these two modes can either collide with one another or not.  If they do not collide, then the ``quasi-sound" is a well-defined excitation for any $k$ (with a decreasing decay rate as $k$ increases), and the transverse momentum diffusion pole is also well-defined for all $k$.   However, if these two modes collide, there is an interesting sequence of two pole collisions.  First, the two ``quasi sound" poles collide with each other on the imaginary axis at a finite $k$, and split into two non-propagating and purely dissipative modes.  One of these dissipative modes becomes the secondary diffusion mode in the quasihydrodynamic limit $k \gg \gamma_{\mathrm{s}}/v_{\mathrm{F}}$, while the other moves up the imaginary axis towards the hydrodynamic momentum diffusion mode.  The hydrodynamic momentum diffusion pole then collides with ``half" of the original ``quasi sound" mode to form the ``quasi sound" mode which will persist throughout the quasihydrodynamic regime.  We emphasize that these two different behaviors occur for the same physical parameters -- the only thing which is changing is the angular orientation of $k$.  A better understanding of the experimental implications of these pole collisions (and/or the feasibility of observing them experimentally) is an interesting future direction to consider.

In a system with long-range Coulomb interactions (including most electronic fluids), the hydrodynamic sound mode described above morphs into a plasmon with a significantly modified dispersion relation \cite{lucas1801}.   Due to the presence of the incoherent conductivity (i.e. the breaking of Galilean invariance), the decay of the plasmon is significantly enhanced \cite{lucasplasma}.  We will not describe this effect in detail in this paper.

\section{Flows in Narrow Channels}\label{sec:channel}

\begin{figure}[t]
\includegraphics[width=.45\textwidth]{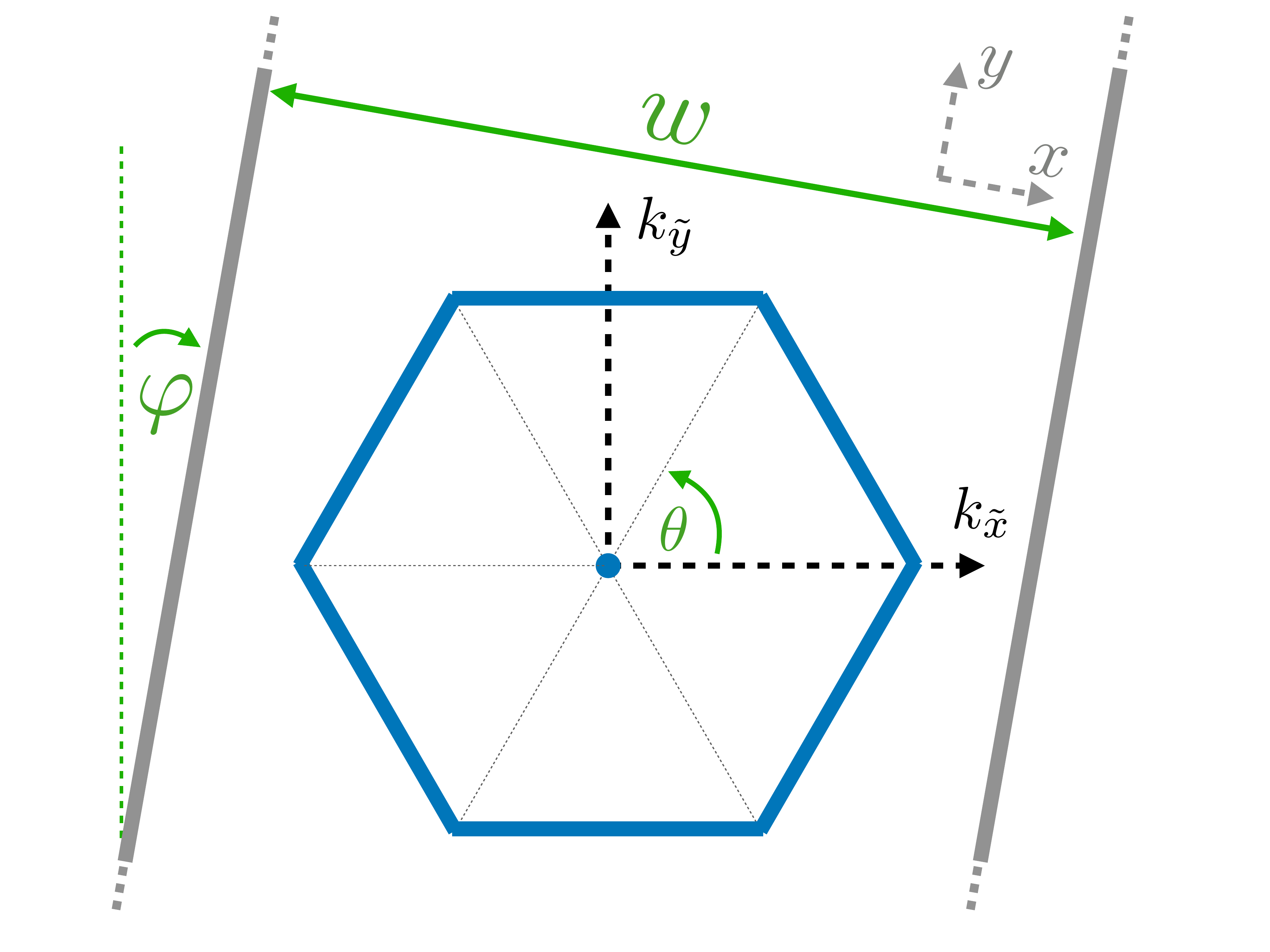}
\centering
\caption{Sketch of the channel flow problem for a channel of width $w$ and a hexagonal Fermi surface. The angle $\varphi$ parameterizes the angular offset of the Fermi surface from the transverse direction of the channel. The spatial coordinates $x,y$ and $\tilde{x},\tilde{y}$ refer to those of the channel and the crystal, respectively.}
\label{fig:hexChannel}
\end{figure}

Our primary application of these kinetic and hydrodynamic equations is their solution in a long and narrow channel: see Figure \ref{fig:hexChannel}.  In particular, we assume that the channel is infinitely long and has a finite width $w$, that electric current is driven by a background electric field applied along the channel, and that the dynamics is independent of time. This is precisely the experimental setup of \cite{mackenzie}, along with the originally proposed test \cite{gurzhi} for hydrodynamic electron flow.  As we go, we will explain the signatures of hydrodynamics we are after, along with how the polygonal models differ from a circular Fermi surface.   We also note that magnetotransport in such channels, which we will not address in this paper, has been studied theoretically in \cite{scaffidi,alekseev18,alekseev19} and experimentally in \cite{haug14,bakarov1810}.

\subsection{The Boltzmann Equation with a Source}
Our starting point is to generalize the linearized Boltzmann equation (\ref{eq:matrixBoltz}) to include a small background electric field.  The background electric field $\mathbf{E}$ will be of the same order as $\Phi$, as it will drive the electronic system out of thermal equilibrium.  The key observation is that starting from the fully nonlinear Boltzmann equation (\ref{eq:fullBoltz}), only a single term $\mathbf{F}\cdot \partial_{\mathbf{p}} f$ depends explicitly on the applied electric field $\mathbf{E}$.   Since \begin{equation}
    \mathbf{F}\cdot \partial_{\mathbf{p}}f = -\frac{\partial f^0}{\partial \epsilon_{\mathbf{p}}} e\mathbf{E}\cdot \mathbf{v}_{\mathbf{p}}  +  \mathrm{O}\left(\Phi\mathbf{E}\right),
\end{equation}
we find that the only change to (\ref{eq:matrixBoltz}) is to add a source term proportional to $\mathbf{E}$: \begin{equation}
    \partial_t |\Phi\rangle + (\mathsf{W}+\mathsf{L})|\Phi\rangle = E_i |\mathsf{J}_i\rangle,  \label{eq:sec4boltz}
\end{equation}
where the current vector \begin{equation}
    |\mathsf{J}_i(\mathbf{x})\rangle = -e\int \mathrm{d}^d\mathbf{p}\; v_i(\mathbf{p}) |\mathbf{xp}\rangle.
\end{equation}
Note that $|\mathsf{J}_i\rangle$ is proportional to $|P^0_i\rangle $.  It will be convenient below to also define the matrices $\mathsf{V}_i$ such that $\mathsf{V}_i|\mathbf{xp}\rangle = v_i(\mathbf{p}) |\mathbf{xp}\rangle$, since 
\begin{equation}
    \mathsf{L} = \mathsf{V}_i \partial_{x_i}.
\end{equation}
As our setup assumes that the electric field does not vary with time and that the electronic system has reached a steady state $t$-independent solution, we will set $\partial_t=0$ henceforth.

We will solve (\ref{eq:sec4boltz}) in the channel geometry given in Figure \ref{fig:hexChannel}.  Note that there are two natural choices of coordinate system to use:  one aligned with the Fermi surface (as in the previous section), and one aligned with the channel.  It is more useful for us to orient our coordinate system with the channel, which is rotated by an angle $\varphi$ from the Fermi surface coordinates.   The fact that this relative orientation of the Fermi surface and the channel is allowed is a key difference between the circular Fermi surface and the polygonal Fermi surface, and we will discuss its consequences below.  For the purposes of solving the Boltzmann equation, we will assume that the channel is homogeneous in the $y$ direction.  Looking for time independent solutions sourced by a constant electric field in the $y$ direction, (\ref{eq:sec4boltz}) reduces to \begin{equation}
    \mathsf{V}_{x} \partial_{x}|\Phi(x)\rangle + \mathsf{W}|\Phi( x)\rangle = E_{ y} |\mathsf{J}_{ y}\rangle. \label{eq:4boltz}
\end{equation}
The conductance of the infinite channel is then defined as follows: \begin{equation}
    GL = \frac{1}{E_{y}}\int\limits_0^w \mathrm{d}x \; \langle \mathsf{J}_{ y} | \Phi(x)\rangle.
\end{equation}
This is simply Ohm's Law:  $GV = I$ ($G=1/R$).

\subsection{Boundary Conditions}
To solve these equations, we must employ suitable boundary conditions.  The boundary conditions that we discuss in this paper take the following form.  Let $|\Phi_>\rangle $ denote the right-moving modes (eigenvectors of $\mathsf{V}_x$ with positive eigenvalues) and $|\Phi_<\rangle $ denote the left-moving modes (eigenvectors of $\mathsf{V}_x$ with negative eigenvalues).  We denote with  $\mathsf{V}_{x>}$ the block submatrix of $\mathsf{V}_x$ which acts on positive eigenvectors, and $\mathsf{V}_{x<}$ the submatrix which acts  on negative eigenvectors.  The boundary conditions will take the form \begin{subequations}\label{eq:bc}\begin{align}
    |\Phi_>(0)\rangle &= \mathsf{M}_{\mathrm{left}}|\Phi_<(0)\rangle, \\
    |\Phi_<(w)\rangle &= \mathsf{M}_{\mathrm{right}}|\Phi_>(w)\rangle.
\end{align}\end{subequations}
The matrices $\mathsf{M}_{\mathrm{left}}$ and $\mathsf{M}_{\mathrm{right}}$ are independent of $w$.  These boundary conditions are easy to understand on physical grounds: at $x=0$, the left-moving modes scatter off of the boundary and become right-moving modes, while at $x=w$ the right-moving modes scatter into left-moving modes .   We will neglect boundary conditions for null vectors of $\mathsf{V}_x$ in this paper: for almost all $\varphi$, $\mathsf{V}_x^{-1}$ is invertible, and we will not present results directly at $\varphi=0$ where the physics becomes singular.

Not all $\mathsf{M}_{\mathrm{left}}$ and $\mathsf{M}_{\mathrm{right}}$ are acceptable.   One constraint on these boundary conditions arises from the demand that the normalized conductance $G\ge 0$:  \begin{align}
    0\le   G &= \int\limits_0^w \mathrm{d}x \;  \langle \Phi|\mathsf{J}_y\rangle = \int\limits_0^w \mathrm{d}x \langle \Phi|\left( \mathsf{V}_x \partial_x + \mathsf{W} \right)|\Phi\rangle \notag \\
    &= \langle \Phi(w)|\mathsf{V}_x|\Phi(w)\rangle - \langle \Phi(0)|\mathsf{V}_x|\Phi(0)\rangle +  \int\limits_0^w \mathrm{d}x \; \langle \Phi|\mathsf{W}|\Phi\rangle \notag \\
    &= \left[\langle \Phi_>(w)|\mathsf{V}_x|\Phi_>(w)\rangle - \langle \Phi_<(0)|\mathsf{V}_x|\Phi_<(0)\rangle +  \int\limits_0^w \mathrm{d}x \; \langle \Phi|\mathsf{W}|\Phi\rangle\right] \notag \\
    &\;\;\;- \left[\langle \Phi_<(0)|\mathsf{M}_{\mathrm{left}}^{\mathsf{T}}\mathsf{V}_{x>}\mathsf{M}_{\mathrm{left}}|\Phi_<(0)\rangle-\langle \Phi_>(w)|\mathsf{M}_{\mathrm{right}}^{\mathsf{T}}\mathsf{V}_{x<}\mathsf{M}_{\mathrm{right}}|\Phi_>(w)\rangle  \right]  . 
\end{align}
In this equation, $|\Phi\rangle$ is to be understood as the solution to (\ref{eq:4boltz}), obeying boundary conditions (\ref{eq:bc}).  In the last line above, the terms inside the square brackets are positive   semidefinite.  Unfortunately, we observe that the second brackets comes with an overall minus sign.  One way to ensure that the boundary conditions are consistent is thus to demand\footnote{For matrices, $\mathsf{A}\ge \mathsf{B}$ if and only if $\langle \Phi|(\mathsf{A}-\mathsf{B})|\Phi\rangle \ge 0$ for any $|\Phi\rangle$.}\begin{subequations}\label{eq:32bcreq}\begin{align}
    \mathsf{V}_{x>} &\ge -\mathsf{M}_{\mathrm{right}}^{\mathsf{T}}\mathsf{V}_{x<}\mathsf{M}_{\mathrm{right}}, \\
    -\mathsf{V}_{x<} &\ge \mathsf{M}_{\mathrm{left}}^{\mathsf{T}}\mathsf{V}_{x>}\mathsf{M}_{\mathrm{right}}.
\end{align}\end{subequations}

In what follows, we will employ the boundary conditions \begin{equation}
    \mathsf{M}_{\mathrm{left}} = \mathsf{M}_{\mathrm{right}} = 0,\label{eq:ourbc}
\end{equation}
which are manifestly compatible with (\ref{eq:32bcreq}).

\subsection{Conductance Across the Ballistic-to-Hydrodynamic Crossover}

\subsubsection{Ballistic Limit}
We first show that for any Fermi surface, $G/w^2$ must be a constant within the Boltzmann framework when the collision integral vanishes and when we employ the generic boundary conditions (\ref{eq:bc}).   If the collision integral vanishes, then (\ref{eq:4boltz}) is solved by \begin{subequations}\label{eq:43phi}\begin{align}
    |\Phi_>(x)\rangle = \mathsf{M}_{\mathrm{left}}|\Phi_<(0)\rangle + E_y  x \mathsf{V}_{x>}^{-1}|\mathsf{J}_y\rangle, \\
    |\Phi_<(x)\rangle = \mathsf{M}_{\mathrm{right}}|\Phi_>(w)\rangle - E_y (w- x) \mathsf{V}_{x<}^{-1}|\mathsf{J}_y\rangle.
\end{align}\end{subequations}
We now analyze \begin{align}
    E_y^2 G &= \langle \Phi_>(w)|\mathsf{V}_x|\Phi_>(w)\rangle + \langle \Phi_>(w)|\mathsf{M}_{\mathrm{right}}^{\mathsf{T}}\mathsf{V}_{x<}\mathsf{M}_{\mathrm{right}}|\Phi_>(w)\rangle \notag \\
    &\;\;\;\; - \langle \Phi_<(0)|\mathsf{V}_x|\Phi_<(0)\rangle - \langle \Phi_<(0)|\mathsf{M}_{\mathrm{left}}^{\mathsf{T}}\mathsf{V}_{x>}\mathsf{M}_{\mathrm{left}}|\Phi_<(0)\rangle.  \label{eq:43G}
\end{align}
Our claim is that $|\Phi_>(0)\rangle $, $|\Phi_>(w)\rangle $, $|\Phi_<(0)\rangle $ and  $|\Phi_<(w)\rangle $ all scale proportionally with $w$.  Suppose that $|\Phi_<(0)\rangle \propto w$.  Then from (\ref{eq:43phi}), we immediately obtain that the other three all scale with $w$.  Manipulating (\ref{eq:43phi}), we obtain \begin{equation}
    \left(1 - \mathsf{M}_{\mathrm{right}}\mathsf{M}_{\mathrm{left}}\right)|\Phi_<(0)\rangle = E_y w \left( \mathsf{M}_{\mathrm{right}} \mathsf{V}_{x>}^{-1} - \mathsf{V}_{x<}^{-1} \right)|\mathsf{J}_y\rangle.
\end{equation}
Since $|\Phi(0)\rangle $ and $|\Phi(w)\rangle$ both scale linearly with $w$, we conclude from (\ref{eq:43G}) that $G \propto w^2$.

With boundary conditions (\ref{eq:ourbc}), the ballistic $w\gamma_\text{f}/v_\text{F}\ll(\varphi\text{ mod }\theta)$ conductance for the even-sided $M$-gon is given by
\begin{equation}
\label{eq:heatBallistic}
    G_{\text{ballistic}}(w,\varphi) = \frac{\mathcal{G}(w)}{M}\sum_{m=0}^{M/2-1}\frac{\cos^{2}\left[m\theta+\left(\varphi\text{ mod }\theta\right)\right]}{\sin\left[m\theta+\left(\varphi\text{ mod }\theta\right)\right]}
\end{equation}
where $\theta=2\mpi/M$ is the symmetry angle of the $M$-gon. In the above we have also introduced the conductance
\begin{equation}
    \mathcal{G}(w)\equiv\left(\frac{\nu e^{2}v_{\text{F}}}{L}\right)w^{2},
\end{equation}
where $L$ is the channel length (assumed to be larger than any other length scale in the problem) and $\nu$ the electronic density of states.

Note that the ballistic conductance (\ref{eq:heatBallistic}) diverges to infinity whenever $(\varphi\text{ mod }\theta)=0$, since in that case two of the Fermi surface edges are exactly transverse to the channel direction, allowing the applied electric field to excite non-decaying electrons that never strike either channel wall.

\subsubsection{Hydrodynamic Limit}
In the hydrodynamic limit $1 \ll w/\ell_\text{s}=w\gamma_\text{s}/v_\text{F}$, we may approximate $G(w)$ by solving the hydrodynamic equations, rather than the full Boltzmann equation.  Because the channel is translation invariant in the $y$-direction, and in the $x$-direction up to boundaries, the only hydrodynamic equation of relevance becomes \begin{equation}
    enE_y = -\hat\eta(\varphi) \partial_x^2 v_y
\end{equation}
where $\hat\eta(\varphi) =\eta_{xyxy}(\varphi)$ is the relevant component of the viscosity tensor in the channel coordinates $x, y$. The solution to this hydrodynamic equation is the classic Poiseuille flow \cite{lucasreview17} \begin{equation}
    v_y(x) = \frac{nE_y}{2\hat\eta(\varphi)} x(w-x),
\end{equation}
from which the total current can be found for a given electric field, leading to  our final result for the conductance: 
\begin{equation}
    G = \frac{n^2e^2 w^3}{12\hat\eta(\varphi)}.  \label{eq:poiseuille}
\end{equation}
With a polygonal Fermi surface, it is possible for $\hat\eta(\varphi)$ to have angular dependence due to the more complicated tensor structures in $\eta_{ijkl}$. Letting $\tilde{x}, \tilde{y}$ denote the crystal coordinates, we calculate the channel coordinate viscosity $\hat{\eta}(\varphi)$ by rotating through the angle $\varphi$ from the crystal coordinates to the channel coordinates:
\begin{equation}
   \eta_{ijkl} =
    R\left(\varphi\right)_{i\tilde{i}}
    R\left(\varphi\right)_{j\tilde{j}}
    R\left(\varphi\right)_{k\tilde{k}}
    R\left(\varphi\right)_{l\tilde{l}}
    \eta_{\tilde{i}\tilde{j}\tilde{k}\tilde{l}}.
\end{equation}
For a square Fermi surface, we find that 
\begin{equation}
    \hat\eta(\varphi) = \tilde\eta + \eta_\perp\cos^2(2\varphi) + \eta_\parallel\sin^2(2\varphi), \label{eq:hateta}
\end{equation}
while for a hexagonal Fermi surface we obtain \begin{equation}
    \hat\eta(\varphi) = \tilde\eta + \eta.
\end{equation}
Note that the hexagonal channel viscosity is independent of the offset angle $\varphi$ due to the enhanced $\mathrm{D}_{12}$ symmetry, as noted before.

\subsubsection{Numerical Results}
In addition to solving the equations in these two extreme limits, we may also numerically solve the Boltzmann equation for arbitrary values of $w/\ell_{\mathrm{s}}$.

\begin{figure}[t]
\centering
\subfloat{
\includegraphics[width=.48\textwidth]{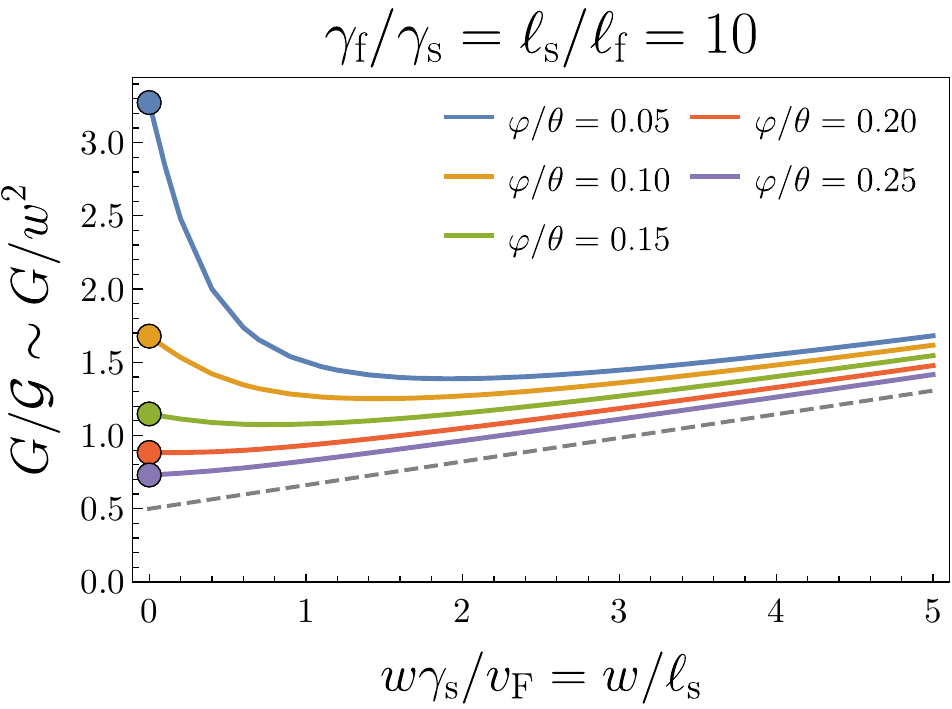}
}
\subfloat{
\includegraphics[width=.48\textwidth]{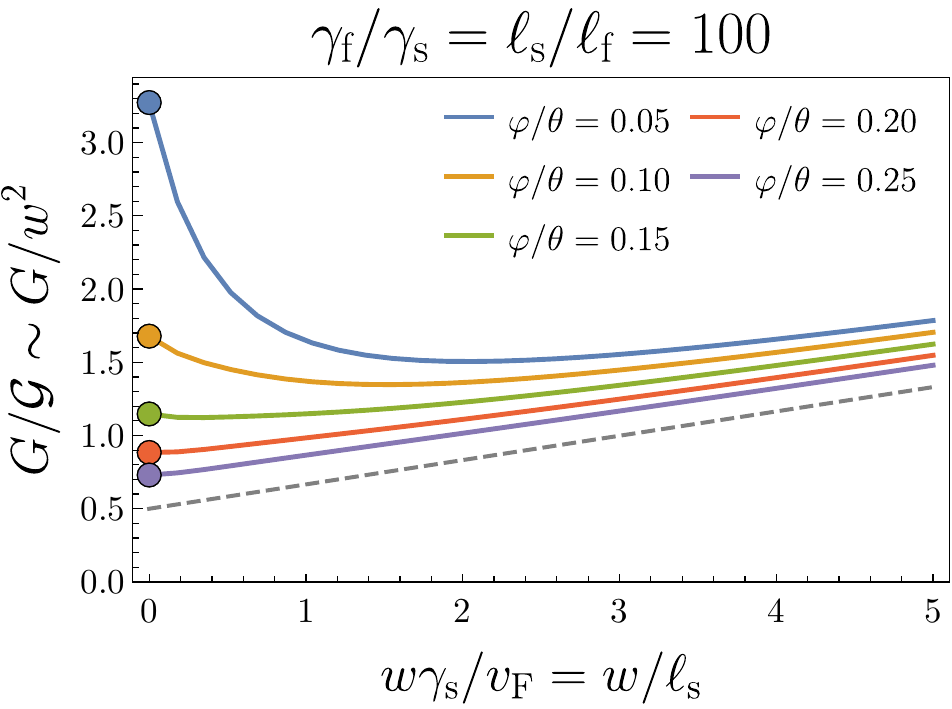}
}
\caption{Numerically computed channel conductance for a hexagonal Fermi surface at different values of $\varphi$ and $\gamma_\text{f}/\gamma_\text{s}$. The slope of viscous hydrodynamic result is shown as a dashed line ($\varphi$-independent in the hexagonal case), and the ballistic result (\ref{eq:heatBallistic}) is plotted as a dot for each channel-offset angle $\varphi$.}
\label{fig:hexNum}
\end{figure}

Our results for a hexagonal Fermi surface ($M=6$) are presented in Figure \ref{fig:hexNum}.   In the hydrodynamic limit of large $w$, we see that all curves are of the form \begin{equation}
    G \approx A(\varphi)w^2 + \frac{n^2e^2 w^3}{12\hat\eta(\varphi)}.
\end{equation} 
As $w\rightarrow \infty$, this precisely matches the predictions of the Navier-Stokes equations.   The coefficient $A(\varphi)$ appears to weakly depend on angle $\varphi$.  Since for the hexagon $\hat\eta(\varphi)$ is angle independent, in the hydrodynamic limit the conductance becomes insensitive to the orientation of the hexagon to leading order in $w$.  Interestingly, the coefficient $A(\varphi)$ is not the same as the ballistic conductance (\ref{eq:heatBallistic}):  as a consequence, we observe a strong non-monotonicity in the $w$ dependence of $G/w^2$ for shallow angles $\varphi$ where the ballistic conductance is large.   In fact, we can understand all of the qualitative features in $G/w^2$.  In our toy model, the current is entirely a quasihydrodynamic mode and only couples at all to $n=1$ modes through quasihydrodynamic decay.   Therefore, we expect $G/w^2$ to depend very weakly on $\gamma_{\mathrm{f}}$.  Indeed, $G/w^2$ essentially only depends on the ratio $w/\ell_{\mathrm{s}}$ -- the length scale over which quasihydrodynamic modes decay determines the ballistic-to-hydrodynamic crossover.  Next, since the lifetime (and therefore the correlation length) of the quasihydrodynamic modes does not depend on $\varphi$, we conclude that the hydrodynamic prediction for $G(w)$ must be quantitatively accurate once $w\gg \ell_{\mathrm{s}}$, independently of whether $G(0)$ is larger or smaller than the hydrodynamic result.   Drawing a curve which smoothly interpolates between (\ref{eq:heatBallistic}) for $w\ll \ell_{\mathrm{s}}$ and (\ref{eq:poiseuille}) for $w\gg \ell_{\mathrm{s}}$, we recover all qualitative features observed in Figure \ref{fig:hexNum}.

\begin{figure}
\centering
\subfloat{
\includegraphics[width=.48\textwidth]{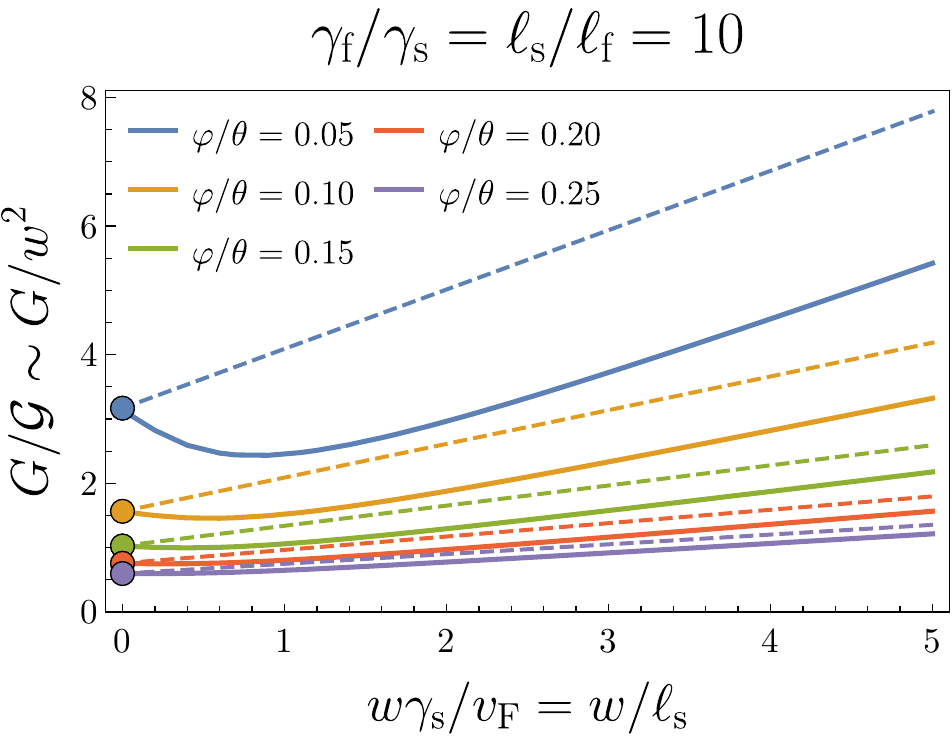}
}
\subfloat{
\includegraphics[width=.48\textwidth]{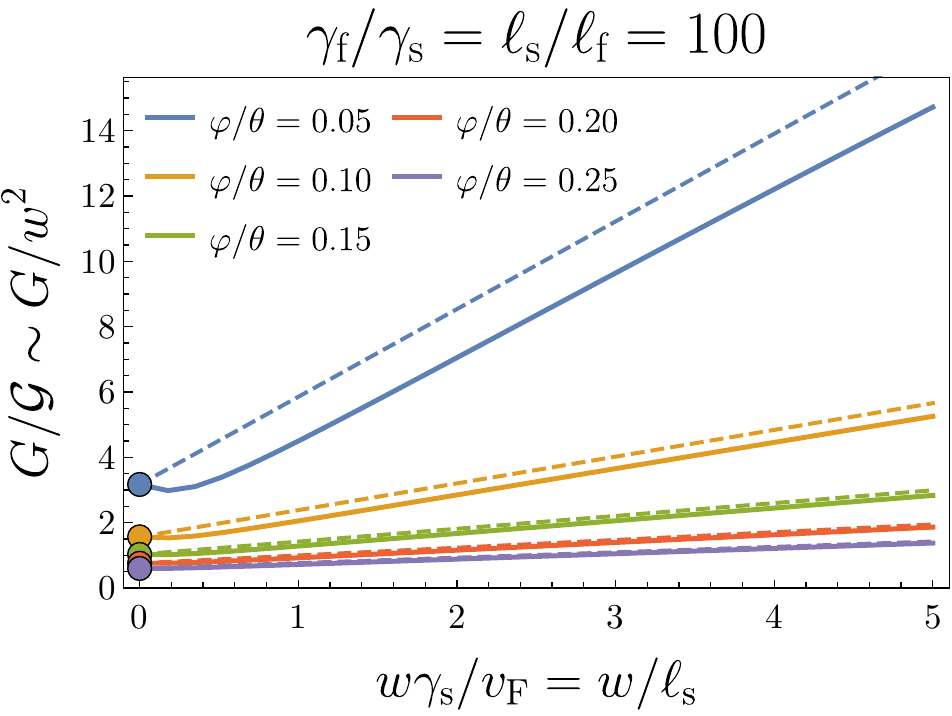}
}
\caption{Numerically computed channel conductance for a square Fermi surface at different values of $\gamma_\text{f}/\gamma_\text{s}$. For each channel-offset angle $\varphi$, the viscous hydrodynamic result is shown as a dashed line (only the slope is physically relevant), and the ballistic result (\ref{eq:heatBallistic}) is plotted as a dot.}
\label{fig:squNum}
\end{figure}

A similar result holds for the channel conductance of a square Fermi surface ($M=4$), shown in Figure \ref{fig:squNum}.  The only important difference here is the angular dependence that arises in $\hat\eta(\varphi)$, as given in (\ref{eq:hateta}).  Using the scalings $\eta_\perp\sim 1/\gamma_{\mathrm{f}}$ and $\eta_\parallel\sim 1/\gamma_{\mathrm{s}}$, along with the result $\gamma_{\mathrm{s}} \ll \gamma_{\mathrm{f}}$, we conclude that as $\varphi\rightarrow 0$ there will be a dramatic reduction in $\hat\eta(\varphi)$, and a correspondingly large enhancement in $G(w)$.  The effect is so strong that it nearly destroys the non-monotonic $w$-dependence in the channel conductance $G(w)$ for small angles $\varphi$.   On the other hand, the angular dependence persists into the hydrodynamic limit.   This strong angular dependence is a clear prediction for experimental studies of viscous flows in materials with square-like Fermi surfaces. 

We also emphasize that in the hydrodynamic limits discussed above, the temperature dependence of the conductance is $G\sim T^2$, since $\gamma_{\mathrm{s,f}}\sim T^2$ (See Appendix \ref{app:coll}).   This persists so long as the dominant source of scattering is two-body electronic collisions.  The temperature dependence (and angular dependence of viscosity) may change if electron-phonon scattering is taken into account:  see e.g. \cite{coulter}.

\section{Conclusion}
We have developed a simple model of the ballistic-to-hydrodynamic crossover in a Fermi liquid with a nearly perfect polygonal Fermi surface.  Qualitative features of the crossover to viscous flow are sensitive to the shape of the Fermi surface.  A particularly simple example of the discrepancy between polygon and circular Fermi surface is found in the crossover between Knudsen and Poiseuille (or Gurzhi) flow in narrow channels.   In a Fermi liquid with a circular Fermi surface, the conductance is a monotonically increasing function of both channel width and temperature.  In contrast, strong non-monotonic width and temperature dependence are possible with the polygon Fermi surface, depending on the relative orientation of the Fermi surface and the channel boundaries.  A common feature of the circular and polygon Fermi surface models is that the width and temperature dependence of conductance are (in the absence of momentum relaxing scattering away from the boundaries) not independent of each other.  This feature of our model casts additional doubt on the proposed hydrodynamic interpretation of the unusual transport data in $\mathrm{PdCoO}_2$, presented in \cite{mackenzie}.  

Even deep in the hydrodynamic limit, the hydrodynamic behavior of the electron fluid changes when the Fermi surface is anisotropic.  The most interesting new phenomenon is the emergence of a new dissipative viscosity, $\tilde\eta$, which arises from the explicit breaking of rotational invariance by the crystal lattice.  It would be interesting if either nonlinear optical response \cite{zaanen} or the vicinity geometry \cite{polini, levitovhydro, torre, levitov1806} used to probe viscous electron flows can also be used to detect a non-vanishing $\tilde\eta$.  As $\tilde\eta > 0$ is expected whenever the Fermi surface is anisotropic, this is a generic new phenomenon in electron fluids in solids.  We hope it can be observed experimentally in the near future.

Looking forward, we encourage looking for conductors with a single, approximately polygonal, small Fermi surface.  One possible candidate is (relatively) low density $\mathrm{SrTiO}_3$ \cite{behnia,stemmer}, with carrier density $n\approx 5\times 10^{17} \; \mathrm{cm}^{-3}$.  Such materials could be natural candidates for hydrodynamic electron flow, and for observing non-universal aspects of the ballistic-to-hydrodynamic crossover,  as we have predicted.

\addcontentsline{toc}{section}{Acknowledgements}
\section*{Acknowledgements}
We thank Steve Kivelson, Andrew Mackenzie and Philip Moll for helpful discussions.  We are especially indebted to the authors of \cite{mackenzie} for their complete experimental data set.  CQC is supported by a Stanford Physics Department Graduate Fellowship.   AL is supported by the Gordon and Betty Moore Foundation's EPiQS Initiative through Grant GBMF4302.


\begin{appendix}

\section{Timescale Separation from Collision Integral}
\label{app:coll}

In this section we schematically evaluate the collision integral for a polygonal Fermi surface. Plugging the distribution function expansion (\ref{eq:Phi}) into the two-body collision integral (\ref{eq:2body}) gives
\begin{align}
\label{eq:linCollision}
    \mathsf{W}|\Phi\rangle
    &\propto
    \frac{1}{T}\int\mathrm{d}^{2}\mathbf{q}\,\mathrm{d}^{2}\mathbf{p}^{\prime}\,\mathrm{d}^{2}\mathbf{q^{\prime}}\,\left|\mathcal{M}_{\mdelta\mathbf{p}}\right|^{2}f_{\mathbf{p^{\prime}}}^{0}f_{\mathbf{q^{\prime}}}^{0}\left(1-f_{\mathbf{p}}^{0}\right)\left(1-f_{\mathbf{q}}^{0}\right)\Big[|\Phi_{\mathbf{p}}\rangle+|\Phi_{\mathbf{q}}\rangle-|\Phi_{\mathbf{p}^{\prime}}\rangle-|\Phi_{\mathbf{q}^{\prime}}\rangle\Big] \times \notag \\
   & \;\;\;\; \mdelta\left(\epsilon_{\mathbf{p}^{\prime}}+\epsilon_{\mathbf{q}^{\prime}}-\epsilon_{\mathbf{p}}-\epsilon_{\mathbf{q}}\right)\mdelta\left(\mathbf{p}^{\prime}+\mathbf{q}^{\prime}-\mathbf{p}-\mathbf{q}\right)
\end{align}
to leading order in $\Phi$, where we have suppressed spatial indices and used $\partial f^{0}/\partial\epsilon=-f^{0}\left(1-f^{0}\right)/T$.  We have set $k_{\mathrm{B}}=1$ in this appendix. Our goal in this section is to calculate $\langle \Phi|\mathsf{W}|\Phi\rangle$,
which represents the rate at which an electronic excitation $\Phi$ relaxes to zero. 

\begin{figure}[t]
\includegraphics[width=.4\textwidth]{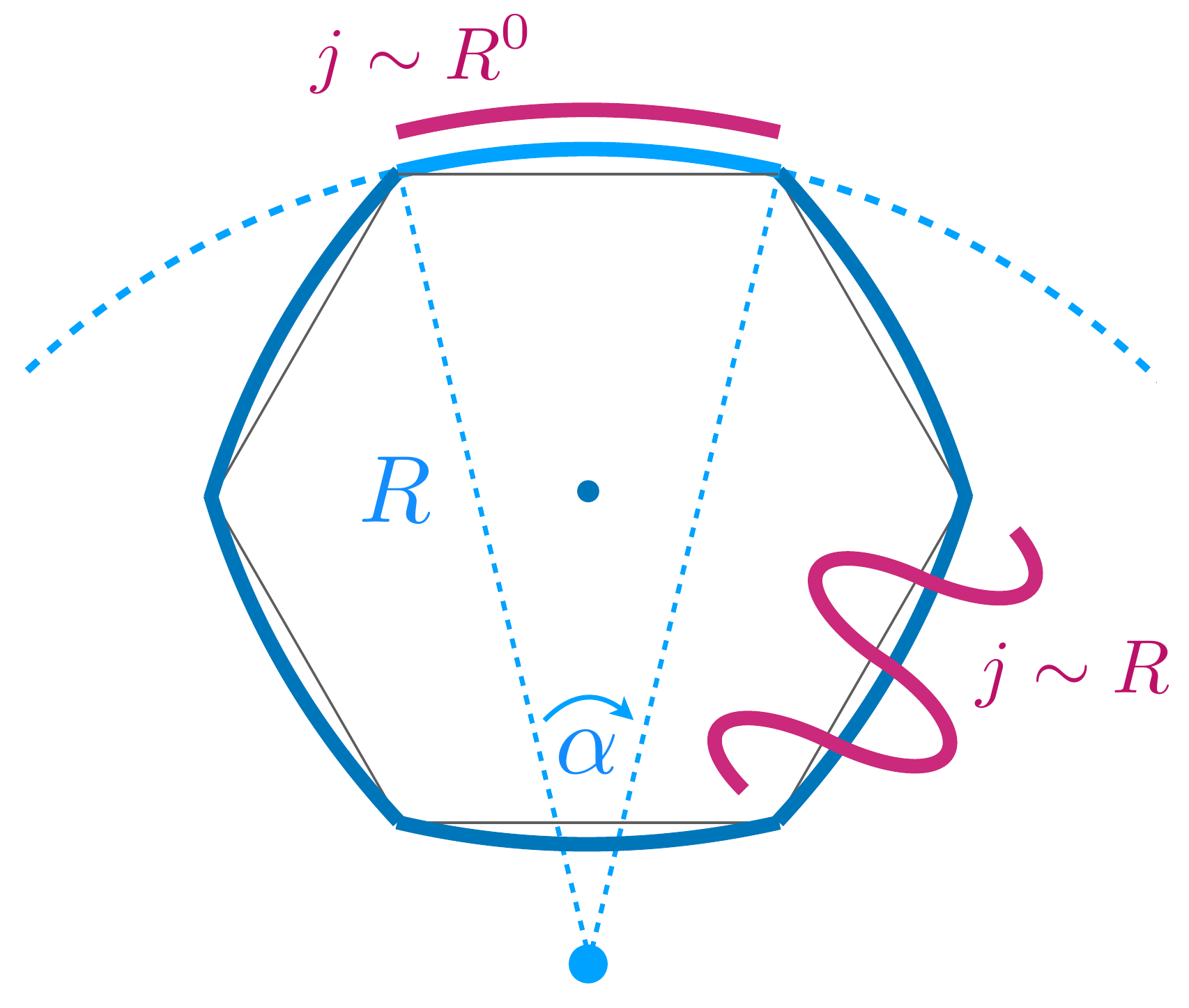}
\centering
\caption{A hexagonal Fermi surface with edges rounded to a radius of curvature $R$ and corresponding degree of curvature $\alpha\sim R^{-1}$. Electronic excitations $\Phi^j_\mathbf{p}\propto \exp\left[\mathrm{i}j\theta_\mathbf{p}\right]$ on rounded edges are shown for $j\sim R^0$ and $j\sim R$ in the limit $R\to\infty$. The $j\sim R$ excitations relax parametrically faster than the $j\sim R^0$ excitations, which contain nonzero particle density on a Fermi surface edge.}
\label{fig:2timesA}
\end{figure}

In the following, we will consider excitations $\Phi$ with support on a single edge of the polygonal Fermi surface. If the Fermi surface is a perfect polygon with flat edges, then the relaxation rate $\langle\Phi|\mathsf{W}|\Phi\rangle$ associated with $\Phi$ will in fact contain singularities in the momentum-conserving $\mdelta$ function  due to the ``sliding" effect discussed in Section \ref{sec:2times}. In order to avoid such singularities, we  regularize the calculation of $\langle\Phi|\mathsf{W}|\Phi\rangle$ by ``rounding out" the Fermi surface edges and replacing them with arcs of large circles; the regularization is then controlled by the radius of curvature $R \propto 1/\alpha$, where the perfect polygonal Fermi surface is recovered in the limit $R\to\infty$ (or equivalently, the limit in which the arc-subtending angle $\alpha\to0$). 
Additionally, any function on a circle  may be represented as a sum of complex exponentials, and so for our purposes it suffices to consider single-edge excitations of the form $\Phi^j\propto \exp\left[\mathrm{i}j\theta_\mathbf{p}\right]$, where $j\in\mathbb{Z}$ and $\theta_\mathbf{p}\in[-\alpha/2,\alpha/2]$ parameterizes momentum along the rounded Fermi surface edge. A sketch of such electronic excitations on a hexagonal Fermi surface regularized in this way is given in Figure \ref{fig:2timesA}.

Following \cite{ledwith1,ledwith2}, we proceed in the calculation of $\langle \Phi^j|\mathsf{W}|\Phi^j\rangle$ by splitting the energy- and momentum-conserving delta functions via an additional integration over energy and momentum transfer:
\begin{subequations}
\begin{align}
    \mdelta\left(\epsilon_{\mathbf{p}^{\prime}}+\epsilon_{\mathbf{q}^{\prime}}-\epsilon_{\mathbf{p}}-\epsilon_{\mathbf{q}}\right)
    & =
    \int\mathrm{d}\omega\,\mdelta\left(\epsilon_{\mathbf{p}}-\epsilon_{\mathbf{p}^{\prime}}-\omega\right)\mdelta\left(\epsilon_{\mathbf{q}}-\epsilon_{\mathbf{q}^{\prime}}+\omega\right),
    \\
    \mdelta\left(\mathbf{p}^{\prime}+\mathbf{q}^{\prime}-\mathbf{p}-\mathbf{q}\right)
    & =
    \int\mathrm{d}^{2}\mathbf{k}\,\mdelta\left(\mathbf{p}-\mathbf{p}^{\prime}-\mathbf{k}\right)\mdelta\left(\mathbf{q}-\mathbf{q}^{\prime}+\mathbf{k}\right).
\end{align}
\end{subequations}
Momentum integrations $\int\mathrm{d}^{2}\mathbf{p}=(\nu/\alpha)\int\mathrm{d}\epsilon_{\mathbf{p}}\int\mathrm{d}\theta_\mathbf{p}$ are then computed by parameterizing momentum-space in the directions perpendicular and parallel to the Fermi surface edge, where $\nu$ is the electronic density of states. Since in our distribution function expansion (\ref{eq:Phi}) we assume that relevant particle energies are very close to the Fermi surface, the integrand in $\langle\Phi|\mathsf{W}|\Phi\rangle$ only has energy-dependence coming from the $f^0$ and $\mdelta_{\sum\epsilon}$. With this assumption, one can then perform the four energy integrals $\int\mathrm{d}\epsilon$ and the integral over energy transfer $\int\mathrm{d}\omega$,
which gives the scaling $\sim T^2$ with  $\mathrm{O}(1)$ constants \cite{ledwith1,ledwith2}.
This is the familiar $T^2$-scaling for quasiparticle scattering rates in Fermi liquids.

It remains then to compute the momenta integrals. In \cite{ledwith1,ledwith2}, the authors calculate these momenta integrals in the case of a circular Fermi surface. Scattering on a single edge of our regularized polygonal Fermi surface differs from the circular case in two important ways: (\emph{i}) our angular
integrations are normalized by a factor of $\alpha$ instead of $2\mpi$,
and (\emph{ii}) the momentum transfer $|\mathbf{k}|$ for us is bounded above by the chord
length $2R\sin(\alpha/2)\approx R\alpha$ instead of the full Fermi surface
diameter $2R$. After taking into account these differences in the result of \cite{ledwith1,ledwith2}, we find in the limit $R\to\infty$ that
\begin{align}
    \gamma\left[\Phi^{j}\right]&\propto\frac{T^{2}}{\alpha^{3}}\int_{\left|\mathbf{k}\right|\lesssim R\alpha}\mathrm{d}^{2}\mathbf{k}\left|\mathcal{M}_{\mathbf{k}}\right|^{2}\frac{1-\cos\left[2j\cos^{-1}\left(k/2R\right)\right]}{k^{2}\left(4R^{2}-k^{2}\right)}\\&\sim T^{2}R\int_{\left|\mathbf{k}\right|\lesssim R\alpha}\mathrm{d}^{2}\mathbf{k}\left|\mathcal{M}_{\mathbf{k}}\right|^{2}\frac{1-\cos\left(jk/R\right)}{k^{2}}\\&\sim T^{2}R \times \left\lbrace \begin{array}{ll} (j/R)^2 &\  \alpha |j| \ll 1 \\ \log |j| &\  \alpha |j| \gg 1 \end{array}\right.,
\end{align}
where we have used the fact that $\alpha\sim R^{-1}$ and taken $j$ even for simplicity \cite{ledwith1}.

Modes that do not vary appreciably on a flat Fermi surface
edge, and thus possess a nonzero net edge particle number, correspond to the limit $j\sim R^{0}$ as $R\to\infty$. In this
limit, $\langle \Phi^j|\mathsf{W}|\Phi^j\rangle \rightarrow 0$. Conversely,
modes that vary appreciably on a flat Fermi surface edge, and thus possess roughly zero particles on the Fermi surface edge,
correspond to the limit $j\sim R$, $R\to\infty$. In this limit,
we see that the decay rate instead diverges $\langle\Phi^j|\mathsf{W}|\Phi^j\rangle\to\infty$. Therefore, in the limit of nearly flat edges, we see a hierachy of timescales emerge: namely, that two-body scattering causes modes that possess roughly zero edge particle number decay at a rate $\gamma_\text{f}$ that is parametrically larger than the decay rate $\gamma_\text{s}\ll \gamma_\text{f}$ of those modes that possess a nonzero edge particle number, with the decay rates satisfying the scaling relation
\begin{equation}
    \frac{\gamma_{\text{f}}}{\gamma_{\text{s}}}
    \sim 
    R
    \sim
    \alpha^{-1},
\end{equation}
as claimed in Section \ref{sec:2times}.

We note that in the above analysis, ``constant" $j\sim R^0$ modes do not decay on a perfectly flat edge, with $\gamma_\text{s}\to 0$ as $R\to\infty$. This is due to the fact that we have only considered scattering within a given Fermi surface edge; once scattering between different edges (i.e. pink scattering pathways in Figure \ref{fig:2times}) is accounted for, such modes are short-circuited by and therefore decay according to the standard $\gamma_\text{s} \sim T^2 R^0$ scaling, which is nonzero and finite even in the limit of perfectly flat edges.

\section{Explicit Presentation of Two-Time Collision Matrix $\mathsf{W}$}
\label{app:Wmats}

For an $M$-gon Fermi surface, the linearized collision operator $\mathsf{W}$ (\ref{eq:Wmatrix}) is a $2M\times 2M$ symmetric matrix. Given the timescale separation property (\ref{eq:decayConstraint}) of $\mathsf{W}$, we see that if we order our Legendre basis $\{|n,m\rangle\}$ as
\begin{equation}
    \beta_{\text{L}}
    =
    \Big\{ \left|0,0\right\rangle ,\left|0,1\right\rangle ,\ldots\left|0,M-1\right\rangle ,\left|1,0\right\rangle ,\left|1,1\right\rangle ,\ldots,\left|1,M-1\right\rangle
    \Big\}, 
\end{equation}
then $\mathsf{W}$ will decompose into slow $(n=0)$ and fast $(n=1)$ sectors as
\begin{equation}
    \mathsf{W}=\left[\begin{array}{cc}
    \mathsf{W}_{\text{s}} & \mathsf{W}_{\text{sf}}\\
    \mathsf{W}_{\text{sf}}^{\text{T}} & \mathsf{W}_{\text{f}}.
\end{array}\right]
\end{equation}
Thus, to specify $\mathsf{W}$ explicitly, it suffices to give the three $M\times M$ matrices $\mathsf{W}_\text{s}$, $\mathsf{W}_\text{sf}$, and $\mathsf{W}_\text{f}$.  Note that $\mathsf{W}$ is a banded and symmetric matrix, which is the (only) constraint arising from dihedral symmetry.

\subsection{Square Fermi Surface}

\begin{subequations}
\begin{align}
    \mathsf{W}_{\text{s}}&=\gamma_{\text{s}}\cdot\frac{1}{8}\left[\begin{array}{cccc}
3 & -2 & 1 & -2\\
-2 & 3 & -2 & 1\\
1 & -2 & 3 & -2\\
-2 & 8 & -2 & 3
\end{array}\right],\\\mathsf{W}_{\text{sf}}&=\gamma_{\text{s}}\cdot\frac{\sqrt{3}}{8}\left[\begin{array}{cccc}
0 & 1 & 0 & -1\\
-1 & 0 & 1 & 0\\
0 & -1 & 0 & 1\\
1 & 0 & -1 & 0
\end{array}\right],\\\mathsf{W}_{\text{f}}&=\frac{1}{8}\left[\begin{array}{cccc}
4\gamma_{\text{f}}+3\gamma_{\text{s}} & 0 & 4\gamma_{\text{f}}-3\gamma_{\text{s}} & 0\\
0 & 4\gamma_{\text{f}}+3\gamma_{\text{s}} & 0 & 4\gamma_{\text{f}}-3\gamma_{\text{s}}\\
4\gamma_{\text{f}}-3\gamma_{\text{s}} & 0 & 4\gamma_{\text{f}}+3\gamma_{\text{s}} & 0\\
0 & 4\gamma_{\text{f}}-3\gamma_{\text{s}} & 0 & 4\gamma_{\text{f}}+3\gamma_{\text{s}}
\end{array}\right].
\end{align}
\end{subequations}

\subsection{Hexagon Fermi Surface}

\begin{subequations}
\begin{align}
    \mathsf{W}_{\text{s}}&=\gamma_{\text{s}}\cdot\frac{1}{60}\left[\begin{array}{cccccc}
32 & -19 & -1 & 8 & -1 & -19\\
-19 & 32 & -19 & -1 & 8 & -1\\
-1 & -19 & 32 & -19 & -1 & 8\\
8 & -1 & -19 & 32 & -19 & -1\\
-1 & 8 & -1 & -19 & 32 & -19\\
-19 & -1 & 8 & -1 & -19 & 32
\end{array}\right],\\\mathsf{W}_{\text{sf}}&=\gamma_{\text{s}}\cdot\frac{\sqrt{3}}{20}\left[\begin{array}{cccccc}
0 & 1 & 1 & 0 & -1 & -1\\
-1 & 0 & 1 & 1 & 0 & -1\\
-1 & -1 & 0 & 1 & 1 & 0\\
0 & -1 & -1 & 0 & 1 & 1\\
1 & 0 & -1 & -1 & 0 & 1\\
1 & 1 & 0 & -1 & -1 & 0
\end{array}\right],\\\mathsf{W}_{\text{f}}&=\frac{1}{60}\left[\begin{array}{cccccc}
40\gamma_{\text{f}}+18\gamma_{\text{s}} & -10\gamma_{\text{f}}+9\gamma_{\text{s}} & 10\gamma_{\text{f}}-9\gamma_{\text{s}} & 20\gamma_{\text{f}}+18\gamma_{\text{s}} & 10\gamma_{\text{f}}-9\gamma_{\text{s}} & -10\gamma_{\text{f}}+9\gamma_{\text{s}}\\
-10\gamma_{\text{f}}+9\gamma_{\text{s}} & 40\gamma_{\text{f}}+18\gamma_{\text{s}} & -10\gamma_{\text{f}}+9\gamma_{\text{s}} & 10\gamma_{\text{f}}-9\gamma_{\text{s}} & 20\gamma_{\text{f}}+18\gamma_{\text{s}} & 10\gamma_{\text{f}}-9\gamma_{\text{s}}\\
10\gamma_{\text{f}}-9\gamma_{\text{s}} & -10\gamma_{\text{f}}+9\gamma_{\text{s}} & 40\gamma_{\text{f}}+18\gamma_{\text{s}} & -10\gamma_{\text{f}}+9\gamma_{\text{s}} & 10\gamma_{\text{f}}-9\gamma_{\text{s}} & 20\gamma_{\text{f}}+18\gamma_{\text{s}}\\
20\gamma_{\text{f}}+18\gamma_{\text{s}} & 10\gamma_{\text{f}}-9\gamma_{\text{s}} & -10\gamma_{\text{f}}+9\gamma_{\text{s}} & 40\gamma_{\text{f}}+18\gamma_{\text{s}} & -10\gamma_{\text{f}}+9\gamma_{\text{s}} & 10\gamma_{\text{f}}-9\gamma_{\text{s}}\\
10\gamma_{\text{f}}-9\gamma_{\text{s}} & 20\gamma_{\text{f}}+18\gamma_{\text{s}} & 10\gamma_{\text{f}}-9\gamma_{\text{s}} & -10\gamma_{\text{f}}+9\gamma_{\text{s}} & 40\gamma_{\text{f}}+18\gamma_{\text{s}} & -10\gamma_{\text{f}}+9\gamma_{\text{s}}\\
-10\gamma_{\text{f}}+9\gamma_{\text{s}} & 10\gamma_{\text{f}}-9\gamma_{\text{s}} & 20\gamma_{\text{f}}+18\gamma_{\text{s}} & 10\gamma_{\text{f}}-9\gamma_{\text{s}} & -10\gamma_{\text{f}}+9\gamma_{\text{s}} & 40\gamma_{\text{f}}+18\gamma_{\text{s}}
\end{array}\right].
\end{align}
\end{subequations}

\section{Group Theory} \label{app:dihedral}
Here we outline the group theory of discrete and continuous rotation groups, and in doing so give a group theoretic explanation for the structure of the quasihydrodynamic equations.  A useful reference for the mathematics below is \cite{tung}.

\subsection{Irreducible Representations of $\mathrm{D}_{2M}$ ($M$ even)}

The dihedral group $\mathrm{D}_{2M}$ of order $2M$ is the group of planar
symmetries of the regular $M$-gon. If we let 
$\rho$ represent a rotation of the $M$-gon about its center by the symmetry
angle $\theta_{M}\equiv2\mpi/M$ and let $r$ represent a reflection
about a fixed symmetry axis, then we may present the group as
\begin{equation}
\mathrm{D}_{2M}=\left\langle r,\rho\left|\right.r^{2}=\rho^{M}=1,r\rho r=\rho^{-1}\right\rangle .
\end{equation}
For even $M$, the irreducible representations of the dihedral group
$\mathrm{D}_{2M}$ are precisely $4$ one-dimensional representations $U_{0}^{\pm},U_{M/2}^{\pm}$
and $\left(M/2-1\right)$ two-dimensional representations $R_{k}$
with $k=1,\ldots,(M/2-1)$. They are given explicitly by specifying
their action on the generators $\rho,r$ of $\mathrm{D}_{2M}$ as follows:
\begin{subequations}\begin{align}
U_{M/2}^{\pm}(r) = U_{0}^{\pm}(r) &=\pm1, \\
U_0^\pm(\rho) &= 1, \\
U_{M/2}^{\pm}(\rho) &= -1, \\
R_k(\rho) &= \left[\begin{array}{cc}
\cos\left(k\theta_{M}\right) & -\sin\left(k\theta_{M}\right)\\
\sin\left(k\theta_{M}\right) & \cos\left(k\theta_{M}\right)
\end{array}\right], \\
R_{k}(r)&=\left[\begin{array}{cc}
-1 & 0\\
0 & 1
\end{array}\right]
\end{align}
\end{subequations}

\subsection{Irreducible Representations of $\mathrm{O}(2)$}

The orthogonal group $\mathrm{O}(2)$ is the group of distance-preserving transformations
of the Euclidean plane that fix the origin. If we let $\rho_{\theta}$
represent a rotation by angle $\theta\in[0,2\mpi)$ about the origin
and let $r$ represent a reflection about some fixed axis through
the origin, then we may present the group as
\begin{equation}
\mathrm{O}(2)=\left\langle r,\left\{ \rho_{\theta}\right\} _{\theta\in[0,2\mpi)}\left|\right.r^{2}=\rho_0=1,\rho_{\theta}\rho_{\phi}=\rho_{\theta+\phi},r\rho_{\theta}r=\rho_{-\theta}\right\rangle .
\end{equation}
We note that $\mathrm{D}_{2M}$ is a subgroup of $\mathrm{O}(2)$ for all $M$.

The irreducible representations of the orthogonal group $\mathrm{O}(2)$ are
precisely $2$ one-dimensional representations $\mathcal{U}_{0}^{\pm}$
and infinitely many two-dimensional representations $\mathcal{R}_{k}$
labeled by positive integers $k\in\mathbb{N}$. They are given
explicitly by specifying their action on the generators $\rho_{\theta},r$
of $\mathrm{O}(2)$ as follows:
\begin{subequations}
\begin{align}
    \mathcal{U}_{0}^{\pm}(\rho_{\theta})&=1,\\
\mathcal{U}_{0}^{\pm}(r)&=\pm1, \\
\mathcal{R}_{k}(\rho_{\theta})&=\left[\begin{array}{cc}
\cos\left(k\theta\right) & -\sin\left(k\theta\right)\\
\sin\left(k\theta\right) & \cos\left(k\theta\right)
\end{array}\right],\\
\mathcal{\mathcal{R}}_{k}(r)&=\left[\begin{array}{cc}
-1 & 0\\
0 & 1
\end{array}\right].
    \end{align}
    \label{eq:O2irreps}
    \end{subequations}
The character table for these irreducible
representations is given in Table \ref{table:characterO2}.
\begin{table}[t]
\begin{center}
\begin{tabular}{c|ccc}
$\mathrm{O}(2)$ & $\left\{ 1\right\} $ & $\{\rho_{\theta},\rho_{-\theta}\}$ & $\left\{ r,r\rho_{\theta}\right\} $\tabularnewline
\hline 
$\chi_{\mathcal{U}_{0}^{+}}$ & $1$ & $1$ & $1$\tabularnewline
$\chi_{\mathcal{U}_{0}^{-}}$ & $1$ & $1$ & $-1$\tabularnewline
$\chi_{\mathcal{R}_{k}}$ & $2$ & $2\cos\left(k\theta\right)$ & $0$\tabularnewline
\end{tabular}
\par\end{center}
\caption{Character table of $\mathrm{O}(2)$.}
\label{table:characterO2}
\end{table}

It is instructive to consider the following question: how do tensor
products of irreducible representations of $\mathrm{O}(2)$ decompose as a
direct sum of said irreducible representations? This question can be answered
by using the orthogonality of irreducible characters and the fact
that, for any group representations $A$ and $B$, we have that $\chi_{A\otimes B}=\chi_{A}\cdot\chi_{B}$.
Thus it is easy to see that
\begin{subequations}
\label{eq:LRO2}
\begin{align}
    \mathcal{U}_{0}^{\eta}\otimes\mathcal{U}_{0}^{\zeta} &= \mathcal{U}_{0}^{\eta\zeta},
    \\
    \mathcal{U}_{0}^{\pm}\otimes\mathcal{R}_{k} &= \mathcal{R}_{k},
    \\
    \mathcal{R}_{k}\otimes\mathcal{R}_{l} &= \mathcal{R}_{|k-l|}\oplus\mathcal{R}_{k+l},
\end{align}
\end{subequations}
where in the last decomposition we have used the trigonometric identity
\begin{equation}
\left(2\cos k\theta\right)\cdot\left(2\cos l\theta\right)=2\cos\left[\left(k-l\right)\theta\right]+2\cos\left[\left(k+l\right)\theta\right]
\end{equation}
and defined the (reducible) representation 
\begin{equation}
\mathcal{R}_{0}\equiv\mathcal{U}_{0}^{+}\oplus\mathcal{U}_{0}^{-}.
\end{equation}
Note that (\ref{eq:LRO2}) give the decomposition rules for tensor products of irreducible representations of $\mathrm{O}(2)$ and thereby determine the so-called \emph{Littlewood-Richardson coefficients} for $\mathrm{O}(2)$.

\subsection{Branching Rules for $\mathrm{D}_{2M}\le\mathrm{O}(2)$}

We note that any representation $\mathcal{A}$ of $\mathrm{O}(2)$ automatically
furnishes a representation $\left.\mathcal{A}\right|_{\mathrm{D}_{2M}}$ of
the subgroup $\mathrm{D}_{2M}\leq \mathrm{O}(2)$ by simply restricting the action of
$\mathcal{A}$ to the subgroup elements. However, it will sometimes
occur that the representation $\left.\mathcal{A}\right|_{D_{2M}}$
of $\mathrm{D}_{2M}$ generated in this way is reducible, even if $\mathcal{A}$
is an irreducible representation of the larger group $\mathrm{O}(2)$:  after all, $\mathrm{D}_{2M}$ has only $\left(M/2+3\right)$
irreducible representations, whereas $\mathrm{O}(2)$ has infinitely many.
More directly, we see that the restricted representation $\left.\mathcal{A}\right|_{\mathrm{D}_{2M}}$
can fail to be irreducible because the subset of matrices $\left\{ \mathcal{A}\left(g\right)\right\} _{g\in \mathrm{D}_{2M}}$
may be simultaneously block-diagonalizable, even if the larger set
of matrices $\left\{ \mathcal{A}\left(g\right)\right\} _{g\in \mathrm{O}(2)}$
is not. 

One-dimensional representations are always irreducible, and so we
note that
\begin{equation}
\left.\mathcal{U}_{0}^{\pm}\right|_{\mathrm{D}_{2M}} =  U_{0}^{\pm}.
\end{equation}
What about the two-dimensional irreducible representations $\mathcal{R}_{k}$
of $\mathrm{O}(2)$? They become reducible when restricted to $\mathrm{D}_{2M}$ precisely
when the matrices
$\mathcal{R}_k(r)$ and $\mathcal{R}_k(\rho)$ as given in (\ref{eq:O2irreps}) are simultaneously diagonalizable. Clearly this is the case if and
only if $\left.\mathcal{R}_{k}\right|_{\mathrm{D}_{2M}}\left(\rho\right)$
is diagonal, i.e. $\sin\left(k\theta_{M}\right)=0$, which occurs when $2k/M \in \mathbb{Z}$.  Defining the reducible representations \begin{subequations}
\begin{align}
R_{0} & \equiv U_{0}^{+}\oplus U_{0}^{-}\\
R_{M/2} & \equiv U_{M/2}^{+}\oplus U_{M/2}^{-},
\end{align}
\end{subequations}we find that 
\begin{subequations}
\label{eq:branching}
\begin{align}
    \left.\mathcal{U}_{0}^{\pm}\right|_{D_{2M}} & \cong U_{0}^{\pm},
    \\
    \left.\mathcal{R}_{k}\right|_{D_{2M}} & \cong R_{f_{M}\left(k\right)}.
\end{align}
\end{subequations}
where 
we have introduced the function
\begin{equation}
    f_{M}\left(k\right)\equiv\frac{\arccos\left[\cos\left(2k\mpi/M\right)\right]}{2\mpi/M}
    =
    M\,\left|\frac{k}{M}-\left\lfloor \frac{k}{M}+\frac{1}{2}\right\rfloor \right|.
\end{equation}
(\ref{eq:branching}) gives the rules for decomposing the restriction of the irreducible representations of $\mathrm{O}(2)$ into direct sums of irreducible representations of the subgroup $D_{2M}\leq O(2)$. In the literature of representation theory, such rules are referred to as \emph{branching rules}.

\subsection{Tensor Representations of $\mathrm{O}(2)$}

The orthogonal group $\mathrm{O}(d)$ has a natural action on real $d$-dimensional, rank-$n$ tensors of the form $T_{i_{1}\cdots i_{n}}$ given by
\begin{equation}
    T_{i_{1}\cdots i_{n}}\xrightarrow{g\in \mathrm{O}(2)}(g\cdot T)_{i_{1}\cdots i_{n}}=T_{j_{1}\cdots j_{n}}\prod_{k=1}^{n}\mathcal{R}_{1}(g)_{i_{k}j_{k}}.
\end{equation}
In this paper we are interested in dimensionality $d=2$, and so for convenience we define $\mathcal{T}_n = (\mathbb{R}^2)^{\otimes n}$ as the vector space of real two-dimensional, rank-$n$ tensors. It is clear then that $\mathrm{O}(2)$ acts on $\mathcal{T}_n$ via the representation
\begin{equation}
    \bigotimes_{k=1}^n \mathcal{R}_1,
\end{equation}
which will reduce into a direct sum of irreducible $\mathrm{O}(2)$-representations via the decomposition rules given in  (\ref{eq:LRO2}).

Consider first the vector space of rank-2 tensors $\mathcal{T}_2$.  Tensors of this type will be especially relevant in our quasihydrodynamic equations, which for example take into account the flux of the momentum density $\mpi_{ij}=\partial_i V_j$. Now, from  (\ref{eq:LRO2})  the action of $\mathrm{O}(2)$ on $\mathcal{T}_2$ is reducible:
\begin{equation}
    \mathcal{R}_1 \otimes \mathcal{R}_1 = \mathcal{U}_0^+ \oplus \mathcal{U}_0^- \otimes \mathcal{R}_2 \label{eq:R1R1}.
\end{equation}
In considering this irreducible decomposition, it will prove useful to explicitly write down a basis of $\mathcal{T}_2$ that block diagonalizes the action of $\mathrm{O}(2)$. Such a basis of $\mathcal{T}_2$ is given by
\begin{equation}
    \left\{ \mdelta_{ij},\epsilon_{ij},\mu_{ij}^{-},\mu_{ij}^{+}\right\} \equiv\left\{ \left[\begin{array}{cc}
1 & 0\\
0 & 1
\end{array}\right]_{ij},\left[\begin{array}{cc}
0 & 1\\
-1 & 0
\end{array}\right]_{ij},\left[\begin{array}{cc}
0 & 1\\
1 & 0
\end{array}\right]_{ij},\left[\begin{array}{cc}
-1 & 0\\
0 & 1
\end{array}\right]_{ij}\right\} 
\end{equation}
which behaves in the following way under the tensor representation $\mathcal{R}_1\otimes\mathcal{R}_1$ of $\mathrm{O}(2)$:
\begin{subequations}
\label{eq:R1R1}
\begin{align}
\left(\mathcal{R}_{1}\otimes\mathcal{R}_{1}\right)\left(\rho_{\theta}\right)\left[\begin{array}{c}
\mdelta_{ij}\\
\epsilon_{ij}\\
\mu_{ij}^{-}\\
\mu_{ij}^{+}
\end{array}\right] & =\left[\begin{array}{cccc}
1 & 0 & 0 & 0\\
0 & 1 & 0 & 0\\
0 & 0 & \cos2\theta & -\sin2\theta\\
0 & 0 & \sin2\theta & \cos2\theta
\end{array}\right]\left[\begin{array}{c}
\mdelta_{ij}\\
\epsilon_{ij}\\
\mu_{ij}^{-}\\
\mu_{ij}^{+}
\end{array}\right]\\
\left(\mathcal{R}_{1}\otimes\mathcal{R}_{1}\right)\left(r\right)\left[\begin{array}{c}
\mdelta_{ij}\\
\epsilon_{ij}\\
\mu_{ij}^{-}\\
\mu_{ij}^{+}
\end{array}\right] & =\left[\begin{array}{cccc}
1 & 0 & 0 & 0\\
0 & -1 & 0 & 0\\
0 & 0 & -1 & 0\\
0 & 0 & 0 & 1
\end{array}\right]\left[\begin{array}{c}
\mdelta_{ij}\\
\epsilon_{ij}\\
\mu_{ij}^{-}\\
\mu_{ij}^{+}
\end{array}\right]
\end{align}
\end{subequations}
These equations demonstrate by comparison with (\ref{eq:O2irreps}) that, within the representation $\mathcal{R}_{1}\otimes\mathcal{R}_{1}$
acting on $\mathcal{T}_{2}$, we have that
\begin{align}
\mdelta_{ij} & \in\mathcal{U}_{0}^{+},\\
\epsilon_{ij} & \in\mathcal{U}_{0}^{-},\\
\left\{ \mu_{ij}^{-},\mu_{ij}^{+}\right\}  & \in\mathcal{R}_{2},\label{eq:muR2}
\end{align}
where $v\in\mathcal{A}$ is understood to mean that the vector
 $v$ lies in the subspace transforming exclusively under the
representation $\mathcal{A}$.  

Similar block diagonalizations will occur for the action of $\mathrm{O}(2)$ on tensor spaces $\mathcal{T}_n$ of higher rank. For the quasihydrodynamical equations corresponding to the hexagonal Fermi surface, we will need to consider rank-3 tensors belonging to the vector space $\mathcal{T}_3$, on which the planar orthogonal group $\mathrm{O}(2)$ acts via the representation
\begin{equation}
    \mathcal{R}_{1}\otimes\mathcal{R}_{1}\otimes\mathcal{R}_{1}=\mathcal{R}_{1}\oplus\mathcal{R}_{1}\oplus\mathcal{R}_{1}\oplus\mathcal{R}_{3}.
\end{equation}
A straightforward calculation shows that the two-dimensional $\mathcal{T}_3$-subspace 
\begin{equation}
   \left\{  \lambda_{ijk}^{-},\lambda_{ijk}^{+}\right\}\in\mathcal{R}_{3}
\end{equation}
transforming exclusively under $\mathcal{R}_3$ is spanned by the rank-3 tensors
\begin{equation}
\label{eq:lambdaR3}
    \lambda_{ijk}^{\pm}=\mu_{ij}^{\pm}\mdelta_{kx} \mp \mu_{ij}^{\mp}\mdelta_{ky},
\end{equation}
analogous to (\ref{eq:muR2}) in the rank-2 case.

\subsection{Representation Theory and the Quasihydrodynamic Equations}

Let $\left|n\right\rangle _{m}$ denote the electronic distribution excitations on the $M$-gon Fermi surface, with $n\in\{0,1\}$ denoting the first two Legendre polynomial modes, and $m\in\{0,1,\ldots,M-1\}$ labeling the edges of the polygonal Fermi surface, with $m=0$ denoting the top edge and $m$ increasing as we move counter-clockwise around the polygon (see Figure \ref{fig:hexModes}). We may consider the dihedral group $\mathrm{D}_{2M}$ as acting on the vector space spanned by the $\left|n\right\rangle _{m}$ in a natural way; namely, as a group element $g\in D_{2M}$ permutes the edges $m$ of the polygon, it shuffles the the electronic excitation vectors $\left|n\right\rangle _{m}$ correspondingly. 

Let $A$ denote the representation of $\mathrm{D}_{2M}$ generated in this way, and in defining this representation let us choose the reflection axis to be the $k_y$-axis. It is clear then that $A$ is in fact the so-called \emph{regular representation} of $\mathrm{D}_{2M}$. The regular representation of any group decomposes as a direct sum of that group's irreducible representations, with each irreducible representation occurring with multiplicity  equal to its dimension. Thus, we have that
\begin{equation}
    A=\Big[U_{0}^{+}\oplus R_{1}\Big]\oplus\left[U_{M/2}^{+}\oplus R_{1}\oplus\left(\oplus_{k=2}^{M/2-1}R_{k}\right)\right]\oplus\left[U_{0}^{-}\oplus U_{M/2}^{-}\oplus \left(\oplus_{k=2}^{M/2-1}R_{k}\right)\right].
    \label{eq:Tdecomp}
\end{equation}
Each bracketed term in Eq. (\ref{eq:Tdecomp}) contains the irreducible subspaces of $\mathrm{D}_{2M}$ whose excitation modes share the same decay rate, with the bracketed terms ordered in increasing decay rate. An explicit construction for the basis $\beta$ that simultaneously block diagonalizes the action of $A$ and diagonalizes the collision integral $\mathsf{W}$ is given in Table \ref{table:dihedralB}. We will henceforth refer to $\beta$ as the \emph{dihedral basis}.

The (quasi)hydrodynamic equations we derive will respect the dihedral symmetry of the Fermi surface. When integrating out modes as described in Section \ref{subsec:integration}, the symmetry of the resulting equations will therefore be most apparent if this computation is carried out in the dihedral basis by writing
\begin{equation}
|\Phi\rangle =\sum_{v\in\beta}v(x,y,t)|v\rangle.
\end{equation}
Integrating out modes in the dihedral basis $\beta$ is in fact quite straightforward due to the fact that $\beta$ diagonalizes the collision integral $\mathsf{W}$.

\begin{table}
\begin{center}
\begin{tabular}{|c|c|c|c|c|}
\hline
dihedral vector  & modes $|n\rangle _{m}$ on edge $m$ & $\mathrm{D}_{2M}$ irrep. & interpretation & decay rate
\tabularnewline
\hline\noalign{\smallskip}
\hline
$|N\rangle $ & $|0\rangle _{m}$ & $U_{0}^{+}$ & charge & \multirow{3}{*}{$0$}
\tabularnewline
\cline{1-4} 
$\left\{\makecell[c]{|P_{x}\rangle \\ |P_{y}\rangle }\right\}$ & $\left\{\makecell[c]{-s_{m\theta}|0\rangle _{m}-g_{M}c_{m\theta}|1\rangle _{m} \\ c_{m\theta}|0\rangle _{m}-g_{M}s_{m\theta}|1\rangle _{m}}\right\}$ & \multirow{1}{*}{$R_{1}$} & \multirow{1}{*}{momentum} & 
\tabularnewline
\hline\noalign{\smallskip}
\hline
$|\widetilde{N}\rangle $ & $(-1)^{m}|0\rangle _{m}$ & $U_{M/2}^{+}$ & orthogonal charge & \multirow{5}{*}{$\gamma_{\text{s}}$}
\tabularnewline
\cline{1-4} 
$\left\{\makecell[c]{|\widetilde{P}_{x}\rangle \\ |\widetilde{P}_{y}\rangle}\right\} $ & $\left\{\makecell[c]{-g_{M}s_{m\theta}|0\rangle _{m}+c_{m\theta}|1\rangle _{m} \\ g_{M}c_{m\theta}|0\rangle _{m}+s_{m\theta}|1\rangle _{m}}\right\}$ & \multirow{1}{*}{$R_{1}$} & \multirow{1}{*}{orthogonal momentum} & \tabularnewline
\cline{1-4} 
$\left\{\makecell[c]{|Q^{k}_{-}\rangle \\ |Q^{k}_{+}\rangle}\right\}$ & $\left\{\makecell[c]{-s_{km\theta}|0\rangle _{m} \\ c_{km\theta}|0\rangle _{m}}\right\}$ & \multirow{1}{*}{$R_{k}$} & \multirow{1}{*}{$n=0$ spin-$k$ } & \tabularnewline
\hline\noalign{\smallskip}
\hline
$\left|F\right\rangle $ & $\left|1\right\rangle _{m}$ & $U_{0}^{-}$ & rotation & \multirow{4}{*}{$\gamma_{\text{f}}$}\tabularnewline
\cline{1-4} 
$|\widetilde{F}\rangle $ & $\left(-1\right)^{m}|1\rangle _{m}$ & $U_{M/2}^{-}$ & orthogonal rotation & \tabularnewline
\cline{1-4} 
$\left\{\makecell[c]{|S^{k}_{-}\rangle \\ |S^{k}_{+}\rangle }\right\}$ & $\left\{\makecell[c]{-c_{km\theta}|1\rangle _{m} \\ -s_{km\theta}|1\rangle _{m}}\right\}$ & \multirow{1}{*}{$R_{k}$} & \multirow{1}{*}{$n=1$ spin-$k$} & \tabularnewline
\hline\noalign{\smallskip}
\end{tabular}
\caption{
Dihedral basis $\beta$ of the $2M$-dimensional vector space of electronic excitations that block diagonalizes the action $A$ of the dihedral group $\mathrm{D}_{2M}$ and diagonalizes the collision integral $\mathsf{W}$. The two-dimensional spin-$k$ subspaces should be understood understood to run between $k=2$ and $k=(M/2-1)$. We have also used the short-hands $s_\alpha\equiv\sin\alpha,  c_\alpha\equiv\cos\alpha$ and introduced the geometrical factor $g_M\equiv\tan(\mpi/M)/\sqrt{3}$ that specifies how differently the $n=0$ and $n=1$ Legendre excitations contribute to the momenta.
}
\label{table:dihedralB}
\end{center}
\end{table}

We emphasize that, although $\left.\mathcal{R}_2\right|_{\mathrm{D}_{2M}}$ is a two-dimensional representation of $\mathrm{D}_{2M}$ acting on a vector space of electronic excitation modes of the form $\{v_-,v_+\}$, when considering the action of $\mathrm{O}(2)$ on tensors it is not natural to write these modes as a rank-1 tensor $v_i$. From Eqs. (\ref{eq:R1R1}, \ref{eq:muR2}), we see that they instead transform naturally under the tensor action of $\mathrm{O}(2)$ when viewed as a traceless symmetric rank-2 tensor
\begin{equation}
    v_{ij}= \frac{v_{-}\mu_{ij}^{-}+v_{+}\mu_{ij}^{+}}{\sqrt{2}}=\frac{1}{\sqrt{2}}\left[\begin{array}{cc}
    -v_{+} & v_{-}\\
    v_{-} & v_{+}
    \end{array}\right]_{ij}.
\label{eq:spin2}
\end{equation}
If $M\geq 6$, the $\mathrm{D}_{2M}$-representation $\left.\mathcal{R}_2\right|_{\mathrm{D}_{2M}}=R_2$ is irreducible, and the rank-2 tensor (\ref{eq:spin2}) represents a single ``spin-2" hydrodynamic mode in our equations of motion. Conversely, for the square Fermi surface $M=4$ we see from Eqs. (\ref{eq:branching}, \ref{eq:R1R1}) that $\left.\mathcal{R}_2\right|_{\mathrm{D}_{8}}=U_2^+\oplus U_2^-$, with the tensor (\ref{eq:spin2}) splitting into one-dimensional irreducible subspaces
\begin{subequations}
\begin{align}
    \widetilde{N}_{ij} & =\mu_{ij}^{+}\widetilde{N},
    \\
    \widetilde{F}_{ij} & =\mu_{ij}^{-}\widetilde{F},
\end{align}
\end{subequations}
that correspond to two distinct spin-2 modes that are even and odd under reflections, respectively.

Similar considerations come into play when studying electronic excitation modes $\{v_-,v_+\}$ that transform under higher spin-$k$ representations $\left.\mathcal{R}_k\right|_{\mathrm{D}_{2M}}$. In particular, the rank-3 tensors that transform exclusively under the representation $\left.\mathcal{R}_3\right|_{\mathrm{D}_{2M}}$ are of the form
\begin{equation}
    v_{ijk}
    \equiv \frac{v_{-}\lambda_{ijk}^{-}+v_{+}\lambda_{ijk}^{+}}{\sqrt{2}},
    \label{eq:spin3}
\end{equation}
where the spin-3 basis tensors $\lambda^\pm_{ijk}$ are given by (\ref{eq:lambdaR3}). The largest $M$-gon we explicitly study is the $M=6$ hexagon, for which $\left.\mathcal{R}_3\right|_{\mathrm{D}_{12}}=U_3^+\oplus U_3^-$ is reducible, and the tensor (\ref{eq:spin3}) splits into one-dimensional irreducible subspaces
\begin{subequations}
\begin{align}
    \widetilde{N}_{ijk} & =\lambda_{ijk}^{+}\widetilde{N},
    \\
    \widetilde{F}_{ijk} & =\lambda_{ijk}^{-}\widetilde{F},
\end{align}
\end{subequations}
that correspond to two distinct spin-3 modes that are even and odd under reflections, respectively.

\end{appendix}

\bibliographystyle{unsrt}
\addcontentsline{toc}{section}{References}
\bibliography{soundbib}

\begin{thebibliography}{10}

\bibitem{molenkamp}
M.~J.~M. de~Jong and L.~W. Molenkamp.
\newblock ``Hydrodynamic electron flow in high-mobility wires",
  \href{http://journals.aps.org/prb/abstract/10.1103/PhysRevB.51.13389}{\textsl{Physical
  Review} \textbf{B51} 11389 (1995)},
  \href{http://arxiv.org/abs/cond-mat/9411067}{\texttt{arXiv:cond-mat/9411067}}.

\bibitem{bandurin}
D.~A.~Bandurin \emph{et al.}
\newblock ``Negative local resistance due to viscous electron backflow in
  graphene",
  \href{http://science.sciencemag.org/content/351/6277/1055}{\textsl{Science}
  \textbf{351} 1055 (2016)},
  \href{http://arxiv.org/abs/1509.04165}{\texttt{arXiv:1509.04165}}.

\bibitem{crossno}
J.~Crossno \emph{et al.}
\newblock ``Observation of the Dirac fluid and the breakdown of the
  Wiedemann-Franz law in graphene",
  \href{http://science.sciencemag.org/content/351/6277/1058}{\textsl{Science}
  \textbf{351} 1058 (2016)},
  \href{http://arxiv.org/abs/1509.04713}{\texttt{arXiv:1509.04713}}.

\bibitem{mackenzie}
P.~J.~W. Moll, P.~Kushwaha, N.~Nandi, B.~Schmidt, and A.~P. Mackenzie.
\newblock ``Evidence for hydrodynamic electron flow in $\mathrm{PdCoO}_2$",
  \href{http://science.sciencemag.org/content/351/6277/1061}{\textsl{Science}
  \textbf{351} 1061 (2016)},
  \href{http://arxiv.org/abs/1509.05691}{\texttt{arXiv:1509.05691}}.

\bibitem{levitov1703}
R.~Krishna~Kumar \emph{et al}.
\newblock ``Super-ballistic flow of viscous electron fluid through graphene
  constrictions", \href{https://doi.org/10.1038/nphys4240}{\textsl{Nature
  Physics} \textbf{13} 1182 (2017)},
  \href{http://arxiv.org/abs/1703.06672}{\texttt{arXiv:1703.06672}}.

\bibitem{felser}
J.~Gooth, F.~Menges, C.~Shekhar, V.~S\"uss, N.~Kumar, Y.~Sun, U.~Drechsler,
  R.~Zierold, C.~Felser, and B.~Gotsmann.
\newblock ``Electrical and thermal transport at the Planckian bound of
  dissipation in the hydrodynamic electron fluid of $\mathrm{WP}_2$",
  \href{http://arxiv.org/abs/1706.05925}{\texttt{arXiv:1706.05925}}.

\bibitem{bakarov}
E.~V. Levinson, G.~M. Gusev, A.~D. Levin, E.~V. Levinson, and A.~K. Bakarov.
\newblock ``Viscous electron flow in mesoscopic two-dimensional electron gas",
  \href{https://doi.org/10.1063/1.5020763}{\textsl{AIP Advances} \textbf{8}
  \texttt{025318} (2018)},
  \href{http://arxiv.org/abs/1802.09619}{\texttt{arXiv:1802.09619}}.

\bibitem{bandurin18}
D.~A. Bandurin, A.~V. Shytov, L.~S. Levitov, R.~K. Kumar, A.~I. Berdyugin,
  M.~Ben Shalom, I.~V. Grigorieva, A.~K. Geim, and G.~Falkovich.
\newblock ``Fluidity onset in graphene",
  \href{https://doi.org/10.1038/s41467-018-07004-4}{\textsl{Nature
  Communications} \textbf{9} \texttt{4533} (2018)},
  \href{http://arxiv.org/abs/1806.03231}{\texttt{arXiv:1806.03231}}.

\bibitem{gurzhi}
R.~N. Gurzhi.
\newblock ``Minimum of resistance in impurity-free conductors",
  \href{http://www.jetp.ac.ru/cgi-bin/e/index/e/17/2/p521?a=list}{\textsl{Journal
  of Experimental and Theoretical Physics} \textbf{17} 521 (1963)}.

\bibitem{lucasreview17}
A.~Lucas and K.~C. Fong.
\newblock ``Hydrodynamics of electrons in graphene",
  \href{http://arxiv.org/abs/1710.08425}{\texttt{arXiv:1710.08425}}.

\bibitem{hkms}
S.~A. Hartnoll, P.~K. Kovtun, M.~M\"uller, and S.~Sachdev.
\newblock ``Theory of the Nernst effect near quantum phase transitions in
  condensed matter, and in dyonic black holes",
  \href{http://journals.aps.org/prb/abstract/10.1103/PhysRevB.76.144502}{\textsl{Physical
  Review} \textbf{B76} \texttt{144502} (2007)},
  \href{http://arxiv.org/abs/0706.3215}{\texttt{arXiv:0706.3215}}.

\bibitem{andreev}
A.~V. Andreev, S.~A. Kivelson, and B.~Spivak.
\newblock ``Hydrodynamic description of transport in strongly correlated
  electron systems",
  \href{http://journals.aps.org/prl/abstract/10.1103/PhysRevLett.106.256804}{\textsl{Physical
  Review Letters} \textbf{106} \texttt{256804} (2011)},
  \href{http://arxiv.org/abs/1011.3068}{\texttt{arXiv:1011.3068}}.

\bibitem{succiturb}
M.~Mendoza, H.~J. Herrmann, and S.~Succi.
\newblock ``Preturbulent regimes in graphene flow",
  \href{http://journals.aps.org/prl/abstract/10.1103/PhysRevLett.106.156601}{\textsl{Physical
  Review Letters} \textbf{106} \texttt{156601} (2011)},
  \href{http://arxiv.org/abs/1201.6590}{\texttt{arXiv:1201.6590}}.

\bibitem{tomadin}
A.~Tomadin, G.~Vignale, and M.~Polini.
\newblock ``A Corbino disk viscometer for 2d quantum electron liquids",
  \href{http://journals.aps.org/prl/abstract/10.1103/PhysRevLett.113.235901}{\textsl{Physical
  Review Letters} \textbf{113} \texttt{235901} (2014)},
  \href{http://arxiv.org/abs/1401.0938}{\texttt{arXiv:1401.0938}}.

\bibitem{lucas3}
A.~Lucas, J.~Crossno, K.~C. Fong, P.~Kim, and S.~Sachdev.
\newblock ``Transport in inhomogeneous quantum critical fluids and in the Dirac
  fluid in graphene",
  \href{http://journals.aps.org/prb/abstract/10.1103/PhysRevB.93.075426}{\textsl{Physical
  Review} \textbf{B93} \texttt{075426} (2016)},
  \href{http://arxiv.org/abs/1510.01738}{\texttt{arXiv:1510.01738}}.

\bibitem{scaffidi}
T.~Scaffidi, N.~Nandi, B.~Schmidt, A.~P. Mackenzie, and J.~E. Moore.
\newblock ``Hydrodynamic electron flow and Hall viscosity",
  \href{https://doi.org/10.1103/PhysRevLett.118.226601}{\textsl{Physical Review
  Letters} \textbf{118} \texttt{226601} (2017)},
  \href{http://arxiv.org/abs/1703.07325}{\texttt{arXiv:1703.07325}}.

\bibitem{hartnoll1704}
A.~Lucas and S.~A. Hartnoll.
\newblock ``Resistivity bound for hydrodynamic bad metals",
  \href{http://arxiv.org/abs/1704.07384}{\texttt{arXiv:1704.07384}}.

\bibitem{polini}
I.~Torre, A.~Tomadin, A.~K. Geim, and M.~Polini.
\newblock ``Non-local transport and the hydrodynamic shear viscosity in
  graphene",
  \href{http://journals.aps.org/prb/abstract/10.1103/PhysRevB.92.165433}{\textsl{Physical
  Review} \textbf{B92} \texttt{165433} (2016)},
  \href{http://arxiv.org/abs/1508.00363}{\texttt{arXiv:1508.00363}}.

\bibitem{levitovhydro}
L.~Levitov and G.~Falkovich.
\newblock ``Electron viscosity, current vortices and negative nonlocal
  resistance in graphene",
  \href{http://www.nature.com/nphys/journal/v12/n7/full/nphys3667.html}{\textsl{Nature
  Physics} \textbf{12} 672 (2016)},
  \href{http://arxiv.org/abs/1508.00836}{\texttt{arXiv:1508.00836}}.

\bibitem{torre}
F.~M.~D. Pellegrino, I.~Torre, A.~K. Geim, and M.~Polini.
\newblock ``Electron hydrodynamics dilemma: whirlpools or no whirlpools",
  \href{http://journals.aps.org/prb/abstract/10.1103/PhysRevB.94.155414}{\textsl{Physical
  Review} \textbf{B94} \texttt{155414} (2016)},
  \href{http://arxiv.org/abs/1607.03726}{\texttt{arXiv:1607.03726}}.

\bibitem{levitov1806}
A.~Shytov, J.~F. Kong, G.~Falkovich, and L.~Levitov.
\newblock ``Particle collisions and negative nonlocal response of ballistic
  electrons",
  \href{https://doi.org/10.1103/PhysRevLett.121.176805}{\textsl{Physical Review
  Letters} \textbf{121} \texttt{176805} (2018)},
  \href{http://arxiv.org/abs/1806.09538}{\texttt{arXiv:1806.09538}}.

\bibitem{levitov1607}
H.~Guo, E.~Ilseven, G.~Falkovich, and L.~Levitov.
\newblock ``Higher-than-ballistic conduction of viscous electron flows",
  \href{http://www.pnas.org/content/114/12/3068.abstract}{\textsl{Proceedings
  of the National Academy of Sciences} \textbf{114} 3068 (2017)},
  \href{http://arxiv.org/abs/1607.07269}{\texttt{arXiv:1607.07269}}.

\bibitem{alekseev}
P.~S. Alekseev.
\newblock ``Negative magnetoresistance in viscous flow of two-dimensional
  electrons",
  \href{http://journals.aps.org/prl/abstract/10.1103/PhysRevLett.117.166601}{\textsl{Physical
  Review Letters} \textbf{117} \texttt{166601} (2016)}.

\bibitem{kwwest}
Q.~Shi, P.~D. Martin, Q.~A. Ebner, M.~A. Zudov, L.~N. Pfeiffer, and K.~W. West.
\newblock ``Colossal negative magnetoresistance in a two-dimensional electron
  gas",
  \href{http://journals.aps.org/prl/abstract/10.1103/PhysRevB.89.201301}{\textsl{Physical
  Review} \textbf{B89} \texttt{201301} (2014)}.

\bibitem{behnia}
X.~Lin, B.~Fauque, and K.~Behnia.
\newblock ``Scalable $T^2$ resistivity in a small single-component Fermi
  surface", \href{https://doi.org/10.1126/science.aaa8655}{\textsl{Science}
  \textbf{349} 945 (2015)},
  \href{http://arxiv.org/abs/1508.07812}{\texttt{arXiv:1508.07812}}.

\bibitem{stemmer}
E.~Mikheev, S.~Raghavan, J.~Y. Zhang, P.~B. Marshall, A.~P. Kajdos, L.~Balents,
  and S.~Stemmer.
\newblock ``Carrier density independent scattering rate in
  $\mathrm{SrTiO}_3$-based electron liquids",
  \href{https://doi.org/10.1038/srep20865}{\textsl{Scientific Reports}
  \textbf{6} \texttt{20865} (2016)},
  \href{http://arxiv.org/abs/1512.02294}{\texttt{arXiv:1512.02294}}.

\bibitem{lucasRFB}
A.~Lucas.
\newblock ``Kinetic theory of electronic transport in random magnetic fields",
  \href{http://arxiv.org/abs/1710.11141}{\texttt{arXiv:1710.11141}}.

\bibitem{ong2010}
K.~P. Ong, J.~Zhang, J.~S. Tse, and P.~Wu.
\newblock ``Origin of anisotropy and metallic behavior in delafossite
  $\mathrm{PdCoO}_2$",
  \href{http://doi.org/10.1103/PhysRevB.81.115120}{\textsl{Physical Review}
  \textbf{B81} \texttt{115120} (2010)}.

\bibitem{mackenzie16}
A.~P. Mackenzie.
\newblock ``The properties of ultrapure delafossite metals",
  \href{http://doi.org/10.1088/1361-6633/aa50e5}{\textsl{Reports on Progress in
  Physics} \textbf{80} \texttt{032501} (2017)},
  \href{http://arxiv.org/abs/1612.04948}{\texttt{arXiv:1612.04948}}.

\bibitem{lucas1810}
S.~Grozdanov, A.~Lucas, and N.~Poovuttikul.
\newblock ``Holography and hydrodynamics with weakly broken symmetries",
  \href{http://arxiv.org/abs/1810.10016}{\texttt{arXiv:1810.10016}}.

\bibitem{avron}
J.~E. Avron.
\newblock ``Odd viscosity",
  \href{https://link.springer.com/article/10.1023/A:1023084404080}{\textsl{Journal
  of Statistical Physics} \textbf{92} 543 (1998)},
  \href{http://arxiv.org/abs/physics/9712050}{\texttt{arXiv:physics/9712050}}.

\bibitem{hartnoll1}
S.~A. Hartnoll.
\newblock ``Theory of universal incoherent metallic transport",
  \href{http://www.nature.com/nphys/journal/v11/n1/full/nphys3174.html}{\textsl{Nature
  Physics} \textbf{11} 54 (2015)},
  \href{http://arxiv.org/abs/1405.3651}{\texttt{arXiv:1405.3651}}.

\bibitem{usui}
H.~Usui \emph{et al.}
\newblock ``Hidden kagome-lattice picture and origin of high conductivity in
  delafossite $\mathrm{PdCoO}_2$",
  \href{http://arxiv.org/abs/1812.07213}{\texttt{arXiv:1812.07213}}.

\bibitem{dgg18}
M.~D. Bachmann, A.~L. Sharpe, A.~W. Barnard, C.~Putzke, M.~K\"onig, S.~Khim,
  D.~Goldhaber-Gordon, A.~P. Mackenzie, and P.~J.~W. Moll.
\newblock ``Super-geometric electron focusing on the hexagonal Fermi surface of
  $\mathrm{PdCoO}_2$",
  \href{http://arxiv.org/abs/1902.03769}{\texttt{arXiv:1902.03769}}.

\bibitem{mazin}
M.~D. Johannes and I.~I. Mazin.
\newblock ``Fermi surface nesting and the origin of charge density waves in
  metals", \href{https://doi.org/10.1103/PhysRevB.77.165135}{\textsl{Physical
  Review} \textbf{B77} \texttt{165135} (2008)},
  \href{http://arxiv.org/abs/0708.1744}{\texttt{arXiv:0708.1744}}.

\bibitem{kamenev}
A.~Kamenev.
\newblock \emph{Field Theory of Non-Equilibrium Systems}
  \href{https://www.amazon.com/Field-Theory-Non-Equilibrium-Systems-Kamenev/dp/0521760828/ref=sr_1_1?ie=UTF8&qid=1478548419&sr=8-1&keywords=field+theory+of+non-equilibrium+systems}{(Cambridge
  University Press, 2011)}.

\bibitem{hartnoll1705}
A.~Lucas and S.~A. Hartnoll.
\newblock ``Kinetic theory of transport for inhomogeneous electron fluids",
  \href{http://arxiv.org/abs/1706.04621}{\texttt{arXiv:1706.04621}}.

\bibitem{ledwith1}
P.~Ledwith, H.~Guo, and L.~Levitov.
\newblock ``Fermion collisions in two dimensions",
  \href{http://arxiv.org/abs/1708.01915}{\texttt{arXiv:1708.01915}}.

\bibitem{ledwith2}
P.~Ledwith, H.~Guo, A.~V. Shytov, and L.~Levitov.
\newblock ``Head-on collisions and scale-dependent viscosity in two-dimensional
  electron systems",
  \href{http://arxiv.org/abs/1708.02376}{\texttt{arXiv:1708.02376}}.

\bibitem{foster}
M.~S. Foster and I.~L. Aleiner.
\newblock ``Slow imbalance relaxation and thermoelectric transport in
  graphene",
  \href{http://journals.aps.org/prb/abstract/10.1103/PhysRevB.79.085415}{\textsl{Physical
  Review} \textbf{B79} \texttt{085415} (2009)},
  \href{http://arxiv.org/abs/0810.4342}{\texttt{arXiv:0810.4342}}.

\bibitem{landau}
L.D. Landau and E.M. Lifshitz.
\newblock \emph{Fluid Mechanics}
  \href{https://www.amazon.com/Fluid-Mechanics-Second-Theoretical-Physics/dp/0750627670/ref=sr_1_1?ie=UTF8&qid=1478217116&sr=8-1&keywords=fluid+mechanics+landau}{(Butterworth
  Heinemann, $2^{\mathrm{nd}}$ ed., 1987)}.

\bibitem{link}
J.~M. Link, B.~N. Narozhny, E.~I. Kiselev, and J.~Schmalian.
\newblock ``Out-of-bounds hydrodynamics in anisotropic Dirac fluids",
  \href{https://doi.org/10.1103/PhysRevLett.120.196801}{\textsl{Physical Review
  Letters} \textbf{120} \texttt{196801} (2018)},
  \href{http://arxiv.org/abs/1708.02759}{\texttt{arXiv:1708.02759}}.

\bibitem{yarom}
K.~Jensen, M.~Kaminski, P.~Kovtun, R.~Meyer, A.~Ritz, and A.~Yarom.
\newblock ``Parity-violating hydrodynamics in 2+1 dimensions",
  \href{https://doi.org/10.1007/JHEP05(2012)102}{\textsl{Journal of High Energy
  Physics} \textbf{05} \texttt{102} (\textbf{2012})},
  \href{http://arxiv.org/abs/1112.4498}{\texttt{arXiv:1112.4498}}.

\bibitem{landauvol7}
L.~D. Landau and E.M. Lifshitz.
\newblock \emph{Theory of Elasticity}
  \href{https://www.amazon.com/Theory-Elasticity-7-Theoretical-Physics/dp/075062633X/ref=sr_1_2?keywords=theory+of+elasticity&qid=1551322028&s=gateway&sr=8-2}{(Butterworth
  Heinemann, $3^{\mathrm{rd}}$ ed., 1986)}.

\bibitem{ericksen}
J.~L. Ericksen.
\newblock ``Conservation laws for liquid crystals",
  \href{https://sor.scitation.org/doi/10.1122/1.548883}{\textsl{Transactions of
  the Society of Rheology} \textbf{5} 23 (1961)}.

\bibitem{leslie}
F.~M. Leslie.
\newblock ``Some constitutive equations for liquid crystals",
  \href{https://doi.org/10.1007/BF00251810}{\textsl{Archive for Rational
  Mechanics and Analysis} \textbf{28} 265 (1968)}.

\bibitem{pershanprl}
D.~Forster, T.~C. Lubensky, P.~C. Martin, J.~Swift, and P.~S. Pershan.
\newblock ``Hydrodynamics of liquid crystals",
  \href{https://doi.org/10.1103/PhysRevLett.26.1016}{\textsl{Physical Review
  Letters} \textbf{26} 1016 (1971)}.

\bibitem{pershan}
P.~C. Martin, O.~Parodi, and P.~S. Pershan.
\newblock ``Unified hydrodynamic theory for crystals, liquid crystals, and
  normal fluids",
  \href{https://doi.org/10.1103/PhysRevA.6.2401}{\textsl{Physical Review}
  \textbf{A6} 2401 (1972)}.

\bibitem{sykes}
J.~Sykes and G.~A. Brooker.
\newblock ``The transport coefficients of a Fermi liquid",
  \href{https://doi.org/10.1016/0003-4916(70)90002-3}{\textsl{Annals of
  Physics} \textbf{56} 1 (1970)}.

\bibitem{lucas1801}
A.~Lucas and S.~Das Sarma.
\newblock ``Electronic sound modes and plasmons in hydrodynamic two-dimensional
  metals", \href{https://doi.org/10.1103/PhysRevB.97.115449}{\textsl{Physical
  Review} \textbf{B97} \texttt{115449} (2018)},
  \href{http://arxiv.org/abs/1801.01495}{\texttt{arXiv:1801.01495}}.

\bibitem{lucasplasma}
A.~Lucas.
\newblock ``Sound waves and resonances in electron-hole plasma",
  \href{http://journals.aps.org/prb/abstract/10.1103/PhysRevB.93.245153}{\textsl{Physical
  Review} \textbf{B93} \texttt{245153} (2016)},
  \href{http://arxiv.org/abs/1604.03955}{\texttt{arXiv:1604.03955}}.

\bibitem{alekseev18}
P.~S. Alekseev and M.~A. Semina.
\newblock ``Ballistic flow of two-dimensional interacting electrons",
  \href{https://doi.org/10.1103/PhysRevB.98.165412}{\textsl{Physical Review}
  \textbf{B98} \texttt{165412} (2018)},
  \href{http://arxiv.org/abs/1801.02879}{\texttt{arXiv:1801.02879}}.

\bibitem{alekseev19}
P.~S. Alekseev and M.~A. Semina.
\newblock ``The Hall effect in ballistic flow of two-dimensional interacting
  particles",
  \href{http://arxiv.org/abs/1903.07925}{\texttt{arXiv:1903.07925}}.

\bibitem{haug14}
L.~Bockhorn, I.~V. Gornyi, D.~Schuh, C.~Reichl, W.~Wegscheider, and R.~J. Haug.
\newblock ``Magnetoresistance induced by rare strong scatterers in a high
  mobility two-dimensional electron gas",
  \href{https://doi.org/10.1103/PhysRevB.90.165434}{\textsl{Physical Review}
  \textbf{B90} \texttt{165434} (2014)},
  \href{http://arxiv.org/abs/1401.7940}{\texttt{arXiv:1401.7940}}.

\bibitem{bakarov1810}
G.~M. Gusev, A.~D. Levin, E.~V. Levinson, and A.~K. Bakarov.
\newblock ``Viscous transport and Hall viscosity in a two-dimensional electron
  system", \href{https://doi.org/10.1103/PhysRevB.98.161303}{\textsl{Physical
  Review} \textbf{B98} \texttt{161303} (2018)},
  \href{http://arxiv.org/abs/1810.06520}{\texttt{arXiv:1810.06520}}.

\bibitem{coulter}
J.~Coulter, R.~Sundararaman, and P.~Narang.
\newblock ``Microscopic origins of hydrodynamic transport in type-II Weyl
  semimetal $\mathrm{WP}_2$",
  \href{https://doi.org/10.1103/PhysRevB.98.115130}{\textsl{Physical Review}
  \textbf{B98} \texttt{115130} (2018)},
  \href{http://arxiv.org/abs/1804.06310}{\texttt{arXiv:1804.06310}}.

\bibitem{zaanen}
D.~Forcella, J.~Zaanen, D.~Valentinis, and D.~van~der Marel.
\newblock ``Electromagnetic properties of viscous charged fluids",
  \href{http://journals.aps.org/prl/abstract/10.1103/PhysRevB.90.035143}{\textsl{Physical
  Review} \textbf{B90} \texttt{035143} (2014)},
  \href{http://arxiv.org/abs/1406.1356}{\texttt{arXiv:1406.1356}}.

\bibitem{tung}
W-K. Tung.
\newblock \emph{Group Theory in Physics},
  \href{https://www.amazon.com/Group-Theory-Physics-Wu-Ki-Tung/dp/9971966565/ref=sr_1_fkmrnull_1?keywords=tung+group+theory+in+physics&qid=1551391272&s=gateway&sr=8-1-fkmrnull}{(World
  Scientific, 1985)}.

\end{thebibliography}

\end{document}